\documentclass[trackchanges, twocolumn]{aastex7}

\usepackage{macro_id}
\usepackage{amsmath}

\newcommand{\NEF}{NASA Einstein Fellow}
\newcommand{\msun}{M$_{\odot}$}

\begin{document}

\title{The JADES Transient Survey II: Volumetric Supernova Rates out to $z$\,$\sim$\,5}

\author[0000-0002-4781-9078]{Christa DeCoursey}
\affiliation{Steward Observatory, University of Arizona, 933 N. Cherry Ave, Tucson, AZ 85721, USA}
\email{cndecoursey@arizona.edu}

\author[0000-0001-5517-6335] {Christian Vassallo}
\affiliation{Tuorla Observatory, Department of Physics and Astronomy, University of Turku, 20014 Turku, Finland}
\email{clvass@utu.fi}

\author[0000-0003-2238-1572]{Louis-Gregory Strolger} 
\affiliation{Space Telescope Science Institute, 3700 San Martin Drive, Baltimore, MD 21218, USA}
\email{strolger@stsci.edu}

\author[0000-0002-2361-7201]{Justin D. R. Pierel} 
\altaffiliation{\NEF}
\affiliation{Space Telescope Science Institute, 3700 San Martin Drive, Baltimore, MD 21218, USA}
\email{jpierel@stsci.edu}

\author[0000-0003-1344-9475]{Eiichi Egami}
\affiliation{Steward Observatory, University of Arizona, 933 N. Cherry Ave, Tucson, AZ 85721, USA}
\email{egami@arizona.edu}

\author[0000-0001-7497-2994] {Seppo Mattila}
\affiliation{Department of Physics and Astronomy, FI-20014 University of Turku, Finland}
\affiliation{School of Sciences, European University Cyprus, Diogenes Street, Engomi, 1516 Nicosia, Cyprus}
\email{sepmat@utu.fi}

\author[0000-0002-4410-5387]{Armin Rest}
\affiliation{Space Telescope Science Institute, 3700 San Martin Drive, Baltimore, MD 21218, USA}
\affiliation{William H. Miller III Department of Physics \& Astronomy, Johns Hopkins University, 3400 N Charles St, Baltimore, MD 21218, USA}
\email{arest@stsci.edu}

\author[0000-0003-4263-2228]{David A. Coulter}
\affiliation{Space Telescope Science Institute, 3700 San Martin Drive, Baltimore, MD 21218, USA}
\affiliation{William H. Miller III Department of Physics \& Astronomy, Johns Hopkins University, 3400 N Charles St, Baltimore, MD 21218, USA}
\email{dcoulter@stsci.edu}

\author[0000-0002-8651-9879]{Andrew J.\ Bunker}
\affiliation{Department of Physics, University of Oxford, Denys Wilkinson Building, Keble Road, Oxford OX1 3RH, UK}
\email{andy.bunker@physics.ox.ac.uk}

\author[0000-0002-0450-7306]{Alex J.\ Cameron}
\affiliation{Cosmic Dawn Center (DAWN), Copenhagen, Denmark}
\affiliation{Niels Bohr Institute, University of Copenhagen, Jagtvej 128, DK-2200, Copenhagen, Denmark}
\email{alex.cameron@nbi.ku.dk}

\author[0000-0002-7566-6080]{James M.\ DerKacy}
\affiliation{Space Telescope Science Institute, 3700 San Martin Drive, Baltimore, MD 21218, USA}
\email{jderkacy@stsci.edu}

\author[0000-0002-2929-3121]{Daniel J.\ Eisenstein}
\affiliation{Center for Astrophysics $|$ Harvard \& Smithsonian, 60 Garden St., Cambridge MA 02138 USA}
\email{deisenstein@cfa.harvard.edu}

\author[0000-0003-0209-674X]{Michael Engesser}
\affiliation{Space Telescope Science Institute, 3700 San Martin Drive, Baltimore, MD 21218, USA}
\email{mengesser@stsci.edu}

\author[0000-0003-2238-1572]{Ori D. Fox}
\affiliation{Space Telescope Science Institute, 3700 San Martin Drive, Baltimore, MD 21218, USA}
\email{ofox@stsci.edu}

\author[0000-0001-6395-6702]{Sebastian Gomez}
\affiliation{Department of Astronomy, The University of Texas at Austin, 2515 Speedway, Stop C1400, Austin, TX 78712, USA}
\email{sebastian.gomez@austin.utexas.edu}

\author[0000-0002-5060-1379]{Massimo Griggio}
\affiliation{Space Telescope Science Institute, 3700 San Martin Drive, Baltimore, MD 21218, USA}
\email{mgriggio@stsci.edu}

\author[0000-0003-4565-8239]{Kevin Hainline}
\affiliation{Steward Observatory, University of Arizona, 933 N. Cherry Ave, Tucson, AZ 85721, USA}
\email{kevinhainline@arizona.edu}

\author[0000-0002-8543-761X]{Ryan Hausen}
\affiliation{William H. Miller III Department of Physics \& Astronomy, Johns Hopkins University, 3400 N Charles St, Baltimore, MD 21218, USA}
\email{rhausen@ucsc.edu}

\author[0000-0001-7673-2257]{Zhiyuan Ji}
\affiliation{Steward Observatory, University of Arizona, 933 N. Cherry Ave, Tucson, AZ 85721, USA}
\email{zhiyuanji@arizona.edu}

\author[0000-0002-9280-7594]{Benjamin D.\ Johnson}
\affiliation{Center for Astrophysics $|$ Harvard \& Smithsonian, 60 Garden St., Cambridge MA 02138 USA}
\email{benjamin.johnson@cfa.harvard.edu}

\author[0000-0002-4985-3819]{Roberto Maiolino}
\affiliation{Kavli Institute for Cosmology, University of Cambridge, Madingley Road, Cambridge, CB3 0HA, UK}
\affiliation{Cavendish Laboratory, University of Cambridge, 19 JJ Thomson Avenue, Cambridge, CB3 0HE, UK}
\affiliation{Department of Physics and Astronomy, University College London, Gower Street, London WC1E 6BT, UK}
\email{rm665@cam.ac.uk}

\author[0000-0003-1169-1954]{Takashi J. Moriya}
\affiliation{National Astronomical Observatory of Japan, National Institutes of Natural Sciences, 2-21-1 Osawa, Mitaka, Tokyo 181-8588, Japan}
\affiliation{Graduate Institute for Advanced Studies, SOKENDAI, 2-21-1 Osawa, Mitaka, Tokyo 181-8588, Japan}
\affiliation{School of Physics and Astronomy, Monash University, Clayton, Victoria 3800, Australia}
\email{takashi.moriya@nao.ac.jp}

\author[0000-0002-4271-0364]{Brant Robertson}
\affiliation{Department of Astronomy and Astrophysics, University of California, Santa Cruz, 1156 High Street, Santa Cruz, CA 95064, USA}
\email{brant@ucsc.edu}

\author[0000-0002-2798-2943]{Koji Shukawa}
\affiliation{Space Telescope Science Institute, 3700 San Martin Drive, Baltimore, MD 21218, USA}
\affiliation{William H. Miller III Department of Physics \& Astronomy, Johns Hopkins University, 3400 N Charles St, Baltimore, MD 21218, USA}
\email{kshukawa@stsci.edu}

\author[0000-0003-2445-3891]{Matthew R. Siebert}
\affiliation{Space Telescope Science Institute, 3700 San Martin Drive, Baltimore, MD 21218, USA}
\email{msiebert@stsci.edu}

\author[0000-0002-4622-6617]{Fengwu Sun}
\affiliation{Center for Astrophysics $|$ Harvard \& Smithsonian, 60 Garden St., Cambridge MA 02138 USA}
\email{fengwu.sun@cfa.harvard.edu}

\author[0000-0002-8224-4505]{Sandro Tacchella}
\affiliation{Kavli Institute for Cosmology, University of Cambridge, Madingley Road, Cambridge, CB3 0HA, UK}
\affiliation{Cavendish Laboratory, University of Cambridge, 19 JJ Thomson Avenue, Cambridge, CB3 0HE, UK}
\email{st578@cam.ac.uk}

\author[0000-0003-2919-7495]{Christina C.\ Williams}
\affiliation{NSF National Optical-Infrared Astronomy Research Laboratory, 950 North Cherry Avenue, Tucson, AZ 85719, USA}
\email{christina.williams@noirlab.edu}

\author[0000-0001-9262-9997]{Christopher N.\ A.\ Willmer}
\affiliation{Steward Observatory, University of Arizona, 933 N. Cherry Ave, Tucson, AZ 85721, USA}
\email{cnaw@as.arizona.edu}

\author[0000-0002-0632-8897]{Yossef Zenati}
\affiliation{{Astrophysics Research Center of the Open University (ARCO), Department of Natural Sciences, Ra’anana 4353701, Israel}}
\affiliation{William H. Miller III Department of Physics \& Astronomy, Johns Hopkins University, 3400 N Charles St, Baltimore, MD 21218, USA}
\affiliation{Space Telescope Science Institute, 3700 San Martin Drive, Baltimore, MD 21218, USA}
\email{yzenati1@jhu.edu}


\begin{abstract}

The JADES Transient Survey (JTS) identified 83 supernova (SN) candidates in the JADES Deep Field, a $\sim$25 arcmin$^2$ region with deep ($\sim$30~mag) multi-band, multi-epoch JWST/NIRCam coverage. We use this sample to derive the first volumetric core-collapse (CC) SN and Type\,Ia (SN\,Ia) rates in the $z$\,$\sim$\,2--5 range. Many of these SNe are photometrically classified from single-epoch photometry (i.e., single spectral energy distributions (SEDs)), so we simulate and classify $\sim$23,000 CC\,SN and SN\,Ia mock SEDs over 0.7\,$\leq$\,$z$\,$\leq$\,5 to quantify single-SED classification accuracy as a function of redshift. We report consistent rates for two samples: (1) the full JTS sample, including single-SED classifications, and (2) the ``gold" sample, restricted to sources classified spectroscopically or with multi-epoch light curves. In units of 10$^{-4}$ CC\,SNe yr$^{-1}$ Mpc$^{-3}$, the full sample CC\,SN rates are 6.5$^{+2.3(0.5)}_{-1.7(0.3)}$ at 2.05\,$<$\,$z$\,$\leq$\,2.78 and 4.1$^{+1.4(1.0)}_{-1.1(0.3)}$ at 2.78\,$<$\,$z$\,$\leq$\,5.20, broadly consistent with the expectations from the galaxy luminosity-based measurements of the cosmic star formation rate density. The rates measured for both samples tentatively exhibit the predicted decline beyond cosmic noon, providing the first direct observational indication of this behavior. A companion paper, \citet{vassallo2026}, presents a more detailed comparison. We measure a full sample SN\,Ia rate of 0.4$^{+0.3(0.1)}_{-0.2(0.2)}$\,$\times$\,10$^{-4}$ SNe\,Ia yr$^{-1}$ Mpc$^{-3}$ at 1.91\,$<$\,$z$\,$\leq$\,3.60. Future high-$z$ SN surveys with JWST and the Roman Space Telescope will expand these samples and provide more robust constraints on SN rates in the high-$z$ Universe.
\end{abstract}

\keywords{\uat{Supernovae}{1668} --- \uat{Core-collapse supernovae}{304}}


\section{Introduction} 
\label{sec:intro}

Supernovae (SNe) are separated into two broad classes based on distinct physical origins and spectrophotometric properties: core-collapse SNe (CC\,SNe) and thermonuclear SNe (i.e., SNe\,Ia; \citealt{filippenko1997}). SNe\,Ia are the thermonuclear explosions of white dwarfs (WDs) in multi-star systems (see \citealt{maoz2014} and \citealt{jha2019} for reviews), whereas CC\,SNe are stellar explosions that occur when massive stars' ($\gtrsim$ 8\msun) iron cores collapse under the force of gravity. 

\subsection{Core-Collapse Supernova Rates} \label{subsec:intro_ccsne}
CC\,SNe are divided into two main types: hydrogen-rich Type II (SNe\,II) and hydrogen-free Type I (SNe\,I). SNe\,I are further divided into SNe\,Ib, characterized by a lack of hydrogen in their spectra, and SNe\,Ic, which exhibit neither hydrogen nor helium in their spectra. SNe\,II are further divided into subtypes based on their unique spectral properties and light curve evolutionary tracks \citep{gal-yam2017}. For example, SNe\,IIP are distinguished by the ``plateau" phase of their light curve evolution, whereas SNe\,IIL light curves exhibit a ``linear" decline \citep{barbon1979}. SNe\,IIn, on the other hand, have diverse light curves powered by interaction with surrounding circumstellar material and exhibit ``narrow" hydrogen lines in their spectra \citep{schlegel1990, kiewe2012, taddia2013, Ransome25}. SNe\,IIb are characterized by early hydrogen features in their spectra that later yield to more dominant helium features \citep[e.g.,][]{woosley1994}. Additionally, superluminous classes of Type I (SLSN-I) and Type II (SLSN-II) SNe have been identified based on their high luminosities \citep{Chen23, Gomez24, Hiramatsu24}. 

Despite the wide variety of spectrophotometric properties exhibited by CC\,SNe, their progenitors all have significantly shorter lifetimes than the timescale of galaxy-scale changes. This means that CC\,SNe effectively trace instantaneous star formation, providing an independent measure of the cosmic star formation rate density (SFRD) history \citep{dahlen1999, madau2014}. 

There are numerous previous studies that have presented volumetric CC\,SN rate measurements, R$_\mathrm{CC}$, at $z$\,$<$\,1 \citep[e.g., ][]{cappellaro1999, dahlen2004, cappellaro2005, botticella2008, bazin2009, graur2011, li2011_rates, botticella2012, mattila2012, melinder2012, taylor2014, cappellaro2015, graur2015, perley2020, frohmaier2021, ma2025, pessi2025}. \citet{dahlen2012} used a sample of 45 CC\,SNe discovered with the Hubble Space Telescope (HST) Advanced Camera for Surveys (ACS) to extend CC\,SN rates out to $z$\,$\sim$\,1 for the first time. After correcting for the fraction of missing SNe in dust-obscured galaxies following \citet{mattila2012}, their CC\,SN rates showed good agreement with expectations based on the independently-measured cosmic SFRD. However, their CC\,SN rates were dominated by large statistical and systematic uncertainties. 

\citet{strolger2015} presented the first measure of CC\,SN rates out to $z$\,$\sim$\,2.5 using the Cosmic Assembly Near-infrared Deep Extragalactic Legacy Survey (CANDELS) and Cluster Lensing And Supernova
survey with Hubble (CLASH) data \citep{grogin2011, koekemoer2011, postman2012}. Their sample included $\sim$44 CC\,SNe spread across six redshift bins within 0.1\,$<$\,$z$\,$<$\,2.5.
Combining their rates with the literature,
they traced a comprehensive CC\,SN rate history with statistical uncertainties
at or below systematic uncertainties. 

\citet{petrushevska2016} used the HAWK-I instrument on the Very Large Telescope to probe CC\,SN rates out to $z$\,$=$\,2.9. Their study harnessed the power of gravitational lensing to discover 5 photometrically-classified CC\,SNe at 0.671\,$\leq$\,$z$\,$\leq$\,1.703 in the Abell 1689 field, although their survey was sensitive to the brightest CC\,SN subtypes out to $z$\,$\sim$\,3. Since their most distant CC\,SN detection was $z$\,$=$\,1.703, they could only set CC\,SN rate upper limits at $z$\,$\sim$\,2--3. 
While their results agree with previous $z$\,$>$\,1 CC\,SN rate measurements, they have very large uncertainties dominated by Poisson statistics due to the small number of CC\,SNe in their sample.

The past CC\,SN rate literature has shown that, within a large uncertainty window, observed CC\,SN rates agree with expectations based on cosmic SFRD history up to the HST SN redshift frontier ($z$\,$\sim$\,2.5). However, in order to understand whether CC\,SN rates trace the full cosmic SFRD history as expected, we must compute CC\,SN rates out to higher redshifts. The cosmic SFRD declines beyond cosmic noon \citep{madau2014}, and it is imperative to observationally verify whether CC\,SN rates follow a similar trend. Discovering disagreement between high-$z$ CC\,SN rates and expectations based on cosmic SFRD history may signal redshift evolution in properties such as mean SN obscuration and initial mass range of their progenitors. This will help us to build a better understanding of how massive stellar populations evolve with cosmic time.


\subsection{Type Ia Supernova Rates} \label{subsec:intro_snia}
SN\,Ia rates have also been studied extensively at $z$\,$\lesssim$\,1 using ground-based facilities \citep[e.g.,][]{cappellaro1999, hardin2000, pain2002, madgwick2003, strolger2003, tonry2003, blanc2004, mannucci2005, neill2006, botticella2008, horesh2008, dilday2010,rodney2010,MennekensN+10,li2011_rates, melinder2012, perrett2012, graur2013, okumura2014, cappellaro2015, MaozD+18, HallakounN_MaozD19, desai2024, desai2026}. 
An SN\,Ia occurs when a WD in a multi-star system explodes via thermonuclear runaway (see \citealt{hillebrandt2000,RuiterA_SeitenzahlI25} for a review). Two main models are currently under debate regarding the identity of the companion in the multi-star system: the single-degenerate (SD) scenario and double-degenerate (DD) scenario. In the SD scenario, the companion star is a main sequence or evolved star that feeds the WD via Roche lobe overflow or stellar winds \citep{whelan1973, nomoto1982}. In the DD scenario, the companion is a second WD whose orbit loses momentum to gravitational wave emission and eventually merges with the primary WD \citep{iben1984, webbink1984,Pakmor12, KashyapR+15}. 

By comparing SN\,Ia rates to the cosmic SFRD history, one can construct the ``delay-time distribution" (DTD) between a hypothetical burst of star formation and subsequent SN\,Ia explosions \citep[e.g.,][]{strolger2020, wiseman2021}. The SD and DD scenarios are expected to exhibit different DTDs, so constructing DTDs based on observed SN\,Ia rates can possibly distinguish between the SD and DD scenarios.

The GOODS+PANS survey \citep{dahlen2004, dahlen2008, kuznetsova2008} and Cluster Supernova Survey (CSS) of the Supernova Cosmology Project \citep{barbary2012} used HST/ACS to measure SN\,Ia rates out to $z$\,$\sim$\,1.5. They found that the SN\,Ia rate peaks around $z$\,$\sim$\,1.2 and declines at higher redshift. On the contrary, \citet{poznanski2007_iarates} and \citet{graur2011} measured the SN\,Ia rate out to similar redshift with the Suprime-Cam on the Subaru Telescope and did not see any high-$z$ decline in SN\,Ia rate. However, their measurements and uncertainties are within the errors of the GOODS+PANS and CSS SN\,Ia rate measurements.

\citet{graur2014} measured SN\,Ia rates out to $z$\,$\sim$\,1.8 and placed the first upper limits on SN\,Ia rates out to $z$\,$\sim$\,2.4 with a sample of $\sim$13 SN\,Ia discovered in the HST CLASH data \citep{postman2012}. Their measurements are consistent within the uncertainties of the previous $z$\,$>$\,1 HST and Subaru measurements \citep{poznanski2007_iarates, dahlen2008, graur2011}. \citet{rodney2014} performed the first SN\,Ia rate measurement out to $z$\,$\sim$\,2.5 with a sample of 65 SNe\,Ia from the HST CANDELS SN program \citep{grogin2011, koekemoer2011}. While their rates are 1--2$\sigma$ lower than other measurements at $z$\,$=$\,1--1.50, their results are consistent with previous measurements at $z$\,$=$\,1.50--2. They provide the first SN\,Ia rate measurement at $z$\,$>$\,2, although there is only one $z$\,$>$\,2 object in their sample. Their SN\,Ia rate at $z$\,$=$\,2--2.50 shows a decline relative to the rate at $z$\,$=$\,1.50--2, but the decline is smaller than the uncertainties. Thus, they can only conclude that SN\,Ia rates steadily rise to $z$\,$\sim$\,1.2 and either steadily decline or flatten at $z$\,$>$1.2.

Despite the immense investment of HST and Subaru in the targeting high-$z$ SNe\,Ia, SN\,Ia rates remain highly uncertain at $z$\,$\gtrsim$\,1. As detailed in \citet{rodney2014}, obtaining a more robust measurement of the $z$\,$>$\,1 SN\,Ia rate function will place tighter constraints on DTD models. These constraints can help to determine the dominant progenitor channel (SD or DD) for SNe\,Ia. It is therefore critical for our understanding of SN\,Ia progenitor systems to build a larger sample of $z$\,$>$\,1 SNe\,Ia and compute well-constrained high-$z$ SN\,Ia rates.


\subsection{The JADES Transient Survey} \label{subsec:intro_jts}
The launch of the James Webb Space Telescope (JWST; \citealt{gardner2023}), with its high spatial resolution and infrared sensitivity, has pushed the transient redshift frontier to $z$\,$\sim$\,5 \citep{decoursey2025_z5, coulter2026_eos, zhuang2026_z5}. There have been many searches for high-$z$ transient/variable sources with JWST data \citep[e.g.,][]{decoursey2023_goodsn, decoursey2023_origins, coulter2024, decoursey2024, coulter2025b_c3d, decoursey2025_jts, decoursey2025_magnif, sammut2025a, sammut2025b, stone2025, tee2025, yan2025, fox2026, zhuang2026_nexus}. The deepest systematic JWST transient survey to date was conducted using data from the JWST Advanced Deep Extragalactic Survey (JADES; \citealt{bunker2020, rieke2020, eisenstein2023}). The JADES program obtained two sets of deep ($\sim$30~mag) 9-band JWST/Near-Infrared Camera (NIRCam) data covering a $\sim$25 arcmin$^2$ portion of the Great Observatories Origins Deep Survey-South (GOODS-S) field \citep{rieke2023}. The two sets of NIRCam images were separated by $\sim$1 observer-frame year, enabling a search for high-$z$ ($z$\,$>$\,2) transients. 

The JADES Transient Survey (JTS) discovered 79 robust and 4 marginal SN candidates out to $z$\,$\approx$\,4.82 \citep{decoursey2025_jts}. Nearly half of the sample ($\sim$38 SNe) is at $z$\,$\gtrsim$\,2, demonstrating JWST's ability to efficiently discover distant SNe. 
Here, we use the JTS SN sample to compute volumetric CC\,SN and SN\,Ia rates out to $z$\,$\sim$\,5 for the first time. \citet{decoursey2025_jts} performed preliminary photometric classification for the JTS sample. We have since improved the classification scheme and present updated photometric classifications. This is essential to distinguish between SNe\,Ia and CC\,SNe. The majority of the JTS SNe were observed with only one light-curve epoch (i.e., one set of 9-band reference and science images), meaning that in these cases, we only have one observed SN spectral energy distribution (SED) to inform our classifications. This makes photometric classification challenging due to a lack of phase constraints. 

In order to quantitatively characterize the accuracy of our classifier when presented with single-epoch SN SEDs, we generated $\sim$23,000 mock SN SEDs with the 9-band JTS filter set and passed them through the classifier. We created mock SN SEDs for several SN subtypes at a variety of redshifts, phases, peak B-band magnitudes (M$_\mathrm{B}$), and color excesses (E(B$-$V)). This is the first study of its kind performed using high-$z$ SNe and JWST/NIRCam filters, in contrast to previous work at lower redshifts \citep[e.g.,][]{poznanski2007_classification}. We constructed CC\,SN vs SN\,Ia confusion matrices as a function of redshift to quantitatively assess our classifier's accuracy.

The primary aim of this paper is to measure volumetric CC\,SN and SN\,Ia rates from the JTS, including a detailed evaluation of classification accuracy. We present two sets of rates: those calculated with the full sample of classified JTS SNe, including SNe that were classified with only one SED, and those calculated with a ``gold" JTS sample, which is restricted to SNe classified either with spectroscopy or with a multi-epoch light curve. The latter is more accurate but limited to a smaller sample suffering from additional selection biases. The physical interpretation of the CC\,SN rates in terms of cosmic SFRD history is presented in a companion paper \citep{vassallo2026}, and the construction of a DTD based on the observed SN\,Ia rates is left to future work. 

Section \ref{sec:obs} describes the observations that were used to discover and characterize the JTS sample. Section \ref{sec:methods_classification} outlines the modified classification scheme, provides a detailed overview of how we generated and classified the mock SN SEDs, and presents the outcome of this classification analysis. In Section \ref{sec:methods_rates}, we explain how we computed the CC\,SN and SN\,Ia rates, and Section \ref{sec:results} presents the resulting JTS rates. We compare our CC\,SN and SN\,Ia rates to the literature, and we briefly compare our CC\,SN rates to expectations arising from galaxy luminosity-based measurements of cosmic SFRD in Section \ref{sec:discussion}. Additionally, we examine the limitations of our classifier and explore potential improvements for future classifiers. Section \ref{sec:conclusions} lists our conclusions.

Throughout this paper, we express magnitudes using the AB system \citep{oke1983} and adopt a flat $\Lambda$CDM cosmology with the following parameters: H$_0$\,$=$\,70 km s$^{-1}$ Mpc$^{-1}$, $\Omega_\Lambda$\,$=$\,0.7, and $\Omega_m$\,$=$\,0.3.


\section{Observations} \label{sec:obs}

The JADES NIRCam Deep Prime Field (hereafter JADES Deep Field) is a $\sim$25 arcmin$^2$ portion of GOODS-S that coincides with the Hubble Ultra Deep Field (HUDF). The JADES Deep Field was imaged with four short-wavelength (SW) wide-band NIRCam filters (F090W, F115W, F150W, and F200W), three long-wavelength (LW) wide-band NIRCam filters (F277W, F356W, and F444W), and two LW medium-band NIRCam filters (F335M and F410M) on two separate occasions with similar observing configurations, separated by an observer-frame year through the JADES Cycle-1 program (program ID: 1180; PI: Eisenstein). The first set of observations (hereafter Epoch1) was taken on UT 2022 September 29 -- October 5, and the second set of observations (hereafter Epoch2) was taken on UT 2023 September 28 -- October 3. \citet{eisenstein2023} details the JADES observing strategy, and \citet{rieke2023} describes the image reduction procedure. 

Subtraction of the Epoch1 and Epoch2 images enabled \citet{decoursey2025_jts} to build the JTS SN sample, with additional NIRCam follow-up observations extending the SN light curves. Observation 19, which is part of the 1180 Medium Depth survey, initially failed and was reexecuted as Observation 219 (hereafter Epoch3) on UT 2023 November 15. Epoch3 used the same 9 NIRCam filters as Epoch1 and Epoch2 (but at a shallower depth) and covered the southern portion of the JADES Deep Field.

JWST program 6541 (PI: Egami) conducted further NIRCam follow-up covering a subset of the JTS SNe \citep{egami2023}. The first set of program 6541 NIRCam observations (hereafter Epoch4) was executed on UT 2023 November 28 with a position angle that maximized coverage of the brightest and highest-redshift SNe discovered in Epoch 2. Epoch4 used F115W, F150W, F200W, F277W, F356W, and F444W. 

Two sets of failed NIRCam observations from program 1180, also part of the JADES Medium Depth Survey, were reexecuted on UT 2024 January 1, along with the second set of program 6541 observations. The first set of failed observations was originally Observation 20 of program 1180 but was reexecuted as Observation 220/222 (hereafter Epoch5.1), using F200W and F277W. The second set of program 6541 observations (hereafter Epoch5.2) was taken shortly after Epoch5.1 and used F150W, F200W, F277W, F356W, and F444W. The second set of failed observations was originally Observation 23 of program 1180 but was reexecuted as Observation 223 (hereafter Epoch5.3). Epoch5.3 used the same 9 NIRCam filters as Epoch1 and Epoch2 but with shallower depths and slightly different observing parameters. Although Epoch5.1, Epoch5.2, and Epoch5.3 were technically all separate observations, they were executed on the same day and thus their photometry is averaged into a single Epoch5 on a source-by-source basis for the light curve fitting, as explained in \citet{decoursey2025_jts}. Tables 1 and 2 of \citet{decoursey2025_jts} present the NIRCam observing log and observing parameters, respectively, for each epoch of the JTS.

Program 6541 additionally followed up a subset of the JTS SNe using the Near-Infrared Spectrograph (NIRSpec) multi-shutter array on JWST. Section 2.2 of \citet{pierel2024} provides a brief overview of the NIRSpec data reduction techniques applied to these observations.


\section{Supernova Classification} \label{sec:methods_classification}


\subsection{Classifying the JTS Supernovae} \label{subsec:jades_sne}

Compared to the light curves of low-$z$ SNe accessible to ground-based telescopes, the light curves of the high-$z$ JTS SNe are sparsely sampled (1--4 epochs), which makes determining the SN type via light curve fitting challenging. Even distinguishing between SNe\,Ia and CC\,SNe becomes difficult with such limited temporal coverage. While we have spectroscopically classified a subset of the JTS SNe (\citealt{pierel2024, siebert2024, coulter2026_tr10}; M. Griggio et al., in preparation; M. Sammut et al., in preparation), most of the SNe are photmetrically classified with sparsely sampled light curves. This naturally raises concerns regarding the accuracy of the JTS sample's SN type classification.

Section 5.1 of \citet{decoursey2025_jts} describes the light curve fitting process in detail. In short, we employ the \texttt{STARDUST2}\footnote{https://github.com/jpierel14/starDust2} Bayesian light curve classification tool \citep{rodney2014}, which measures likelihoods over the SN simulation parameter space with a nested sampling algorithm \citep{skilling2004}. We represent SNe\,Ia with the \texttt{SALT3-NIR} model \citep{guy2007, kenworthy2021, pierel2022} and CC\,SNe with 25 SN\,II and 15 SN\,Ib/c spectrophotometric time series templates. These CC\,SN templates were developed for the SN analysis software \texttt{SNANA} \citep{kessler2009} and were based on SN samples from the Sloan Digital Sky Survey \citep{frieman2008, sako2008, dandrea2010}, Supernova Legacy Survey \citep{astier2006}, and Carnegie Supernova Project \citep{hamuy2006, stritzinger2009, morrell2012}. \citet{pierel2018} extended these models to the near-infrared in anticipation of next-generation space telescope observations.

In this analysis, we have reclassified the JTS SNe after making some changes to the classification method described in \citet{decoursey2025_jts}. We list the changes below.

\begin{enumerate}

\item We restricted the SN\,Ia \texttt{x1} (``stretch") and \texttt{c} (``color") parameters to physically reasonable bounds ($-$2\,$\leq$\,\texttt{x1}\,$\leq$\,2, $-$1\,$\leq$\,\texttt{c}\,$\leq$\,1; \citealt{scolnic2018}). 

\item For sources with only one SED, we did not fit for \texttt{x1} since the light curve rise/fall shape cannot be constrained without multi-epoch observations. Rather, we fixed \texttt{x1} at 0 for these cases.

\item We set redshift bounds of [$z$\,$-$\,3$\sigma_z$, $z$\,$+$\,3$\sigma_z$], where $z$ is the input redshift and $\sigma_z$ is the input redshift uncertainty. The input redshift either comes from SN spectroscopy or from host galaxy redshift measurements (see Tables \ref{tab:jades23} and \ref{tab:jades22} in Appendix \ref{appendix:updated_classifications}).


\item We removed the \texttt{sncosmo}\footnote{https://github.com/sncosmo/sncosmo} extrapolation feature that fixes the template flux at a constant non-zero value beyond the template's late-time phase bound. Rather, we set the template flux to \texttt{NaN} beyond each template's late-time phase bound to avoid fitting observed fluxes to extrapolated non-zero values outside of the template phase range. However, we retained the extrapolation feature that sets the template flux to 0 prior to the early-time phase bound (i.e., pre-explosion).

\item In the Bayesian likelihood calculation, we only consider photometry from filters that lie within the rest-frame spectral coverage of the \texttt{SALT3-NIR} model. \texttt{SALT3-NIR} is the sole SN\,Ia model in the template library, but the template library contains many CC\,SN templates with varying spectral coverage, so this restriction ensures a fair comparison between the CC\,SN and SN\,Ia fits. To enforce this consistency, any CC\,SN template that does not fully cover the \texttt{SALT3-NIR} wavelength range is excluded from the likelihood calculation. Conversely, for CC\,SN templates that extend beyond the \texttt{SALT3-NIR} coverage, photometry outside the shared wavelength range is not included in the likelihood calculation. This requirement has important implications for the redshift range over which SNe can be classified, which is discussed later in this section.

\item We set a maximum fitting time of 30 minutes per template to reduce computing time while still providing sufficient time for each template to converge to a reasonable fit if possible. We found that the templates would either converge in 30 minutes or be unable to produce a reasonable fit. Therefore, we set the 30 minute upper limit and excluded non-convergent templates from the likelihood calculation. 

\item We removed the Type IIn \texttt{snana-2006ez} template from the library because it produced unrealistic light curves for the JWST/NIRCam LW bands. See Appendix \ref{appendix:tr16} for details.

\item We adopt no prior on the relative probabilities of an input SN being classified as an SN\,Ia, SN\,II, or SN\,Ib/c. 

\end{enumerate}

When \texttt{STARDUST2} is supplied with input SN photometry, it iterates through the library of SN\,Ia, SN\,II, and SN\,Ib/c templates and then assigns classification ``probabilities" for each SN type based on their relative goodness of fits. We denote these output probabilities as P$_\mathrm{SD}$(Ia), P$_\mathrm{SD}$(II), and P$_\mathrm{SD}$(Ib/c), where ``SD" stands for \texttt{STARDUST2}. It is important to note, however, that these probabilities should not be interpreted literally as, for example, a source having 35\% chance of being an SN\,II, 25\% chance of being an SN\,Ia, and a 40\% chance of being an SN\,Ib/c. Input SN photometry may be poorly fit by the SN\,II, SN\,Ib/c, and SN\,Ia templates, but if the SN\,II templates provide comparatively better fits, the SN\,II ``probability" would be very high despite all fits being poor. So, P$_\mathrm{SD}$(II)\,$=$\,1 does not mean that there is a 100\% chance that the input source is an SN\,II. Rather, it means that the SN\,II templates produced comparatively better fits than the SN\,Ib/c and SN\,Ia templates. The probabilities indicate which SN type best fits the data but do not reflect the quality of the fit. If a source is classified with spectroscopy, however, then its spectroscopic classification is treated with 100\% certainty and is prioritized over its photometric classification.

With our current template library and modified classification scheme, we cannot reliably classify sources that have host redshifts of $z$\,$\lesssim$\,0.7, have host redshift lower limits of $z$\,$-$\,3$\sigma_z$\,$\lesssim$\,0.7, or have unknown redshifts. 
At $z$\,$\lesssim$\,0.7, the rest-frame spectral coverage of the \texttt{SALT3-NIR} model (i.e., the only SN\,Ia model in our library) does not extend red enough (beyond $\lambda_\mathrm{rest}$\,$=$\,2$\mu$m) to fit rest-frame mid-infrared observations. This means that at $z$\,$\lesssim$\,0.7, the number of JTS filters covered by \texttt{SALT3-NIR} is equal to or below the five free parameters in the fitting scheme, rendering the fits underconstrained and susceptible to overfitting. We therefore restrict our CC\,SN and SN\,Ia rate analysis to JTS sources at $z$\,$>$\,0.7. 

\begin{figure*}
    \centering
 {\includegraphics[width=0.8\linewidth]{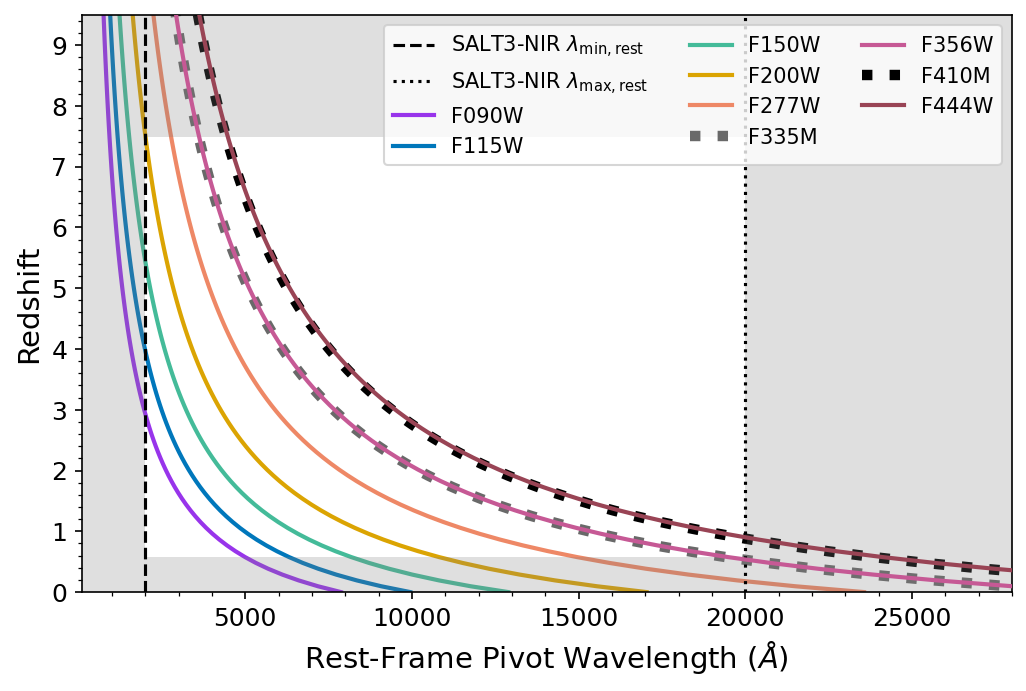}}
    \caption{Redshift as a function of rest-frame wavelength ($\mathrm{\AA}$) for the 9 JTS filters. The minimum (2,000$\mathrm{\AA}$) and maximum (20,000$\mathrm{\AA}$) wavelengths covered by the \texttt{SALT3-NIR} model are denoted with vertical dashed and dotted lines, respectively. In order to classify SNe via light curve fitting, the number of datapoints within a model's spectral coverage must exceed the number of free parameters. The shaded gray regions show the redshift and rest-frame wavelength regimes where this criterion is not met for the JTS filter set and \texttt{SALT3-NIR} model. 
    }
    \label{fig:salt3_wavelength_bounds}
\end{figure*}

We do not encounter this same issue at our upper redshift bound of $z$\,$\sim$\,5 because the \texttt{SALT3-NIR} rest-frame spectral coverage overlaps with enough JTS filters such that the number of datapoints exceeds the number of free parameters. Figure \ref{fig:salt3_wavelength_bounds} displays redshift as a function of rest-frame wavelength for the nine JTS filters, highlighting the redshift and rest-frame wavelength regimes where the JTS filters fall out of the \texttt{SALT3-NIR} coverage.

We present the updated classifications for the 2023 and 2022 JTS SN samples in Tables \ref{tab:jades23} and \ref{tab:jades22}, respectively, in Appendix \ref{appendix:updated_classifications}. We also list the associated host redshifts. Refer to \citet{decoursey2025_jts} for an explanation of the host assignment and redshift determination methods. 

We classified 66 of the 83 JTS SNe. Of the 17 sources that lack classifications, nine lie at $z$\,$<$\,0.7, three have $z-3\sigma_z$\,$<$\,0.7, and three have unknown redshifts. The two remaining unclassified sources are \tr{102} and \tr{16}. \tr{102} fell in a portion of the JADES Deep Field with incomplete Epoch1 NIRCam coverage and therefore lacked sufficient photometric datapoints to be classified. \tr{16} remains unclassified because removing the unphysical Type\,IIn \texttt{snana-2006ez} template from the library caused its classification to switch from P$_\mathrm{SD}$(CC)\,$=$\,1 to P$_\mathrm{SD}$(Ia)\,$=$\,1. In both cases, the ``best fits" poorly matched the data, making the classification ambiguous. See Appendix \ref{appendix:tr16} for more details. We only include the JTS sources with classifications in the CC\,SN and SN\,Ia rate calculations. See Sections \ref{subsec:csfrd_comparison} and \ref{subsec:ia_comparison} for discussion on how the omission of non-classified sources affects the derived CC\,SN and SN\,Ia rates. 


\subsubsection{Validating ``Single-SED" JTS Classifications with Spectroscopically-Classified SNe} \label{subsubsec:single_epoch_test}

JWST program 6541 obtained NIRSpec spectra for a small subset of the JTS sample \citep{egami2023}. With this dataset, we spectroscopically classified SN 2023adsy as an SN\,Ia \citep{pierel2024}, SN 2023adta as an SN\,Ic-BL \citep{siebert2024}, SN 2023adtd as an SN\,Ib/c (M. Griggio et al., in preparation), and SN 2023adto and SN 2023adtu as SNe\,IIP (M. Sammut et al., in preparation). We also classified SN 2023adsv as a likely SN\,II, although the spectrum contains very little SN light \citep{coulter2026_tr10}. 

We tested the validity of single-SED photometric classifications by isolating single-epoch photometry for these six spectroscopically-classified JTS SNe and performing light curve fitting for each individual epoch with the modified \texttt{STARDUST2} code. In $\sim$70\% of the cases, \texttt{STARDUST2} correctly identified CC\,SNe with only one SED (5--9 bands of coverage), but it generally failed to distinguish CC\,SN types (i.e., SNe\,II vs SNe\,Ib/c). \texttt{STARDUST2} misclassified SN 2023adsy (the only SN\,Ia in the spectroscopically-classified JTS sample) as a CC\,SN for three of the four time-separated SEDs. However, this is not surprising because the \texttt{SALT3-NIR} model was constructed with spectrophotometric data of typical low-$z$ SNe\,Ia, and SN 2023adsy is an abnormally red high-$z$ SN\,Ia \citep{pierel2024}. 

We further tested \texttt{STARDUST2's} ability to identify SNe\,Ia with only one SED by performing the same test for SN 2023aeax, a $z$\,$=$\,2.15 SN\,Ia published by \citet{pierel2025}. \texttt{STARDUST2} correctly classified SN 2023aeax as an SN\,Ia for each of its three time-separated SEDs. \texttt{STARDUST2} also correctly identified SN\,2025ogs, a spectroscopically-classified SN\,Ia at $z$\,$=$\,2.05, as an SN\,Ia for each of its three time-separated SEDs \citep{siebert2025}.

Though conducted on a small sample, this initial test indicated that \texttt{STARDUST2} can generally distinguish between CC\,SNe and \textit{normal} SNe\,Ia when provided with only one SED but cannot consistently disentangle the CC\,SN types. \texttt{STARDUST2} also struggles to identify abnormal SNe\,Ia. 


\subsection{Testing \texttt{STARDUST2} with Mock SN SEDs} \label{subsec:mock_sne}

Despite the generally positive results of the JTS single-SED classification validation test (Section 
\ref{subsubsec:single_epoch_test}), there remains uncertainty regarding the accuracy of single-SED classifications because this test was performed on a small number of bright SNe with spectroscopic data. To identify combinations of SN subtype, redshift, phase, peak M$_\mathrm{B}$, and \mbox{E(B$-$V)} that are most susceptible to misclassification in single-SED classifications, we used \texttt{sncosmo} to generate single-epoch SEDs for thousands of mock SNe spanning the main SN subtypes (i.e., SN\,Ia, SN\,Ib/c, SN\,IIP, SN\,IIL, SN\,IIn) across a range of these parameters. We then applied the \texttt{STARDUST2} classification scheme to this mock SN sample and compared the best-fit models to the known input SN parameters. 



\subsubsection{Determining the Parameter Grids}

With limited computational resources, we were unable to sample all possible redshift, phase, peak M$_\mathrm{B}$, and \mbox{E(B$-$V)} values for the mock SN SEDs. Instead, we strategically sampled parameter values to maximize coverage of the different SED shapes and absolute brightnesses that the various SN subtypes assume throughout their light curve evolution. Table \ref{tab:mock_sn_parameters} lists the parameter grids used to generate the mock SNe. 

\begin{deluxetable*}{ccccc}
\tablecaption{Mock Supernova Model Parameter Grids}
\tablehead{
\colhead{Subtype} & \colhead{Model} & \colhead{Rest-Frame Phase (Days)} & \colhead{M$_\mathrm{B,peak}$} & \colhead{E(B$-$V)}
} 
\label{tab:mock_sn_parameters}
\startdata
SN\,Ia   & \texttt{hsiao}        & $-$10, 0, 10, 20, 30, 40, 50, 60, 70, 80  & $-$18.84, $-$19.35, $-$19.86 & 0, 0.1, 0.2, 0.3, 0.4 \\
SN\,Ib/c & \texttt{nugent-sn1bc} & 5, 10, 20, 30, 40, 50, 60, 70             & $-$16.99, $-$17.69, $-$18.39 & 0, 0.1, 0.2, 0.3, 0.4 \\
SN\,IIP  & \texttt{nugent-sn2p}  & 5, 10, 30, 50, 70, 90, 110, 130           & $-$15.92, $-$16.89, $-$17.86 & 0, 0.1, 0.2, 0.3, 0.4 \\
SN\,IIL  & \texttt{nugent-sn2l}  & 5, 10, 20, 30, 40, 50, 60, 70             & $-$17.16, $-$18.07, $-$18.97 & 0, 0.1, 0.2, 0.3, 0.4 \\
SN\,IIn  & \texttt{nugent-sn2n}  & 5, 10, 20, 30, 40, 50, 60, 70             & $-$17.23, $-$18.71, $-$20.19 & 0, 0.1, 0.2, 0.3, 0.4 \\
\enddata
\tablecomments{For the SN\,Ia model, phase is relative to B-band peak. For the CC\,SN models, phase is relative to explosion. Each SN subtype's redshift grid was $z$\,$=$0.7--5.0, $\delta_\mathrm{z}$\,=\,0.1}
\end{deluxetable*}

\begin{figure*}
    \centering
 {\includegraphics[width=0.49\linewidth]{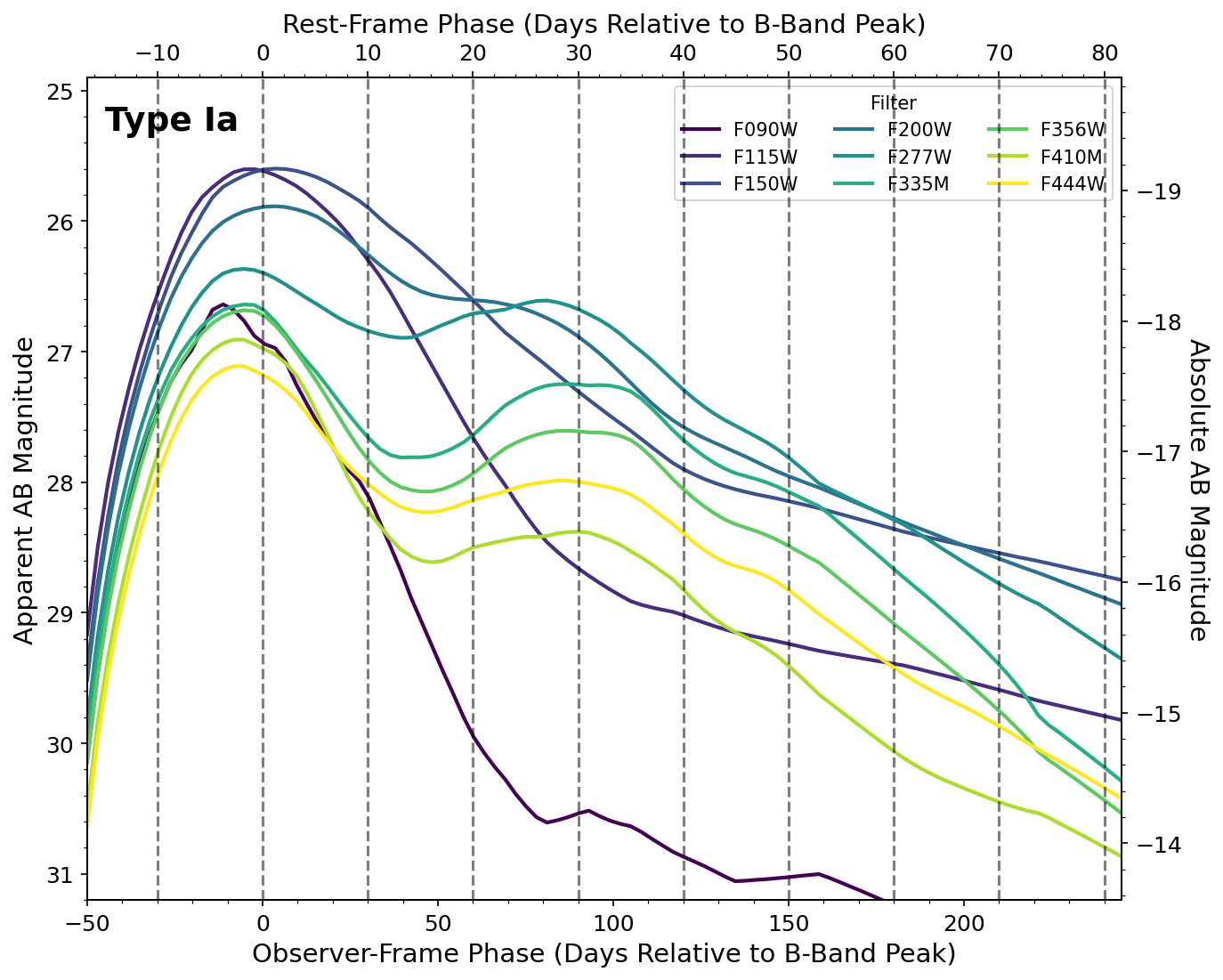}}
{\includegraphics[width=0.49\linewidth]{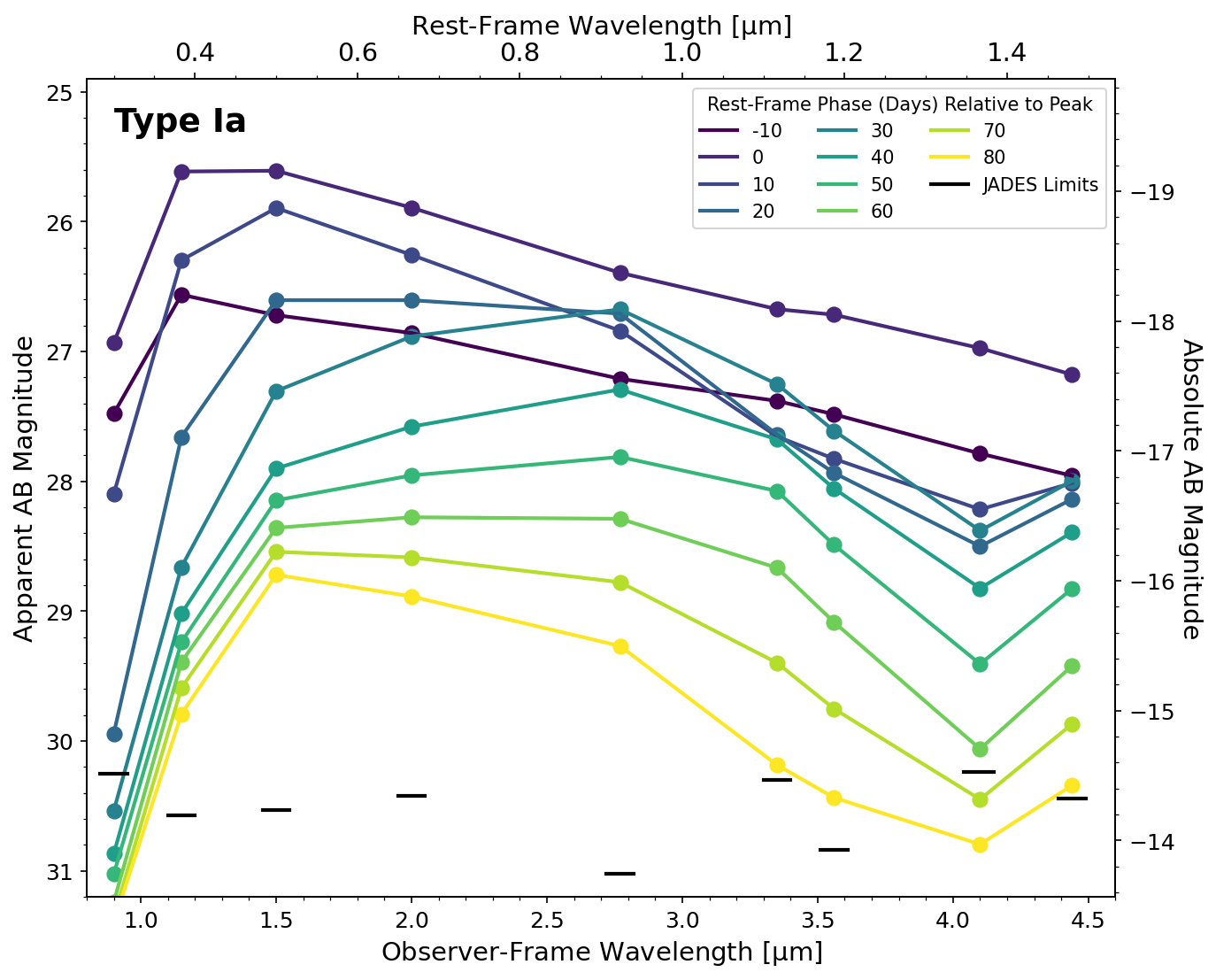} }
    \caption{\textit{Left:} The light curves of a dust-free \texttt{hsiao} SN\,Ia at $z$\,$=$\,$2$, with each light curve corresponding to a JTS filter. The dashed gray vertical lines mark sampled phases (Table \ref{tab:mock_sn_parameters}). \textit{Right:} The SEDs at the sampled phases for the dust-free \texttt{hsiao} SN\,Ia at $z$\,$=$\,2. The short horizontal black lines denote the 2$\sigma$ detection limits for each JTS filter.  
    }
    \label{fig:ia_sed_lc}
\end{figure*}

We used the \texttt{hsiao v3.0} SN\,Ia model to generate the mock SN\,Ia SEDs \citep{hsiao2007}, as this model was not in the \texttt{STARDUST2} template library. To define the SN\,Ia phase grid, we identified phases corresponding to distinct SED shapes throughout SN\,Ia light curve evolution. The right panel of Figure \ref{fig:ia_sed_lc} shows $z$\,$=$\,2 \texttt{hsiao} SN\,Ia SEDs evolving through time, with each SED representing a sampled phase. The left panel of Figure \ref{fig:ia_sed_lc} shows the corresponding \texttt{hsiao} light curves for the nine JTS filters, with the sampled phases marked by gray dashed vertical lines. 

We represented the SNe\,IIP, SNe\,IIL, SNe\,IIn, and SNe\,Ib/c with the \texttt{nugent-sn2p}, \texttt{nugent-sn2l}, \texttt{nugent-sn2n}, and \texttt{nugent-sn1bc} templates, respectively \citep{gilliland1999, levan2005}. These templates, not in the \texttt{STARDUST2} library, were used to avoid biasing the fitting procedure. SNe\,IIb are excluded from our mock CC\,SN sample due to the lack of a \texttt{nugent} SN\,IIb template. Generating such a template in a manner consistent with the existing templates is beyond the scope of this work. Given that SNe\,IIb represent only $\sim$10\% of the CC\,SN population \citep{li2011_fractions, graur2017}, we do not expect this omission to meaningfully impact our results. 

\begin{figure*}
    \centering
{\includegraphics[width=7cm]{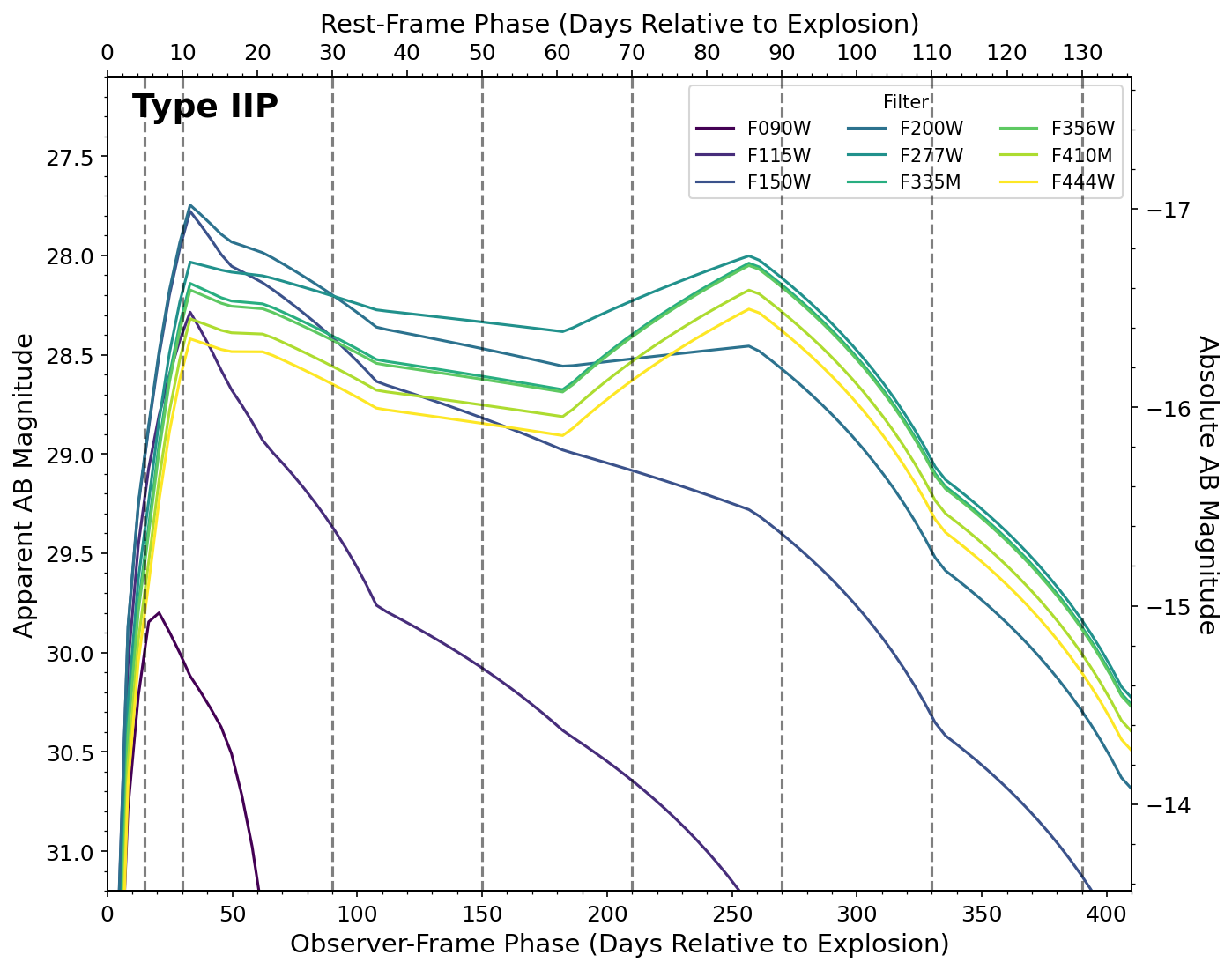}}
    \qquad
{\includegraphics[width=7cm]{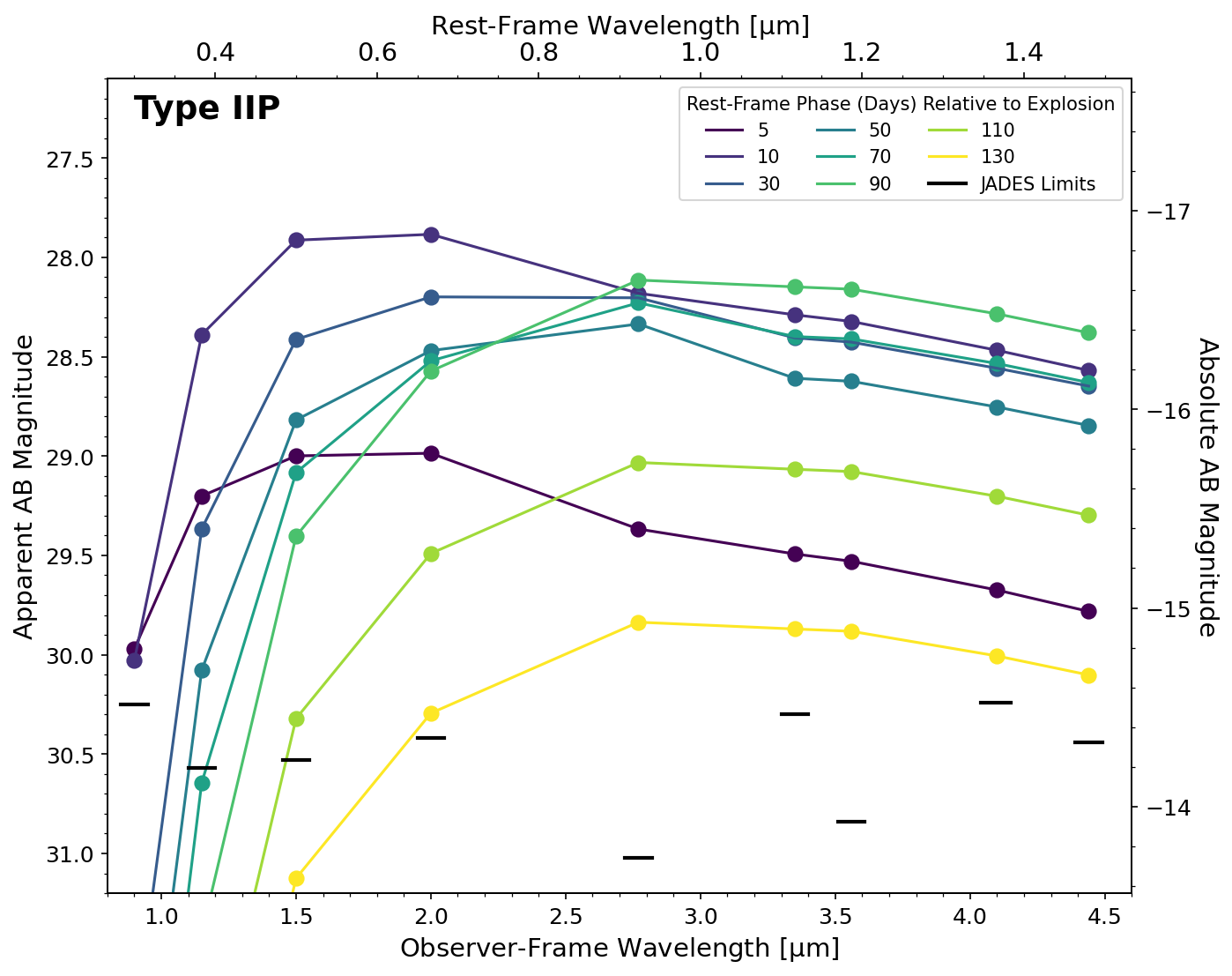}}
    \qquad
{\includegraphics[width=7cm]{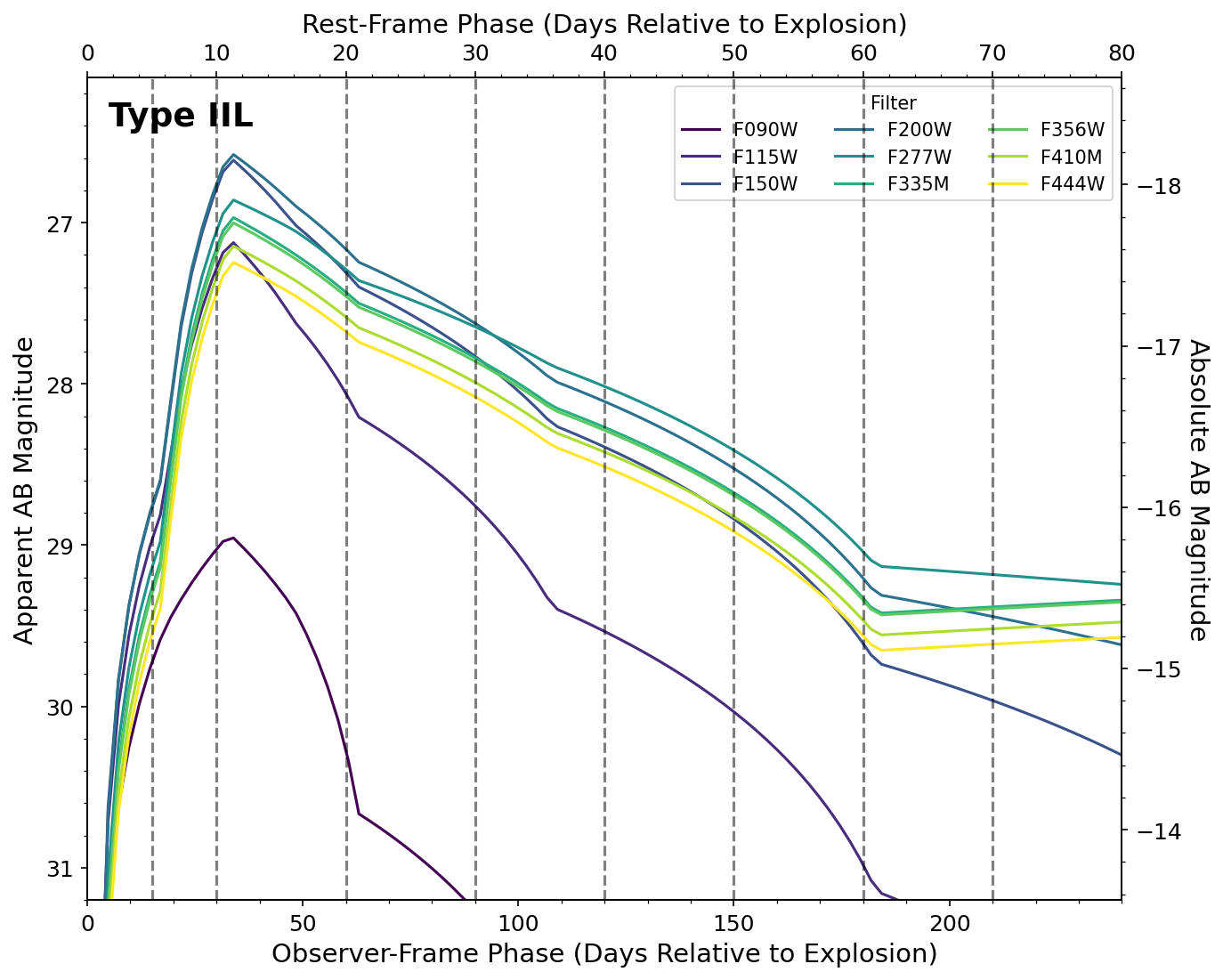}}
    \qquad
{\includegraphics[width=7cm]{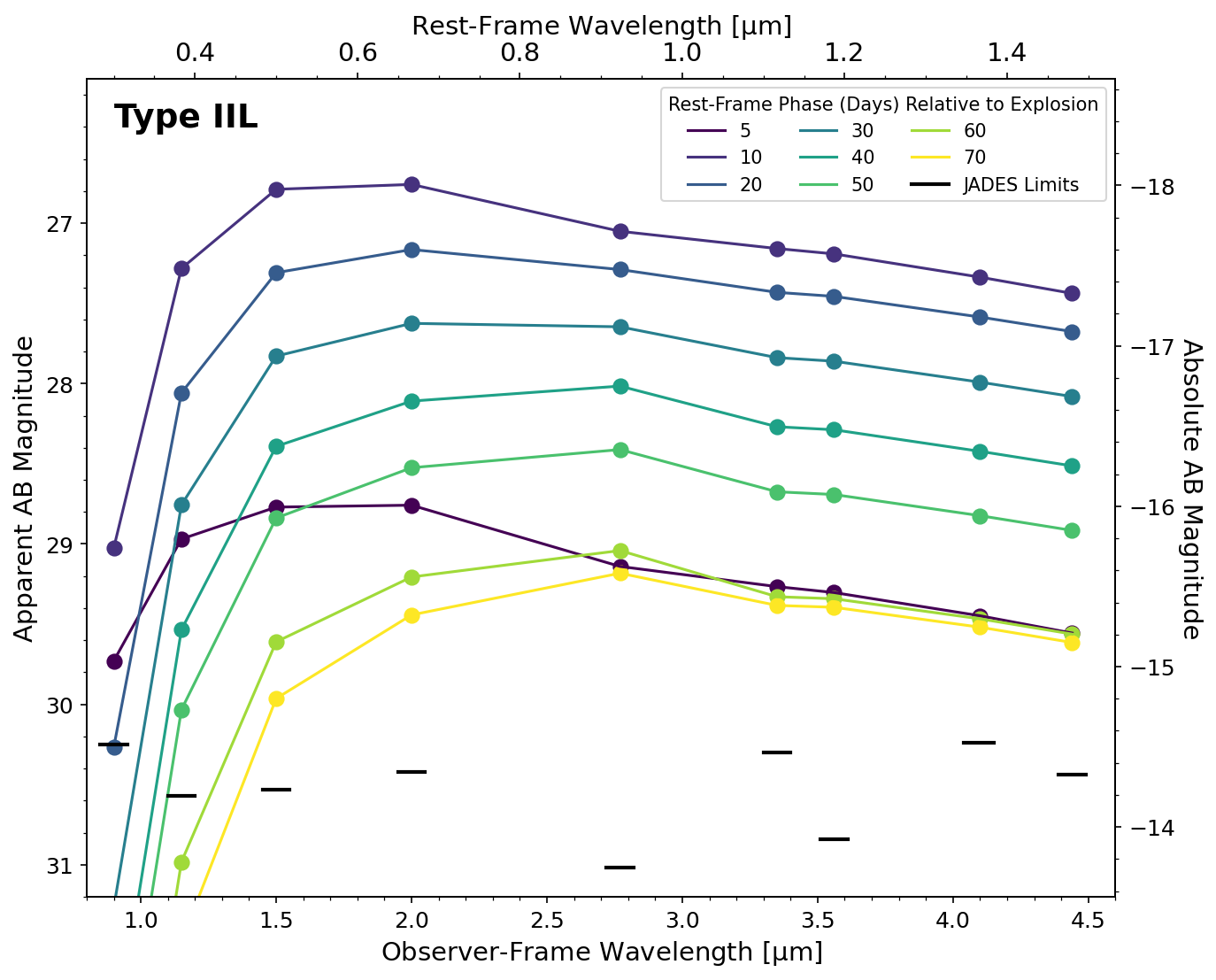}}    
    \qquad
{\includegraphics[width=7cm]{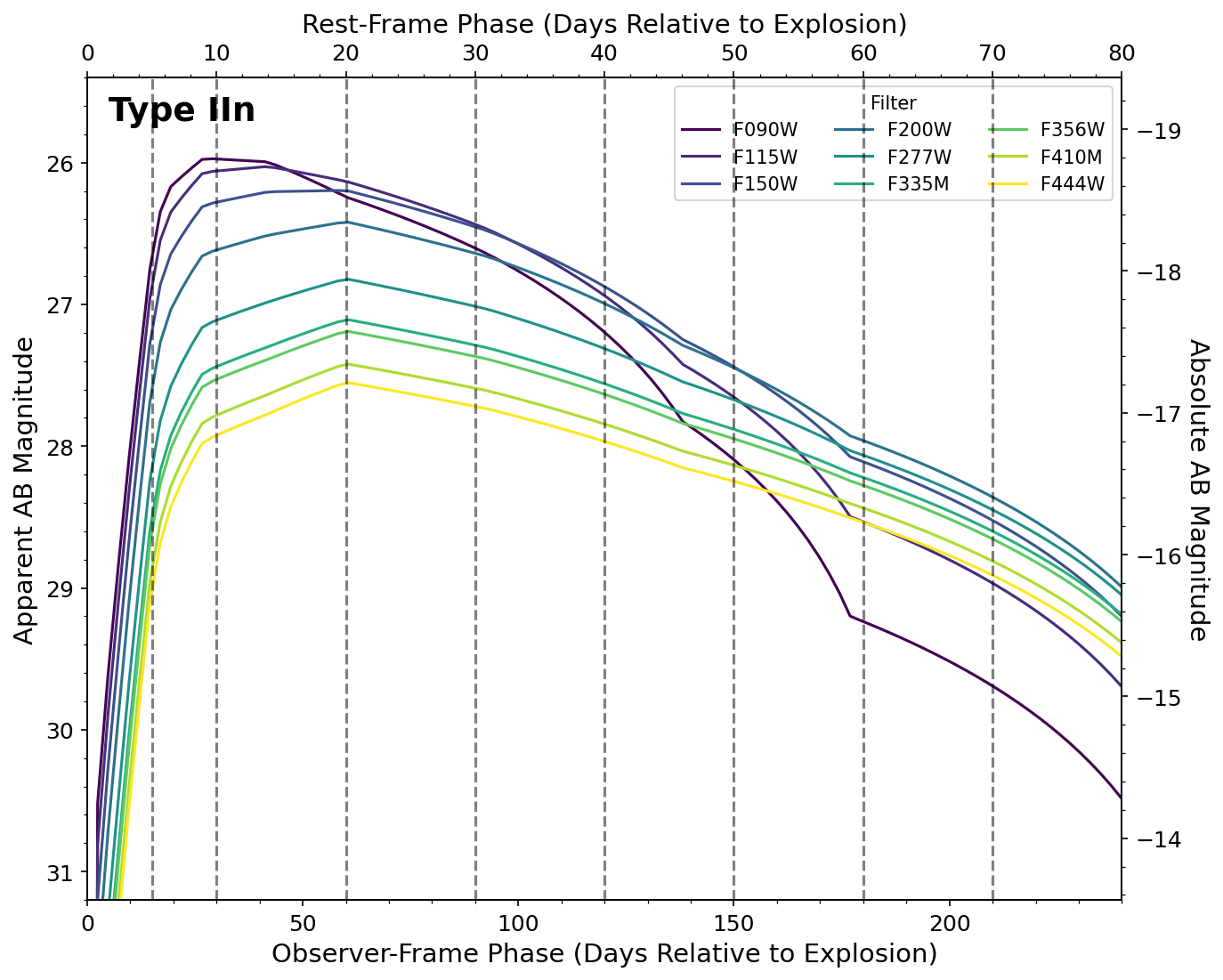}}
    \qquad
{\includegraphics[width=7cm]{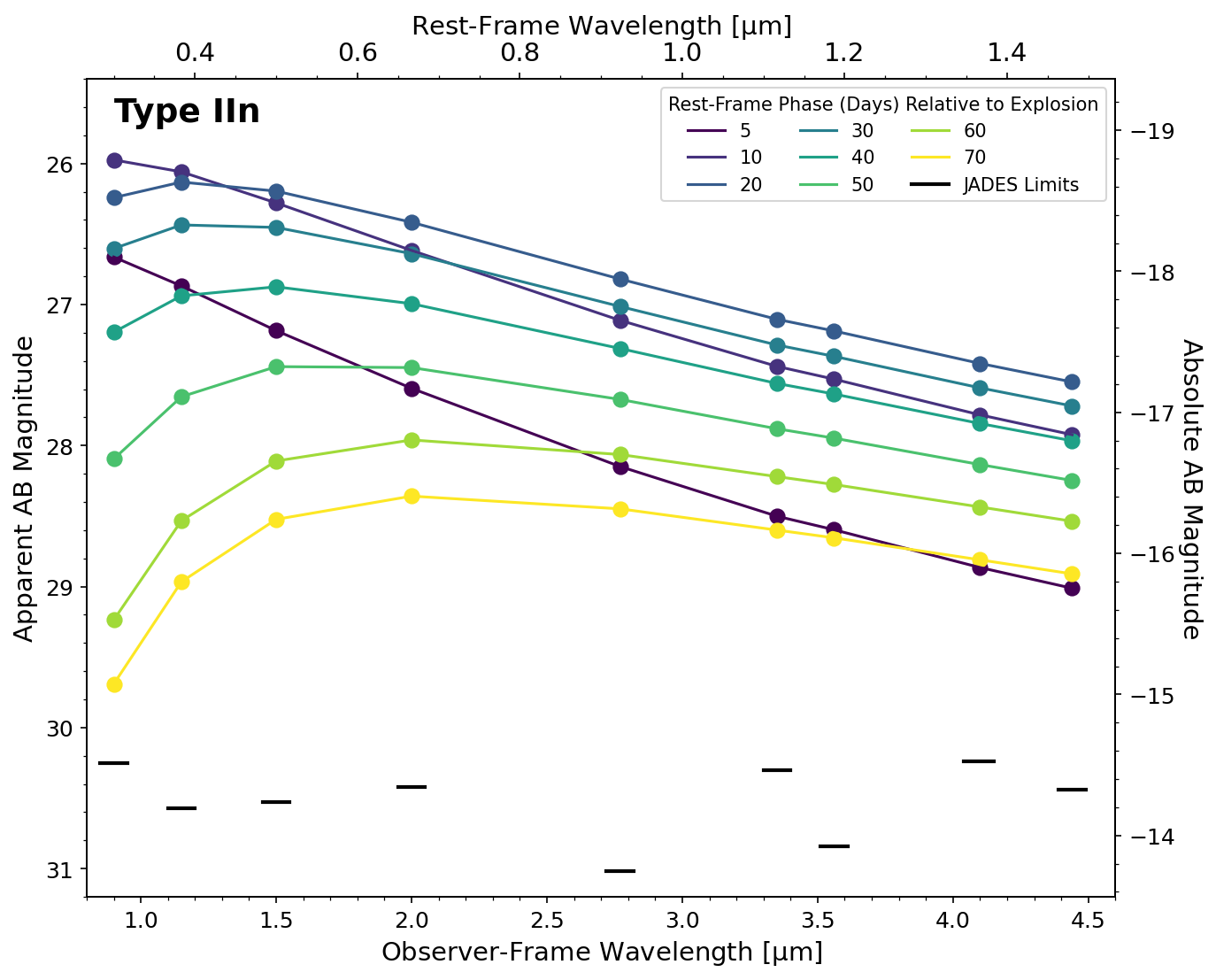}}
    \qquad
{\includegraphics[width=7cm]{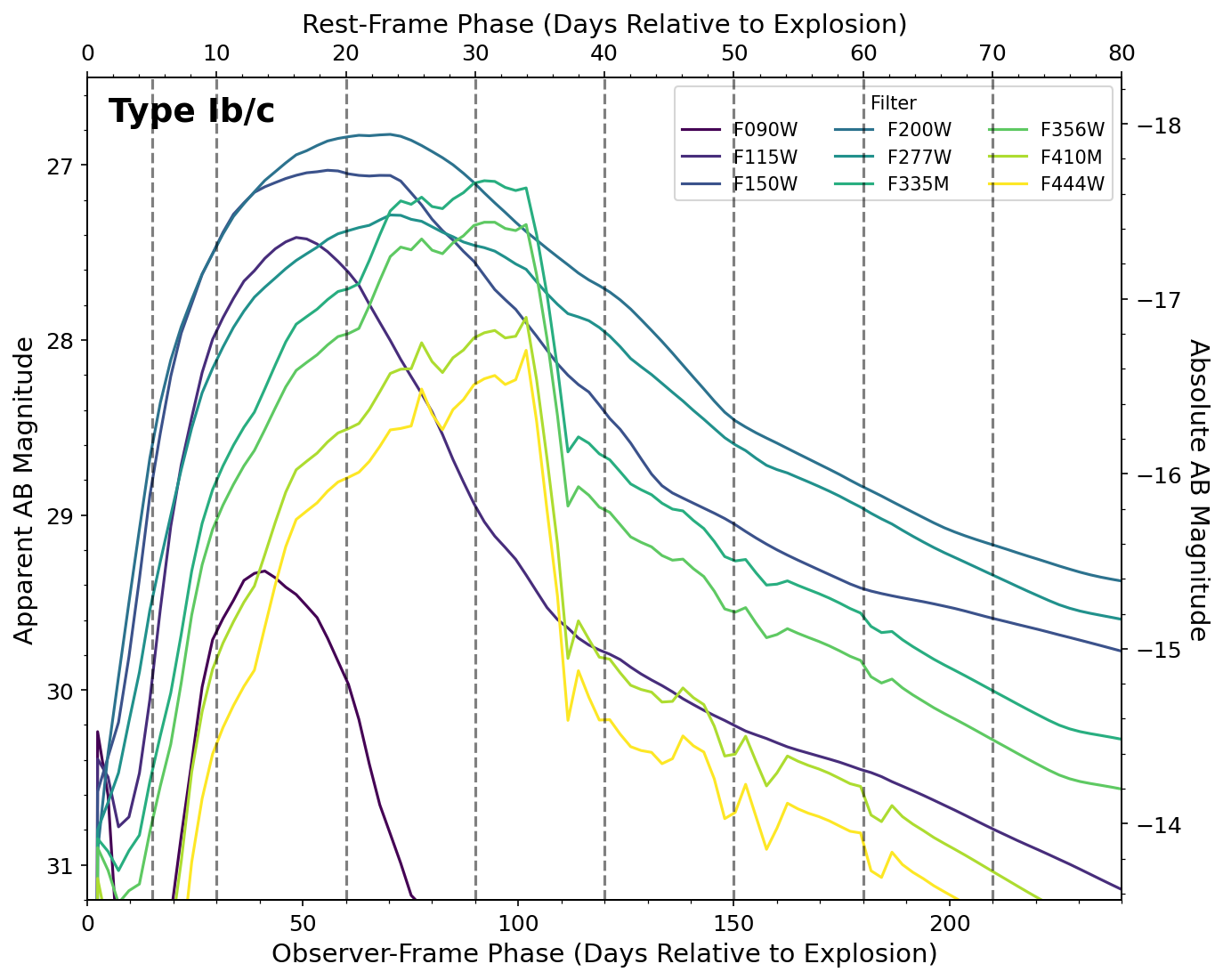}}
    \qquad
{\includegraphics[width=7cm]{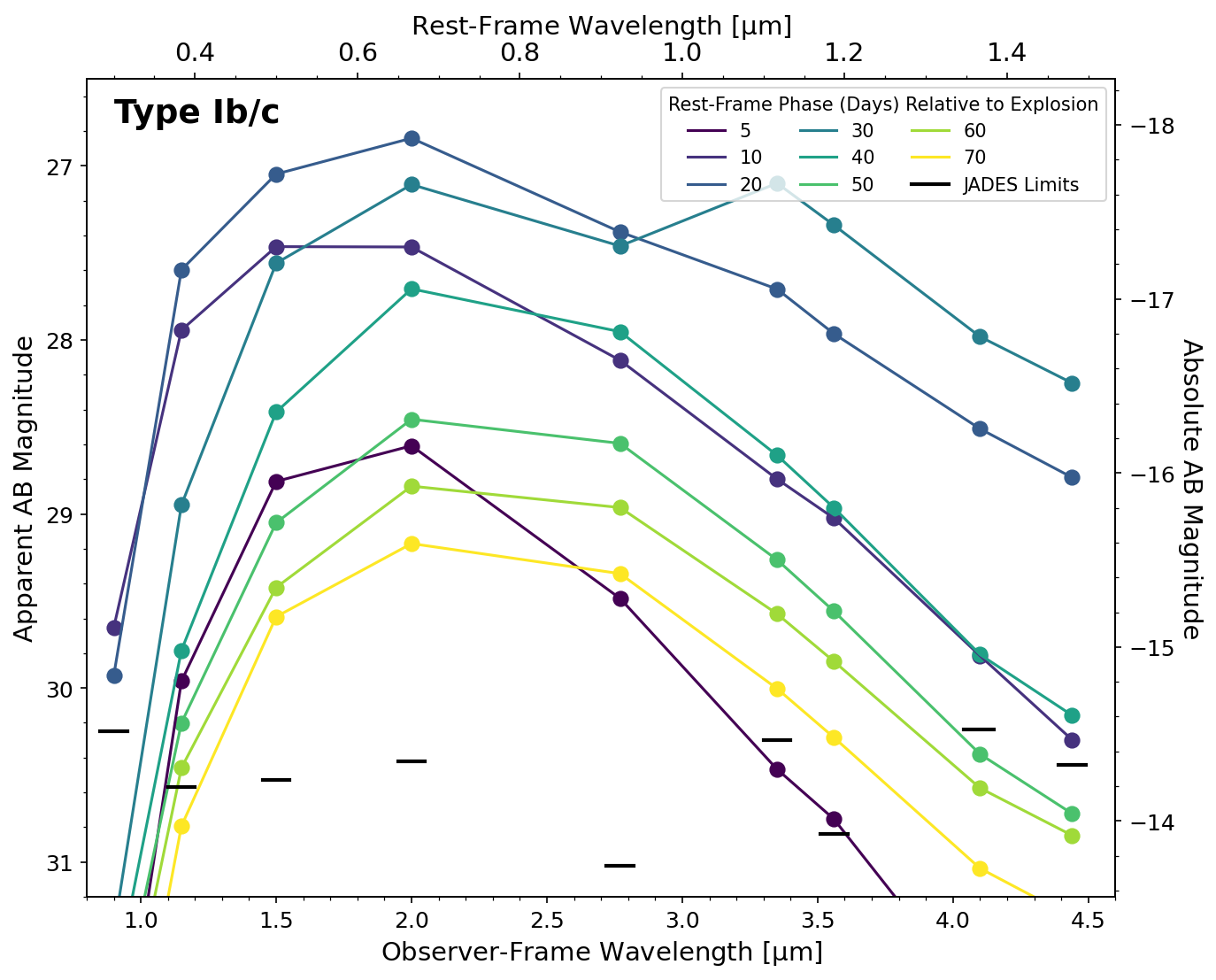}}  
    \caption{From top to bottom, we show light curves (left) and SEDs (right) for a dust-free SN\,IIP, SN\,IIL, SN\,IIn, and SN\,Ibc at $z$\,$=$\,2. \textit{Left}: The light curves corresponding to the nine JTS filters, with dashed gray vertical lines marking the phases that are sampled in each subtype's phase grid (Table \ref{tab:mock_sn_parameters}). \textit{Right}: The SEDs corresponding to each sampled phase, with short horizontal black lines denoting the 2$\sigma$ detection limits for each JTS filter.  
    }
    \label{fig:cc_sed_lc}
\end{figure*}

Following the same approach as for the SNe\,Ia, we assigned a phase grid to each CC\,SN subtype such that each sampled phase captured a distinct SED shape that the SN assumes through its phase evolution. The right panels of Figure \ref{fig:cc_sed_lc} show the time-evolving SEDs for a $z$\,$=$\,2 SN of each subtype, with each SED representing a sampled phase. The left panels show the corresponding light curves in the nine JTS filters, with dashed gray vertical lines marking the sampled phases. 

To select the grid of peak M$_\mathrm{B}$ values, we referred to the volume-limited peak M$_\mathrm{B}$ Gaussian distributions presented in \citet{richardson2014}, converted to AB magnitudes with \citet{blanton2007}. For each subtype, we sampled the mean peak M$_\mathrm{B}$ ($\mu_\mathrm{M_B}$) and the $\mu_\mathrm{M_B}$\,$\pm$\,1$\sigma_\mathrm{M_B}$ absolute magnitudes, as listed in Table \ref{tab:mock_sn_parameters}. This allowed us to explore the classification accuracy of systematically bright, systematically faint, and typical SNe for each subtype. \citet{richardson2014} presents individual peak M$_\mathrm{B}$ Gaussian distributions for SNe\,Ib and SNe\,Ic, so we averaged the two distributions to set the M$_\mathrm{B}$ grid for our mock SNe\,Ib/c.

For the color excess E(B$-$V), we selected from a grid of \mbox{E(B$-$V)}\,$=$\,0--0.4 in steps of $\delta_{\mathrm{E(B-V)}}$\,$=$\,0.1 and adopted the Fitzpatrick extinction law \citep{fitzpatrick1999}. We did not consider \mbox{E(B$-$V)}\,$>$\,0.4 because highly dust-obscured SNe are more likely to evade detection and thereby less likely to exist in the JTS sample. \mbox{E(B$-$V)}\,$=$\,0--0.4 encompasses the best-fit values for the majority of the JTS SNe with 3$+$ light curve epochs (Table \ref{tab:jades23}).

For each SN subtype, we sampled redshift in the $z$\,$=$\,0.7--5.0 range with $\delta_z$\,$=$\,0.1. Section \ref{subsec:jades_sne} justifies this redshift range.
We classified each mock SN under two different redshift conditions: a spec-$z$ case and a ``photo-$z$" case. In both cases, to reduce computation cost, we bounded the redshift to [$z$\,$-$\,3$\sigma_z$, $z$\,$+$\,3$\sigma_z$], where $z$ was the input redshift and $\sigma_z$\,$=$\,0.001. These tight bounds effectively forced \texttt{STARDUST2} to treat redshift as a fixed parameter, even in the ``photo-$z$" case. The key difference between the spec-$z$ and photo-$z$ cases was the accuracy of the input redshift. In the spec-$z$ cases, \texttt{STARDUST2} was supplied with the true SN redshifts. However, in the photo-$z$ cases, the input redshifts were randomly drawn from Gaussian distributions centered on the true SN redshifts. The standard deviations of these redshift distributions were randomly drawn from a Gaussian distribution with a mean of $\mu$\,$=$\,0 and a standard deviation of $\sigma$\,$=$\,0.18, where $\sigma$\,$=$\,0.18 corresponds to the median photo-$z$ uncertainty for galaxies with $0.7$\,$\leq$\,$z$\,$\leq$\,5 and S/N$_\mathrm{F200W}$\,$\geq$\,5 in version 0.9.5 of the JADES photo-$z$ catalog. We took the absolute value of the random draw and imposed a minimum value of 0.01.  


\subsubsection{Generating the Mock SN SEDs} \label{subsubsec:model_generation}

For each combination of gridded parameters listed in Table \ref{tab:mock_sn_parameters}, we used \texttt{sncosmo} to generate an SN SED spanning the nine JTS filters. To more closely replicate the observed JTS SEDs, we then replaced any photometric point falling below its filter's respective JTS detection threshold with the corresponding upper limit \citep{decoursey2025_jts}. The current version of \texttt{STARDUST2} cannot properly incorporate upper limits, however, so we used the following approximation: for each non-detection, we set the input flux to 0 and the input flux uncertainty to the JTS 1$\sigma$ upper limit. This effectively allowed \texttt{STARDUST2} to fit fluxes up to the 3$\sigma$ upper limit at the input wavelength, but it introduced an unphysical bias towards zero flux. Although this treatment of upper limits was not ideal, we preferred it over discarding all non-detections so that we could retain all available information in the SED fitting process.

We then calculated the uncertainties associated with the photometric detections. We set a 1\% uncertainty floor to account for the zeropoint uncertainty. This 1\% uncertainty floor was then added in quadrature with an estimate of the background noise
and an estimate of Poisson noise. Next, we slightly perturbed each photometric detection by replacing it with a randomly drawn value from a Gaussian distribution with the mean set as the initial photometric value and the standard deviation set as the associated photometric uncertainty.
It was possible that an initial photometric value was just barely above its respective JTS detection threshold, and the perturbed value fell below the detection threshold. We therefore compared the perturbed photometry against the respective JTS detection thresholds and replaced any non-detection with the associated upper limits, as previously described. The finalized mock SEDs were then tabulated.











Although we generated single-epoch SEDs for every SN parameter combination in Table \ref{tab:mock_sn_parameters}, we did not attempt to classify every single one of them. This is because certain parameter combinations generated SEDs that were mostly or entirely below the JTS detection thresholds (e.g., a dust-extincted SN\,IIP with a relatively faint peak M$_\mathrm{B}$ at $z$\,$=$\,4). In these cases, we either would not be able to detect these SNe with the JADES observations, or they would not have passed the rigorous JTS selection criteria (Section 3.2 in \citealt{decoursey2025_jts}). 
To ensure that our mock SN sample was observationally realistic, we required the mock SN photometry to exceed the 2$\sigma$ JTS detection thresholds in at least three filters in order to be eligible for classification. 

We then passed the sources that fulfilled this requirement through the modified \texttt{STARDUST2} classifier.
With only one input SED per source, we did not expect to distinguish specific CC\,SN subtypes in our classifications. However, our primary goal was to measure CC\,SN and SN\,Ia rates, so we were only concerned with accurately distinguishing SNe\,Ia from CC\,SNe. Thus, we summed the SN\,II and SN\,Ib/c outputs from \texttt{STARDUST2}, P$_\mathrm{SD}$(II) and P$_\mathrm{SD}$(Ib/c), into a generalized CC\,SN output, P$_\mathrm{SD}$(CC). 



\subsection{Single-SED Classification Accuracy} \label{subsec:results_classification}


Following the parameter grids listed in Table \ref{tab:mock_sn_parameters}, we generated mock SN SEDs for 6,600 SNe\,Ia and 5,280 SNe\,IIP, SNe\,IIn, SNe\,IIL, and SNe\,Ib/c. 
Of these, 6,035 SNe\,Ia, 3,716 SNe\,IIP, 4,737 SNe\,IIn, 4,200 SNe\,IIL, and 4,209 SNe\,Ib/c fulfilled the selection criteria (Section \ref{subsubsec:model_generation}) and were passed to the \texttt{STARDUST2} classifier.


\subsubsection{CC\,SN vs SN\,Ia} \label{subsec:results_cc_vs_ia}

Using the known mock SN input types and their resulting output CC\,SN and SN\,Ia classification probabilities, we constructed CC\,SN vs SN\,Ia confusion matrices. For each SN subtype, we tracked two quantities: the true positive rate (TPR), defined as the fraction of input SNe of a given type that were correctly classified as that type, and the false negative rate (FNR), defined as the fraction of input SNe of a given type that were incorrectly classified as a different type. In constructing the confusion matrices, we counted the total number of SNe of each subtype that were correctly and incorrectly classified, where partial classifications were allowed for sources that were not classified with either P$_\mathrm{SD}$(CC)\,$=$\,1 or P$_\mathrm{SD}$(Ia)\,$=$\,1. For example, if \texttt{STARDUST2} assigned an input \texttt{nugent-sn2p} model with P$_\mathrm{SD}$(CC)\,$=$\,0.75 and P$_\mathrm{SD}$(Ia)\,$=$\,0.25, then we added 0.75 to the TPR$_\mathrm{CC}$ component and 0.25 to the FNR$_\mathrm{CC}$ component. The same approach was applied to the SN\,Ia classifications, yielding TPR$_\mathrm{Ia}$ and FNR$_\mathrm{Ia}$ confusion matrix components.



\begin{figure*}
    \centering
{\includegraphics[width=0.49\linewidth]{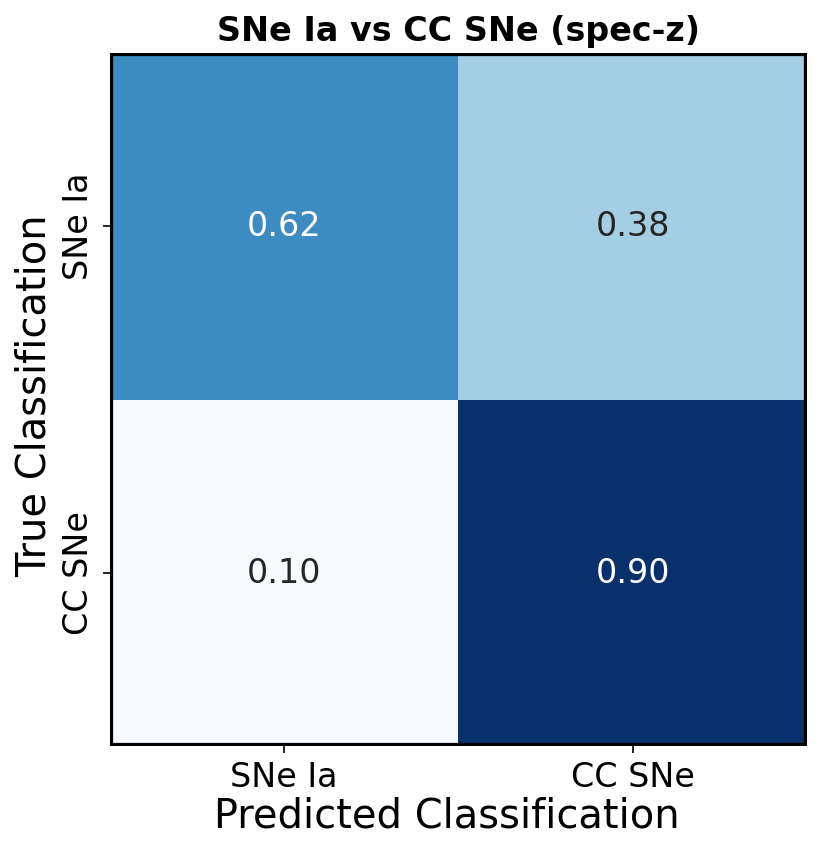}}
{\includegraphics[width=0.49\linewidth]{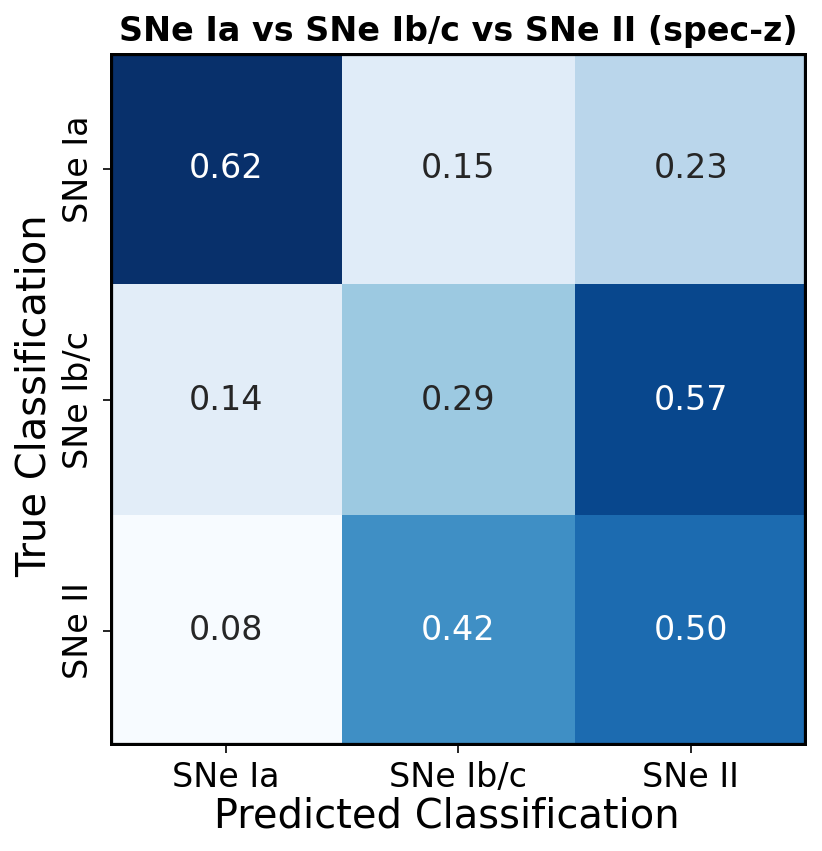} }
{\includegraphics[width=0.60\linewidth]{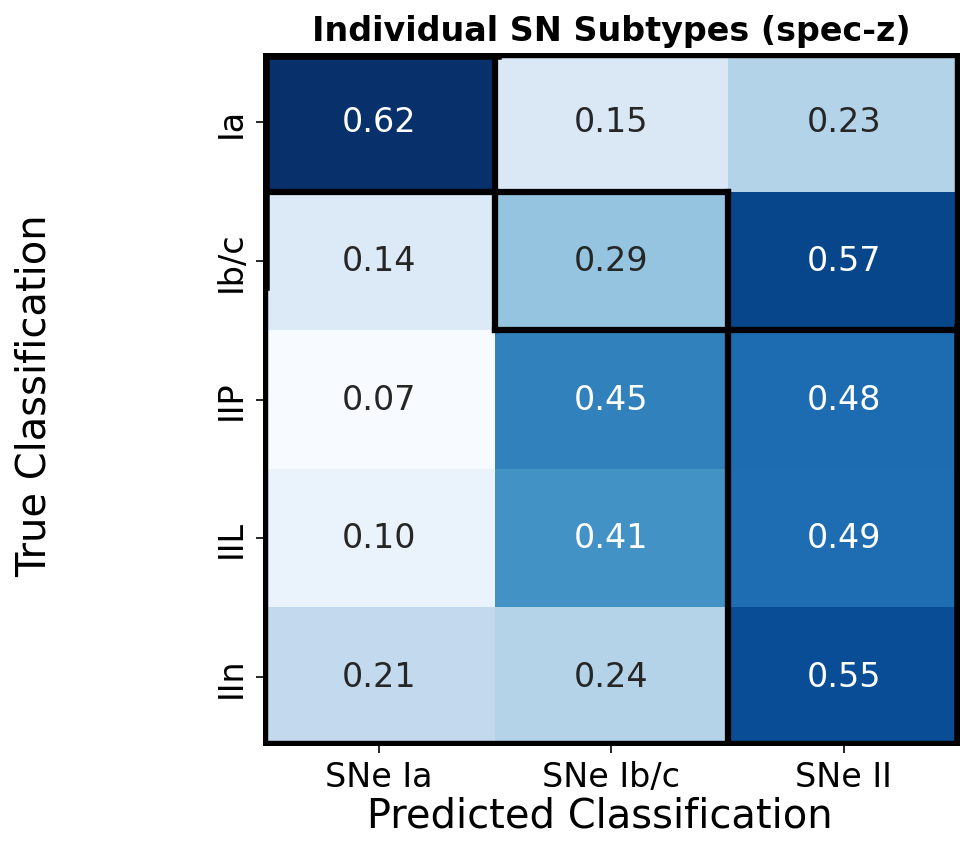} }
    \caption{Confusion matrices for mock SNe with spectroscopic redshifts. \textit{Top Left:} The CC\,SN versus SN\,Ia confusion matrix. Here, every mock CC\,SN subtype of each parameter combination is combined into a general ``CC\,SN" class. 
    \textit{Top Right:} The SN\,Ia versus SN\,II versus SN\,Ib/c confusion matrix. The mock SNe\,IIP, SNe\,IIL, and SNe\,IIn of every parameter combination are combined into the ``SN\,II" class. 
    \textit{Bottom:} A confusion matrix where every input CC\,SN subtype is shown individually, but their predicted classifications are divided into SNe\,Ia, SNe\,II, and SNe\,Ib/c. 
    }
    \label{fig:confusion_matrices_zspec}
\end{figure*}

\begin{figure*}
    \centering
{\includegraphics[width=0.49\linewidth]{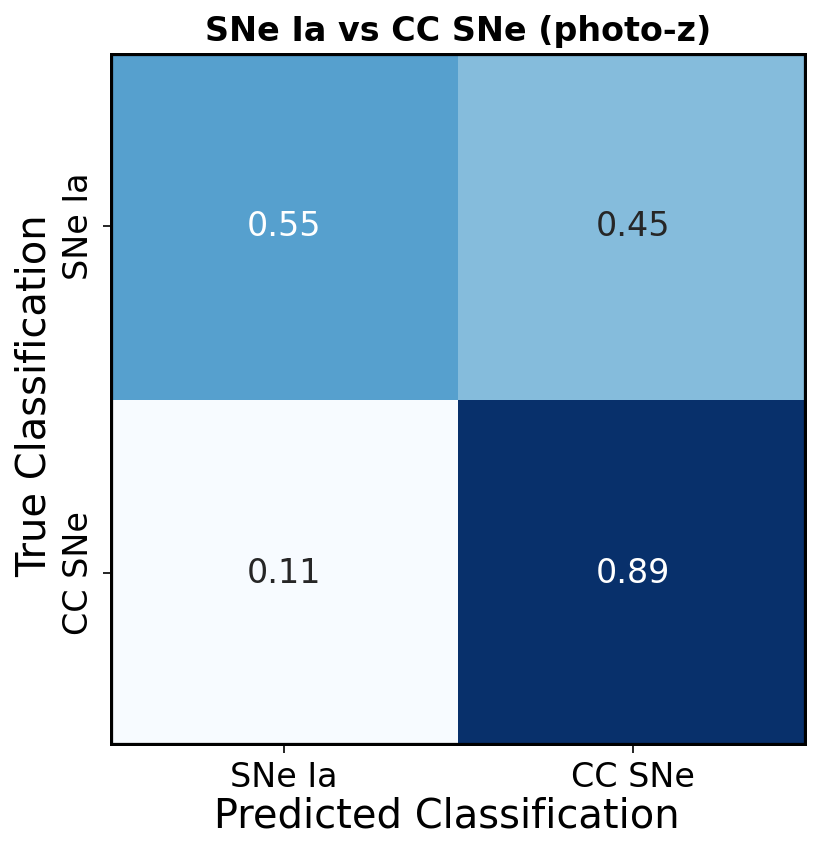}}
{\includegraphics[width=0.49\linewidth]{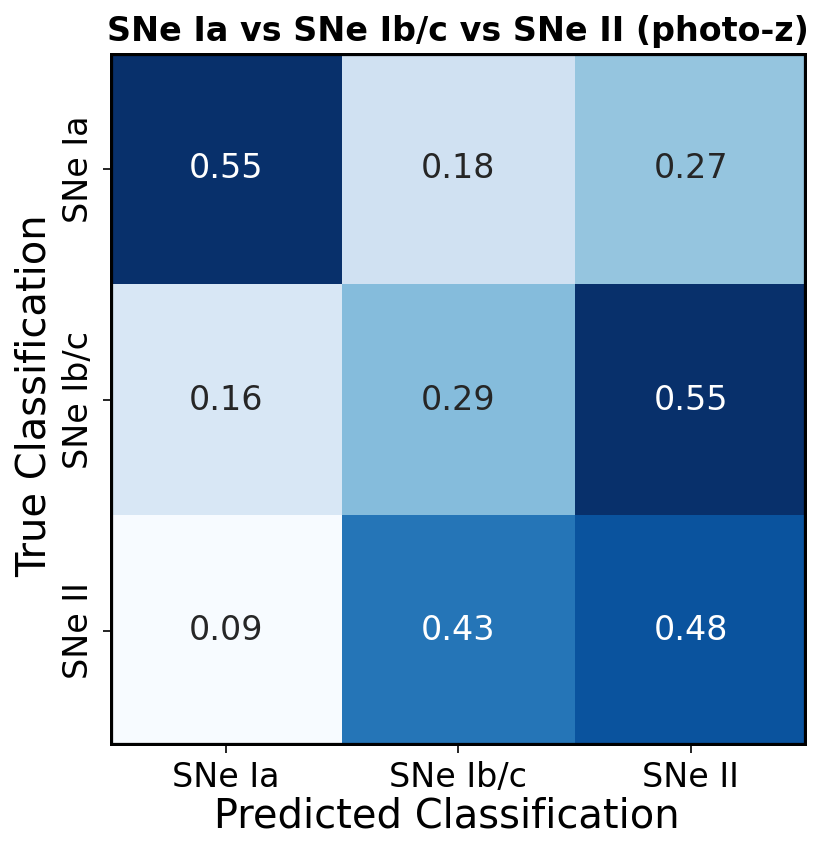} }
{\includegraphics[width=0.60\linewidth]{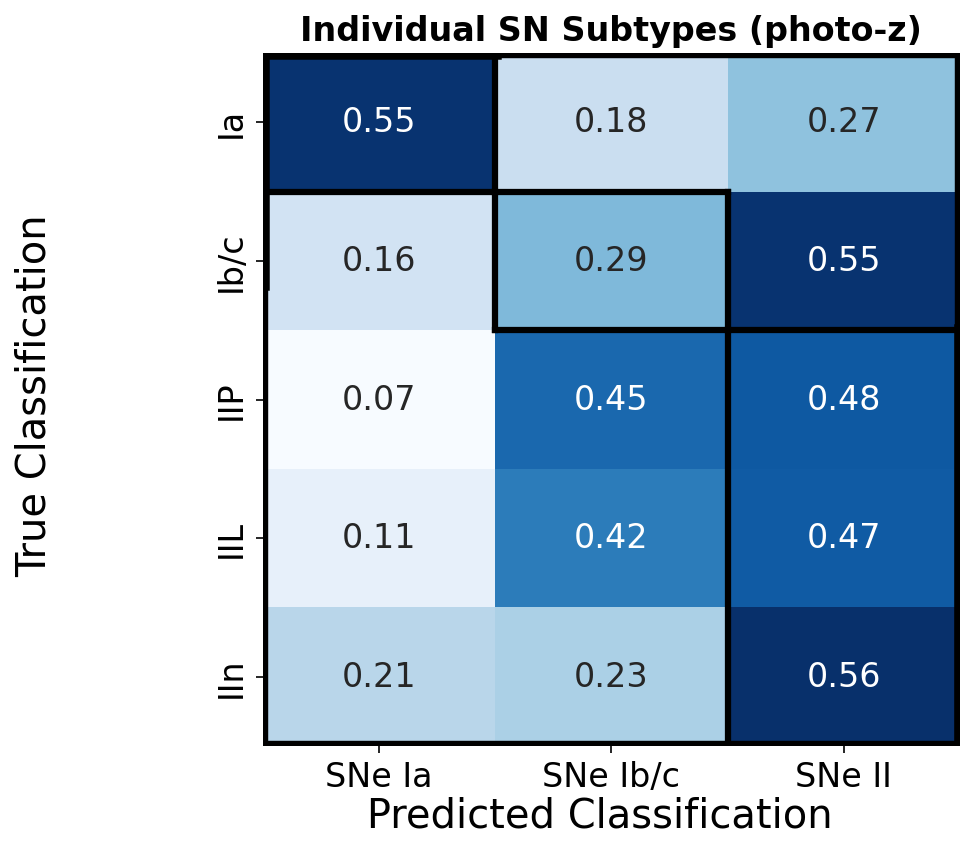} }
    \caption{Confusion matrices for mock SNe with photometric redshifts. \textit{Top Left:} The CC\,SN versus SN\,Ia confusion matrix. Every mock CC\,SN subtype of each parameter combination is combined into a general ``CC\,SN" class. 
    \textit{Top Right:} The SN\,Ia versus SN\,II versus SN\,Ib/c confusion matrix. The mock SNe\,IIP, SNe\,IIL, and SNe\,IIn of every parameter combination are combined into the ``SN\,II" class. 
    \textit{Bottom:} A confusion matrix where every true input CC\,SN subtype is shown individually but their predicted classifications are divided into SNe\,Ia, SNe\,II, and SNe\,Ib/c. 
    }
    \label{fig:confusion_matrices_zphot}
\end{figure*}

The top left panel of Figures \ref{fig:confusion_matrices_zspec} and \ref{fig:confusion_matrices_zphot} present the CC\,SN vs SN\,Ia confusion matrices for the spec-$z$ and photo-$z$ mock SN samples, respectively. The true CC\,SN components include the mock SNe\,IIP, IIn, IIL, and Ib/c of every parameter combination.  When computing TPR$_\mathrm{CC}$ and FNR$_\mathrm{CC}$, we weighted the contribution of each CC\,SN subtype by its fractional representation in the total CC\,SN population, adopting the volume-limited fractions from \citet{li2011_fractions}. Note, however, that the volume-limited fractions from \citet{graur2017}, which details the relative CC\,SN subtype fractions in low- and high-mass galaxies, have superseded those from \citet{li2011_fractions}.
Because we did not include SNe\,IIb in our analysis ($\sim$10\% of the CC\,SN population), we renormalized the fractions of SNe\,IIP, SNe\,IIn, SNe\,IIL, and SNe\,Ib/c to sum to 1. It is currently unknown how the CC\,SN subtype fractions evolve at high-$z$, so we adopted the fractions observed in the local Universe.

Across the full redshift range ($z$\,$=$\,0.7--5), the average TPR$_\mathrm{Ia}$ for input SNe\,Ia with spectroscopic (photometric) redshifts is P$_\mathrm{SD}$(Ia)\,$=$\,0.62 (0.55). This does not mean that, for example, 62 out of 100 SNe\,Ia with spec-$z$s will be correctly identified as SNe\,Ia with P$_\mathrm{SD}$(Ia)\,$=$\,1. Rather, this means that, on average, an input SN\,Ia with a spec-$z$ will be classified as an SN\,Ia with P$_\mathrm{SD}$(Ia)\,$=$\,0.62. The TPR$_\mathrm{CC}$ averaged across the full redshift range is P$_\mathrm{SD}$(CC)\,$=$\,0.90 (0.89) for CC\,SNe with spectroscopic (photometric) redshifts. \texttt{STARDUST2} is significantly more accurate in its classification of single-SED CC\,SNe than SNe\,Ia. We explore various reasons why \texttt{STARDUST2} struggles to accurately classify single-SED SNe\,Ia in Section \ref{subsec:discussion_stardust_performance}.

\subsubsection{SN\,II vs SN\,Ib/c} \label{subsec:results_ia_ii_ibc}

We also studied the average P$_\mathrm{SD}$(Ia), P$_\mathrm{SD}$(II), and P$_\mathrm{SD}$(Ib/c) outputs for each mock SN type to evaluate whether \texttt{STARDUST2} can distinguish CC\,SN types with one input SED (also explored in  Section \ref{subsubsec:single_epoch_test} for the spectroscopically-classified JTS sources). 
The SN\,Ia versus SN\,II versus SN\,Ib/c confusion matrices for the spec-$z$ and photo-$z$ mock SN samples are shown in the top right panel of Figures \ref{fig:confusion_matrices_zspec} and \ref{fig:confusion_matrices_zphot}, respectively. In both the spec-$z$ and photo-$z$ cases, \texttt{STARDUST2} correctly classifies SNe\,Ib/c with 
P$_\mathrm{SD}$(Ib/c)\,$=$\,0.29 on average. On the contrary, \texttt{STARDUST2} misclassifies input SNe\,Ib/c as SNe\,II with 
P$_\mathrm{SD}$(II)\,$=$\,0.57 (0.55) for the spec-$z$ (photo-$z$) case. \texttt{STARDUST2} correctly classifies SNe\,II with 
P$_\mathrm{SD}$(II)\,$=$\,0.50 (0.48) on average. However, it misclassifies input SNe\,II as SNe\,Ib/c with 
P$_\mathrm{SD}$(Ib/c)\,$=$\,0.42 (0.43) on average. Given this high degree of SN\,II and SN\,Ib/c confusion, we do not attempt to compute rates for the individual CC\,SN types. We explore some common SNe\,II vs SNe\,Ib/c misclassification scenarios in Appendix \ref{appendix:stardust_success_failure}.



\subsubsection{Individual SN Subtypes} \label{subsec:results_individual_subtype}

\begin{figure*}
    \centering
 {\includegraphics[width=0.49\linewidth]{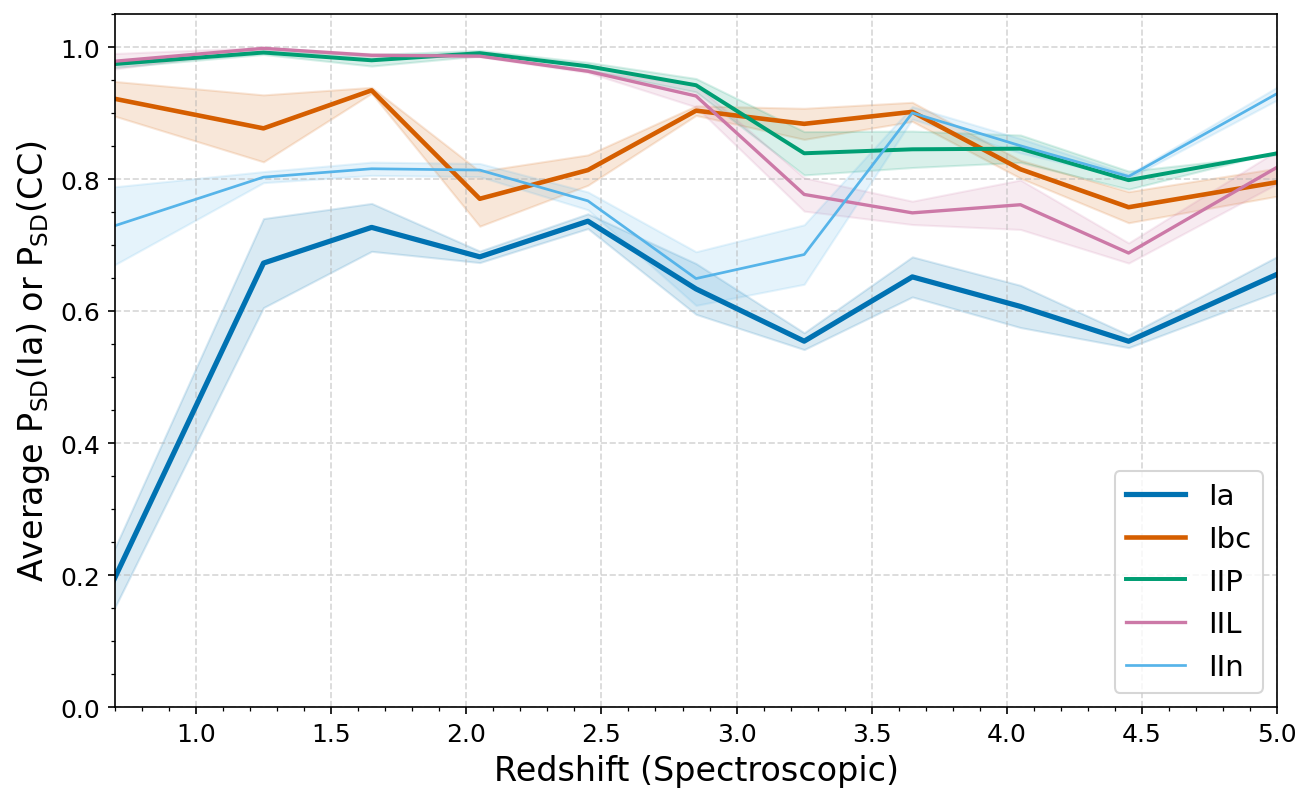}}
{\includegraphics[width=0.49\linewidth]{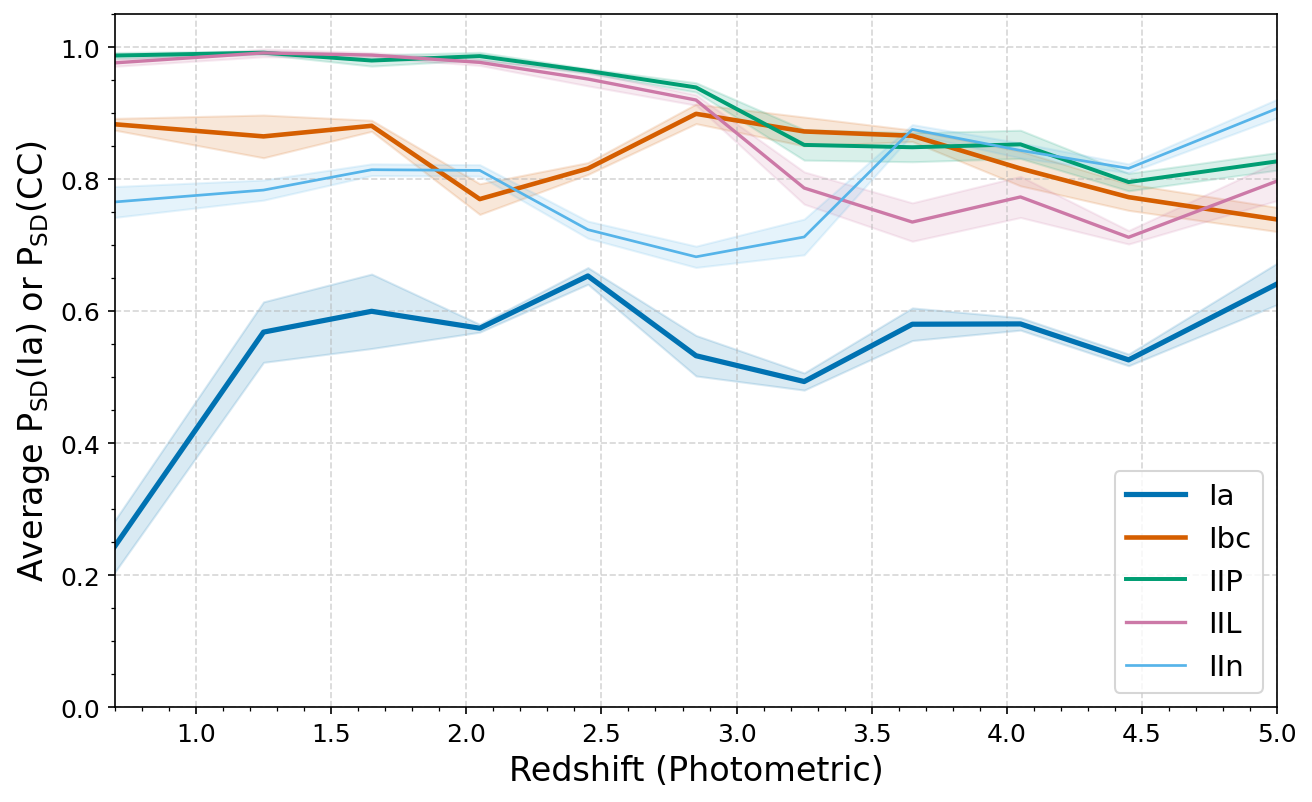} }
    \caption{Average P$_\mathrm{SD}$(Ia) and P$_\mathrm{SD}$(CC) output for input SNe\,Ia and CC\,SNe, respectively, as a function of redshift. The y-axis shows the true positive classification rate for each input SN subtype. The left (right) plot shows results for the spec-$z$ (photo-$z$) mock SN samples. The curves are colored according to input SN subtype. We have binned the redshifts from $z$\,$=$\,0.7 to $z$\,$=$\,5 with $\delta_z$\,$=$\,0.4, and the shaded regions show the standard error on the mean.
    }
    \label{fig:classification_accuracy_vs_z}
\end{figure*}

We also constructed confusion matrices showing the average P$_\mathrm{SD}$(Ia), P$_\mathrm{SD}$(II), and P$_\mathrm{SD}$(Ib/c) predictions for each individual input SN subtype. These confusion matrices are presented in the bottom panel of Figures \ref{fig:confusion_matrices_zspec} and \ref{fig:confusion_matrices_zphot} for the spec-$z$ and photo-$z$ cases, respectively. 
Figure \ref{fig:classification_accuracy_vs_z} shows, as a function of redshift, the average P$_\mathrm{SD}$(CC) output for each input CC\,SN subtype and the average P$_\mathrm{SD}$(Ia) output for input SNe\,Ia. In other words, the y-axis represents the average TPR for each input SN subtype, where we have combined P$_\mathrm{SD}$(II) and P$_\mathrm{SD}$(Ib/c) into P$_\mathrm{SD}$(CC).

For SNe with both spectroscopic and photometric redshifts, SNe\,Ib/c are correctly classified as CC\,SNe with P$_\mathrm{SD}$(CC)\,$\sim$\,0.75--0.90 across the 0.7\,$\leq$\,$z$\,$\leq$\,5 range, and 
SNe\,IIP are correctly classified as CC\,SNe with P$_\mathrm{SD}$(CC)\,$\gtrsim$\,0.80 across the full redshift range. SNe\,IIL follow a similar trend as SNe\,IIP, though with slightly lower P$_\mathrm{SD}$(CC) average outputs at $z$\,$\gtrsim$\,2.5. \texttt{STARDUST2} struggles to identify SNe\,IIn as CC\,SNe more than other CC\,SN subtypes, likely due to the removal of the Type\,IIn \texttt{snana-2006ez} template from the template library (see Appendix \ref{appendix:tr16}). We explore the common misclassification scenarios of the individual SN\,II subtypes in Appendix \ref{appendix:stardust_success_failure}.

For each CC\,SN subtype, P$_\mathrm{SD}$(CC)\,$\gtrsim$\,0.60 at every redshift.
The same is not true for P$_\mathrm{SD}$(Ia) with respect to input SNe\,Ia. At $z$\,$=$\,0.7, input SNe\,Ia are only correctly classified as SNe\,Ia with P$_\mathrm{SD}$(Ia)\,$\sim$\,0.20 (0.25) for the spec-$z$ (photo-$z$) case. This rises to P$_\mathrm{SD}$(Ia)\,$\sim$0.60--0.70 (0.50--0.65) in the $z$\,$\sim$\,1.3--5.0 range for the spec-$z$ (photo-$z$) case. Section \ref{subsec:discussion_stardust_performance} discusses the reason for this redshift dependence on SN\,Ia classification accuracy.


\subsubsection{Misclassification Correction Attempts} \label{subsec:results_correction_factors}


In order to account for potential misclassification of the single-SED JTS sources in our CC\,SN and SN\,Ia rate calculations, we attempted to apply correction factors to their P$_\mathrm{SD}$(Ia) and P$_\mathrm{SD}$(CC) outputs. We used Bayes's Theorem to compute P(Ia$\vert$Ia$^\prime$) and P(CC$\vert$CC$^\prime$). P(Ia$\vert$Ia$^\prime$) is the probability that a source is truly an SN\,Ia given that \texttt{STARDUST2} assigns it as an SN\,Ia, and P(CC$\vert$CC$^\prime$) is the probability that a source is truly a CC\,SN given that \texttt{STARDUST2} assigns it as a CC\,SN. We describe the details of this misclassification correction method in Appendix \ref{appendix:bayesian_method}. However, this correction method was rather ineffective because most single-SED JTS sources had output \texttt{STARDUST2} probabilities of P$_\mathrm{SD}$(CC)\,$=$\,1 and P$_\mathrm{SD}$(Ia)\,$=$\,0. Applying Bayes’s Theorem to a 100\% (0\%) prior probability yields an output probability of 100\% (0\%), so most single-SED JTS classification
probabilities were completely unchanged. 

As an alternative, we tested a frequentist method of single-SED misclassification correction, described in Appendix \ref{appendix:frequentist_method}. With this method, we used the observed number of single-SED JTS CC\,SNe and SNe\,Ia per redshift bin, along with average redshift-binned CC\,SN and SN\,Ia TPRs, to estimate the true number of single-SED CC\,SNe and SNe\,Ia that would be observed in the absence of misclassification. Unfortunately, the statistical and systematic uncertainties associated with the corrected CC\,SN and SN\,Ia counts were too large to yield meaningful constraints on the classifications. 

Following the failures of these misclassification correction methods, we chose not to apply them in our CC\,SN and SN\,Ia rate calculations. Rather, we simply present the full-sample CC\,SN and SN\,Ia rates with the caveat that they are not corrected for potential misclassification of the single-SED sources.

\section{Determining Volumetric Rates} \label{sec:methods_rates}

\subsection{CC\,SN Rate Calculation} \label{subsec:methods_ccsn_rates}


The analysis employs a probabilistic ``control time" formalism that rigorously accounts for survey cadence, depth, area, classification uncertainty, redshift uncertainty, and intrinsic diversity in SN properties. For a given SN subtype, redshift, and survey depth, we define the SN's ``visibility window" ($w_\mathrm{vis}$) as the average duration for which its light curve remains above the survey's detection threshold. The SN's control time is then its effective detectability window for the full survey after folding in survey cadence. For a sparsely-cadenced survey (cadence\,$>$\,visibility window), an SN's control time is its visibility window multiplied by the number of discovery epochs in the survey. However, for surveys with a cadence shorter than the visibility window, an SN's control time is the survey duration. If SNe that appear in the first discovery epoch are also included in the sample (i.e., SNe that exploded prior to the start of the survey, which we call ``pre-survey SNe"), then the SN's control time is the survey duration summed with its visibility window. In defining these terms, we assume that all discovery epochs are compared against the same template epoch taken well before the start of the survey. 

Since the JADES observations were not specifically designed for SN science, they do not exactly follow the conventions defined above. A template image is inherently defined to contain no SN light, but for the JTS, Epoch1 acts as a template for Epoch2 and vice-versa. This dual template and discovery role for the Epoch1 and Epoch2 images works, however, because they are separated by one observer-frame year, which is longer than the visibility window for the majority of SN types at any redshift. This scenario approximates the sparsely-cadenced survey scenario, in which an SN's control time is its visibility window multiplied by the number of discovery epochs.

{\bf (i) Estimating the volumetric rates by assuming intrinsic subtype fractions ({\boldmath $f_{\mathrm{int,}i}$})} --- From the total volumetric CC\,SN rate, $R_\mathrm{CC}$, within a given redshift interval, the rate of $i$th subtype CC\,SN, $R_{\mathrm{CC,}i}$, is derived as,
\begin{equation} \label{eq:rate_derivation_eq1}
  R_{\mathrm{CC,}i} = R_\mathrm{CC}\,\cdot\,f_{\mathrm{int,}i},
\end{equation}
where $f_{\mathrm{int,}i}$ is the $i$th subtype's intrinsic volume-limited fractional contribution to the CC\,SN population. Then, the observed number of the $i$th CC\,SN subtype, $N_{\mathrm{obs,CC,}i}$, can be expressed as,
\begin{equation} \label{eq:rate_derivation_eq2}
  N_{\mathrm{obs,CC,}i} = R_{\mathrm{CC,}i}\,\cdot\,t_{\mathrm{c,}i}\,\cdot\,\Delta V,
\end{equation}
where $t_{\mathrm{c,}i}$ is the $i$th subtype's control time and $\Delta V$ is the comoving survey volume. By inserting Equation~\ref{eq:rate_derivation_eq1}, this becomes,
\begin{equation}
    N_{\mathrm{obs,CC,}i} = R_\mathrm{CC}\,\cdot\,f_{\mathrm{int,}i}\,\cdot\,t_{\mathrm{c,}i}\,\cdot\,\Delta V. 
\end{equation}
The observed total number of CC\,SNe, $N_\mathrm{obs,CC}$, can then be written as,
\begin{eqnarray}
\label{eq:rate_derivation_eq4}
  N_\mathrm{obs,CC} &=& \sum_i{N_{\mathrm{obs,CC,}i}} \nonumber \\ \label{eq:rate_derivation_eq4}
  &=& R_\mathrm{CC}\,\cdot\,\Delta V \,\sum_i{f_{\mathrm{int,}i}\,\cdot\,t_{\mathrm{c,}i}}. 
\end{eqnarray}
Rearranging this equation will lead to the following expression for the total volumetric CC\,SN rate, $R_\mathrm{CC}$:
\begin{equation} \label{eq:cc_rate_fvol}
  R_\mathrm{CC} = \frac{N_\mathrm{obs,CC}}{\Delta V \sum_i t_{\mathrm{c,}i} \cdot f_{\mathrm{int,}i}}.
\end{equation}
To calculate $R_\mathrm{CC}$, we will iterate over CC\,SN subtypes, $i$, within a given redshift interval.

In our calculation, we assume that the intrinsic volume-limited CC\,SN subtype fractions do not evolve with redshift. This is unlikely to be true, but we currently have no measurements of the intrinsic or even the observed CC\,SN subtype fractions at high-$z$, so we adopt this assumption in this work.

{\bf (ii) Deriving the volumetric rates by measuring observed subtype fractions ({\boldmath $f_{\mathrm{obs,}i}$})} --- Future high-$z$ SN surveys, such as the High-Latitude Time Domain Survey (HLTDS) with the Nancy Grace Roman Space Telescope (Roman) and the JADES Extended Transient Survey \citep{egami2025}, will provide the first measurements of CC\,SN subtype fractions in the high-$z$ Universe (see Section \ref{subsec:roman}). For these future surveys, the $i$th SN subtype's fractional contribution to the {\it observed} SN sample, $f_{\mathrm{obs,}i}$, can be used instead of $f_{\mathrm{int,}i}$, which we do not know a priori for the high-redshift CC\,SN population. 

With $f_{\mathrm{obs,}i}$, $N_{\mathrm{obs,CC,}i}$ relates to $N_\mathrm{obs,CC}$ simply as,
\begin{equation}
\label{eq:cc_rate_fobs_derivation1}
N_{\mathrm{obs,CC},i} = N_\mathrm{obs,CC} \cdot f_{\mathrm{obs},i}.
\end{equation}
Inserting this into Equation~\ref{eq:rate_derivation_eq2}, we obtain,
\begin{equation} \label{eq:cc_rate_fobs_derivation2}
  R_{\mathrm{CC,}i} = \frac{N_\mathrm{obs,CC}\cdot f_{\mathrm{obs,}i}}{\Delta V \cdot t_{\mathrm{c},i}}.
\end{equation}
Since $R_\mathrm{CC}$ is simply,
\begin{equation}
\label{eq:cc_rate_fobs_derivation3}
R_\mathrm{CC} = \sum_i R_{\mathrm{CC},i},
\end{equation}
it can be written as follows by inserting Equation~\ref{eq:cc_rate_fobs_derivation2}:
\begin{equation} \label{eq:cc_rate_fobs}
  R_\mathrm{CC} = \frac{N_\mathrm{obs,CC}}{\Delta V}\sum_i\frac{f_{\mathrm{obs,}i}}{t_{\mathrm{c,}i}}
\end{equation}
The value of $f_{\mathrm{obs},i}$ would be determined via robust photometric and/or spectroscopic classification of the entire observed SN sample, which was not possible for the JTS sample.

Note that Equation~\ref{eq:cc_rate_fobs} is equivalent to Equation~\ref{eq:cc_rate_fvol} since,
\begin{equation}
   f_{\mathrm{obs,}i} = \frac{t_{\mathrm{c},i} \cdot f_{\mathrm{int},i}}{\sum_j t_{\mathrm{c},j} \cdot f_{\mathrm{int},j}},
\end{equation}
with the caveat that the observationally determined SN subtype fraction ($f_{\mathrm{obs},i}$) with a limited sample size will likely miss SN populations with a short control time ($t_{\mathrm{c},i}$)
and/or a small intrinsic fraction ($f_{\mathrm{int},i}$).

\subsection{Methodology}

Most JTS SNe lack spectroscopic classifications and some lack spectroscopic host redshifts, so both redshift and type classifications are treated probabilistically in calculating the observed (or ``effective") number of CC\,SNe in each redshift bin. For each candidate event $k$, the contribution to the CC\,SN count is weighted by the product of its host galaxy's redshift probability distribution, $P_k(z_\mathrm{host})$, and its CC\,SN classification probability, $P_k(CC)$. For sources whose hosts have spectroscopic redshifts, P$_k(z_\mathrm{host})$ is simply a delta function at the host galaxy spec-$z$. Sources that have been spectroscopically classified as CC\,SNe are assigned $P_k(CC)$\,$=$\,1, whereas for sources without spectroscopic classifications, $P_k(CC)$ is obtained from \texttt{STARDUST2} (see Section \ref{subsec:jades_sne}). The observed number of CC\,SNe in a redshift bin is then
\begin{equation} \label{eq:cc_num}
N_\mathrm{obs,CC}(z_1,z_2) = \int_{z_1}^{z_2}\sum\limits_{k=1}^{N} [{P}_{k}(z_\mathrm{host})\times\,{P}_k(CC)]\,dz,
\end{equation}
where $N$ is the total number of SNe in the sample, regardless of subtype and redshift.

The visibility window quantifies the time during which a CC\,SN of a given subtype, redshift, peak luminosity, age, and extinction would be detectable by a survey with a given depth in one epoch. Synthetic light curves are generated using the \texttt{nugent} models \citep{gilliland1999, levan2005} and evaluated at discrete rest-frame epochs. The synthetic light curves are generated in only one filter per redshift bin, so we select an observed JTS NIRCam filter for each redshift bin that closely corresponds to a rest-frame SDSS filter for which SN light curve evolution is well-understood. See Appendix \ref{appendix:boomrate} for more details.

The \texttt{nugent} models include SNe\,IIP, SNe\,IIL, SNe\,IIn, and SNe\,Ib/c. Although SNe\,Ib and SNe\,Ic exhibit similar light curve evolution and are both represented by the \texttt{nugent-sn1bc} model, they differ in their peak absolute magnitude distributions and volume-limited subtype fractions. We therefore generate separate synthetic light curves for the two subtypes, adopting distinct peak magnitude distributions and subtype fractions. We apply the intrinsic volume-limited subtype fractions from \citet{li2011_fractions}. There is no \texttt{nugent} model for SNe\,IIb and generating an SN\,IIb model is beyond the scope of this paper, so we do not directly include SNe\,IIb in the rate calculation. However, we add the SN\,IIb volume-limited fraction ($\sim$10\%) to the SN\,Ib fraction to increase the relative contribution of SNe\,Ib to the ``weighted" control time in the denominator of Equation \ref{eq:cc_rate_fvol}. This indirectly accounts for SNe\,IIb in the rate calculation, as SNe\,IIb and SNe\,Ib exhibit similar light curve properties and thus would have similar visibility windows \citep{arcavi2012}.

Model fluxes are compared against detection efficiency curves derived from point source injection and recovery tests. Appendix \ref{appendix:boomrate} presents the details of how we derived the detection efficiency curves. By comparing the model fluxes to the survey efficiency in detecting objects of similar flux, we determine the likelihood, $\epsilon$, of detecting a CC\,SN model at a given light-curve age. We integrate $\epsilon$ over the full rest-frame light-curve evolution to yield the subtype- and redshift-dependent visibility window:
 \begin{equation}\label{eq:visibility_window}
     w_\mathrm{vis}(\beta) = \int_{t_\mathrm{min}}^{t_\mathrm{max}} \epsilon(\beta, t) dt,
 \end{equation}
where $t_\mathrm{min}$ and $t_\mathrm{max}$ are the rest-frame light curve bounds, and $\beta$ contains adjustable parameters of the SN model (subtype and redshift) and the survey depth. 

To account for intrinsic diversity within each CC\,SN subtype, visibility windows are integrated over peak luminosity probability distributions, expressed in B-band absolute magnitudes. They are also integrated over dust extinction internal to the CC\,SN host galaxies:
 \begin{equation}\label{eq:visibility_window_dm_da}
 \begin{split}
     w_\mathrm{vis}(\beta) = \int_{M_\mathrm{B,max}}^{M_\mathrm{B,min}} \int_0^{A_\mathrm{V,max}} \int_{t_\mathrm{min}}^{t_\mathrm{max}} \epsilon(\beta;t,M_\mathrm{B},A_\mathrm{V})P(M_\mathrm{B}) \\ \times P(A_\mathrm{V})\,dM_\mathrm{B}\,dA_\mathrm{V}\,dt
\end{split}
 \end{equation}
where $M_\mathrm{B,max}$ and $M_\mathrm{B,min}$ refer to the maximum (faintest) and minimum (brightest) peak B-band SN magnitudes that we consider, and $A_{\mathrm{V,\,max}}$ refers to the maximum V-band host extinction that we consider. For the $M_\mathrm{B}$ bounds, we use the $\mu$\,$\pm$\,3$\sigma$ peak B-band SN brightnesses from \citet{richardson2014}. We apply host extinction in the SN rest-frame ($A_{\lambda\mathrm{,rest}}$\,$=$\,[0,10]), but we compute the likelihood of each amount of host extinction by converting it to the rest-frame V-band and comparing it to the $A_\mathrm{V}$ host extinction distribution from \citet{kelly2012}. We convert the host extinction from the SN rest-frame to the rest-frame V-band using the Calzetti law \citep{calzetti2000}. Equation \ref{eq:visibility_window_dm_da} is used to compute the visibility window for each CC\,SN subtype at seven equally-spaced redshifts between $z_1$ and $z_2$, and we adopt the volume-weighted average of those values as that subtype's visibility window for the [$z_1$,$z_2$] redshift bin. The redshift corresponding to this volume-weighted visibility window is the redshift bin's ``effective redshift," $z_\mathrm{eff}$.

The control time calculation determines the total time for which the SN is visible to our survey in the rest frame, which is dependent on the survey cadence, number of discovery epochs, and SN visibility window:

 \begin{equation}\label{eq:control_time}
t_\mathrm{c} = 
\begin{cases} 
N_\mathrm{epoch} \times w_\mathrm{vis} & \text{if } w_\mathrm{vis} \leq t_\mathrm{cad}, \\[+8pt]
N_\mathrm{epoch} \times t_\mathrm{cad} & \text{if } w_\mathrm{vis} > t_\mathrm{cad}, \\[-4pt]
                                       & \phantom{\text{if }}\mathrm{pre\mathrm{-}survey\,SNe\,excluded} \\[-2pt]
N_\mathrm{epoch} \times t_\mathrm{cad} + w_\mathrm{vis} & \text{if } w_\mathrm{vis} > t_\mathrm{cad}, \\[-4pt]
                                       & \phantom{\text{if }}\mathrm{pre\mathrm{-}survey\,SNe\,included} \\
\end{cases}
 \end{equation}
where $t_\mathrm{cad}$ is the rest-frame survey cadence evaluated at the effective redshift, and $N_\mathrm{epoch}$ is the number of discovery epochs. These equations assume the survey cadence is the same between each epoch, but different components of Equation \ref{eq:control_time} can be combined for a survey with variable cadence. Furthermore, the various SN subtypes will have different visibility windows, so different components of Equation \ref{eq:control_time} may be required for the various subtypes. These equations are only applicable for computing rates of normal CC\,SNe. In order to calculate the rates of exotic SNe that evolve slowly over periods of time that may exceed the survey duration (e.g., pair-instability SNe), more complicated processes are required \citep[e.g.,][]{gabrielli2024}.



\subsection{SN\,Ia Rate Calculation} \label{subsec:methods_snia_rates}

A similar control time procedure is applied to calculate the volumetric SN\,Ia rate.
The volumetric SN Ia rate in a redshift interval $[z_1,z_2]$ is
\begin{equation}\label{eq:Ia_rate}
  R_{\rm Ia} = \frac{N_{\rm obs,Ia}}{t_\mathrm{c}\,\Delta V} ,
\end{equation}
where $N_\mathrm{obs}$ is the observed number of SNe\,Ia in the redshift interval, $\Delta V$ is the comoving volume subtended by the survey solid angle over the redshift interval, and $t_\mathrm{c}$ is volume-weighted average of the SN Ia control times in the [$z_1$, $z_2$] redshift bin. As with CC\,SNe, both redshift and type are handled probabilistically for SNe\,Ia.

To compute the SN\,Ia visibility window, we generated synthetic SN\,Ia light curves using the \texttt{nugent-sn1a} model \citep{nugent2002}. We used the peak magnitude distribution from \citet{richardson2014} and the host galaxy extinction distribution from \citet{jha2007}. 




\section{Results} \label{sec:results}

\subsection{Volumetric Core-Collapse Supernova Rates} \label{subsec:results_cc_rates}

We computed CC\,SN rates based on (1) the full JTS CC\,SN sample, including sources that were classified with just one SED and (2) the sample of JTS sources that were classified with either a multi-epoch light curve or a spectrum. We consider this second sample to be our ``gold sample," as the classifications based on multi-epoch light curve fitting and spectroscopy are significantly more robust than the photometric classifications based on just one SED. Because we did not correct the full-sample CC\,SN rates for potential single-SED misclassification, we expect that SN\,Ia contamination affects those rates.

For each sample, we define the redshift bins such that the total number of observed SNe (N$_\mathrm{obs,tot}$) is distributed as evenly as possible among the bins. This ensures comparable statistical uncertainties for each rate within the sample. Because each SN contributes to the counts according to its host redshift probability distribution, exact balance is not possible. Each bin excludes its lower edge and includes its upper edge.

\begin{figure*}
    \centering
 {\includegraphics[width=1\linewidth]{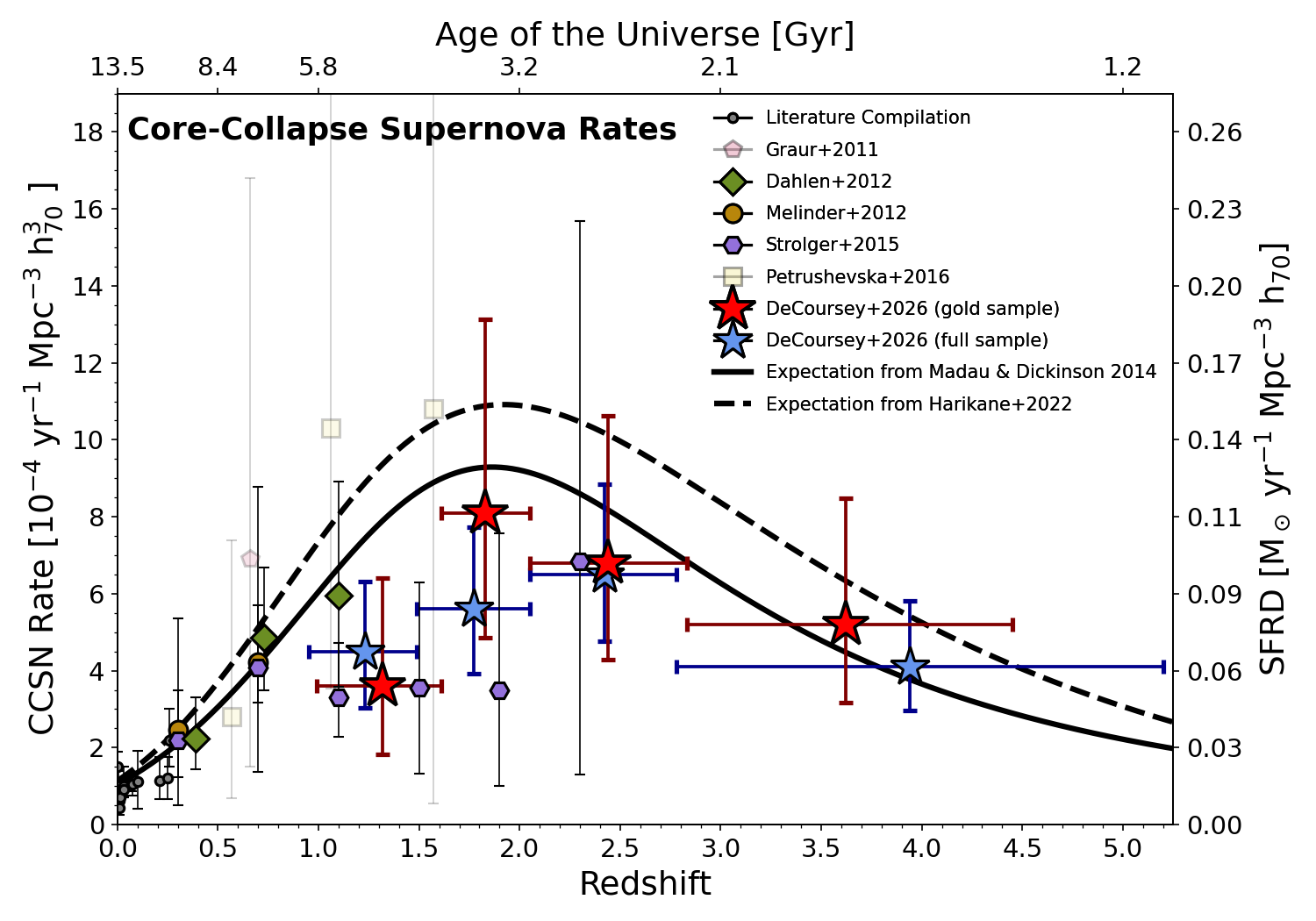}}
    \caption{The redshift-binned volumetric JTS CC\,SN rates. The red stars and blue stars are the volumetric CC\,SN rates arising from the JTS gold sample and full sample, respectively. The x-axis errorbars define the redshift bins, and the y-axis errorbars show the combined statistical and systematic CC\,SN rate uncertainties. As a comparison, we display rates from \citet{petrushevska2016} as faint yellow squares, \citet{strolger2015} as purple hexagons, \citet{melinder2012} as bronze circles, \citet{dahlen2012} as green diamonds, and \citet{graur2011} as faint pink pentagons. Lower redshift CC\,SN rates are compiled from the literature and shown as small gray circles \citep{cappellaro1999, cappellaro2005, botticella2008, li2011_rates, botticella2012, mattila2012, taylor2014, cappellaro2015, graur2015, perley2020, frohmaier2021, ma2025, pessi2025}. The right-side y-axis maps cosmic SFRD to the CC\,SN rate assuming that every 8\msun\,$\leq$\,$M$\,$\leq$\,50\msun\ star explodes as a CC\,SN and that star formation follows a Salpeter IMF \citep{salpeter1955}. Following these assumptions, the solid and dashed black lines show the expected CC\,SN rates based on the cosmic SFRD from \citet{madau2014} and \citet{harikane2022}, respectively.}
    \label{fig:cc_rates}
\end{figure*}

The full sample and gold sample redshift-binned JTS volumetric CC\,SN rates are presented in Figure \ref{fig:cc_rates} and Table \ref{tab:ccsn_rates}. The statistical uncertainties on the rates listed in Table \ref{tab:ccsn_rates} are based on Poisson and binomial statistics for small samples \citep{gehrels1986}. These statistical uncertainties are calculated with N$_\mathrm{obs,filt}$. This quantity represents the observed number of CC\,SNe in each redshift bin that are detected in the NIRCam filter selected for that bin's visibility window calculation. As detailed in Appendix \ref{appendix:boomrate}, the visibility window calculation requires the selection of one representative filter for each redshift bin. The selected filter for each bin is also listed in Table \ref{tab:ccsn_rates}. We also provide an estimate of the systematic CC\,SN rate uncertainties arising from misclassification. These are listed in parentheses in Table \ref{tab:ccsn_rates}. Refer to Appendix \ref{appendix:systematics} for an explanation of how these systematic uncertainties were estimated.

As seen in Figure~\ref{fig:cc_rates}, the gold and full samples show consistent rates over the full redshift range probed by the JTS ($z$\,$\sim$\,1--5) within the measurement uncertainties. Taken together, they show enhanced rates at $z$\,$\sim$\,2, which is consistent with the cosmic SFRD histories shown as the solid and dashed black curves in the figure (see Section~\ref{subsec:csfrd_comparison} for how these curves were produced by converting SFRDs to CC\,SN rates).
For each redshift bin, the gold- and full-sample rates show $\sim$1--2$\sigma$ agreement with the SFRD expectations although they are consistently lower in all redshift bins except for the highest-$z$ one. Examined in detail, the two samples show slightly different trends: the gold sample shows a clear peak in the $z$\,$=$\,1.61--2.05 bin, whereas the full sample shows a broad peak over $z$\,$=$\,1.49--2.05 and $z$\,$=$\,2.05--2.78. Both samples, however, show a decline in rate towards their highest-redshift bins. 

Figure \ref{fig:cc_rates} also plots a collection of literature rates. There are three sets of literature rates that can be directly compared to our CC\,SN rates in the 1\,$\lesssim$\,$z$\,$\lesssim$\,2.5 regime: those from \citet{dahlen2012} (green diamonds),  \citet{strolger2015} (purple hexagons), and \citet{petrushevska2016} (faint yellow squares). 
The large uncertainties with some of the measurements make a meaningful comparison difficult, but our measurements are broadly consistent with the previous results covering the same redshift range.
At $z$\,$>$\,3, this work provides the first CC\,SN rate measurements, so there is no existing result to compare directly.



\begin{deluxetable*}{cccccc}
\tablecaption{JTS Volumetric Core-Collapse Supernova Rates}
\tablehead{
\colhead{Redshift Range} & \colhead{z$^a_\mathrm{eff}$} & \colhead{Filter$^b$} & \colhead{N$^c_\mathrm{obs,tot}$} & \colhead{N$^d_\mathrm{obs,filt}$} & \colhead{R$^e_\mathrm{CC}$}
} 
\label{tab:ccsn_rates}
\startdata
\multicolumn{3}{l}{\textbf{Full Sample}} \\
0.95\,$<$\,$z$\,$\leq$\,1.49 & 1.23 & F115W & 13.7 & 10.7 & 4.5$^{+1.8(0.3)}_{-1.3(0.7)}$ \\ 
1.49\,$<$\,$z$\,$\leq$\,2.05 & 1.77 & F150W & 14.8 & 12.2 & 5.6$^{+2.1(0.4)}_{-1.6(0.5)}$ \\ 
2.05\,$<$\,$z$\,$\leq$\,2.78 & 2.42 & F200W & 14.6 & 13.6 & 6.5$^{+2.3(0.5)}_{-1.7(0.3)}$ \\ 
2.78\,$<$\,$z$\,$\leq$\,5.20 & 3.94 & F277W & 14.4 & 13.4 & 4.1$^{+1.4(1.0)}_{-1.1(0.3)}$ \\ 
\hline
\multicolumn{3}{l}{\textbf{Gold Sample}} \\
0.99\,$<$\,$z$\,$\leq$\,1.61 & 1.32 & F115W & 6.0 & 4.0 & 3.6$^{+2.8(0.2)}_{-1.7(0.5)}$ \\ 
1.61\,$<$\,$z$\,$\leq$\,2.05 & 1.83 & F150W & 5.8 & 5.8 & 8.1$^{+5.0(0.5)}_{-3.2(0.5)}$ \\ 
2.05\,$<$\,$z$\,$\leq$\,2.83 & 2.44 & F200W & 6.5 & 6.5 & 6.8$^{+3.8(0.4)}_{-2.5(0.2)}$ \\ 
2.83\,$<$\,$z$\,$\leq$\,4.45 & 3.62 & F277W & 6.1 & 6.1 & 5.2$^{+3.1(1.1)}_{-2.0(0.4)}$ \\ 
\enddata
\tablecomments{$^a$$z_\mathrm{eff}$ is the effective volume-weighted redshift of the redshift bin.}
\tablecomments{$^b$Observed NIRCam filter used for the visibility window calculation. See Appendix \ref{appendix:boomrate} for more details.}
\tablecomments{$^c$N$_\mathrm{obs,tot}$ is the total number of observed CC\,SNe in that redshift bin, including sources not detected in the selected filter.}
\tablecomments{$^d$N$_\mathrm{obs,filt}$ is the number of observed CC\,SNe in that redshift bin that were detected in the selected filter. This value is used in the rate calculation (see Equation \ref{eq:cc_rate_fvol}).}
\tablecomments{$^e$In units of 10$^{-4}$ CC\,SNe year$^{-1}$ comoving Mpc$^{-3}$ with statistical uncertainties listed first and systematic uncertainties listed in parentheses.}
\end{deluxetable*}

\subsection{Volumetric Type Ia Supernova Rates} \label{subsec:results_ia_rates}

As with  the CC\,SN rates, we computed JTS SN\,Ia rates with two different samples: the full sample and ``gold sample." 
We expect that the full sample SN\,Ia rates suffer from two competing biases, with one arising from true CC\,SNe being misclassified as SNe\,Ia (false positives) and the other arising from true SNe\,Ia being misclassified as CC\,SNe (false negatives). 

\begin{figure*}
    \centering
 {\includegraphics[width=1\linewidth]{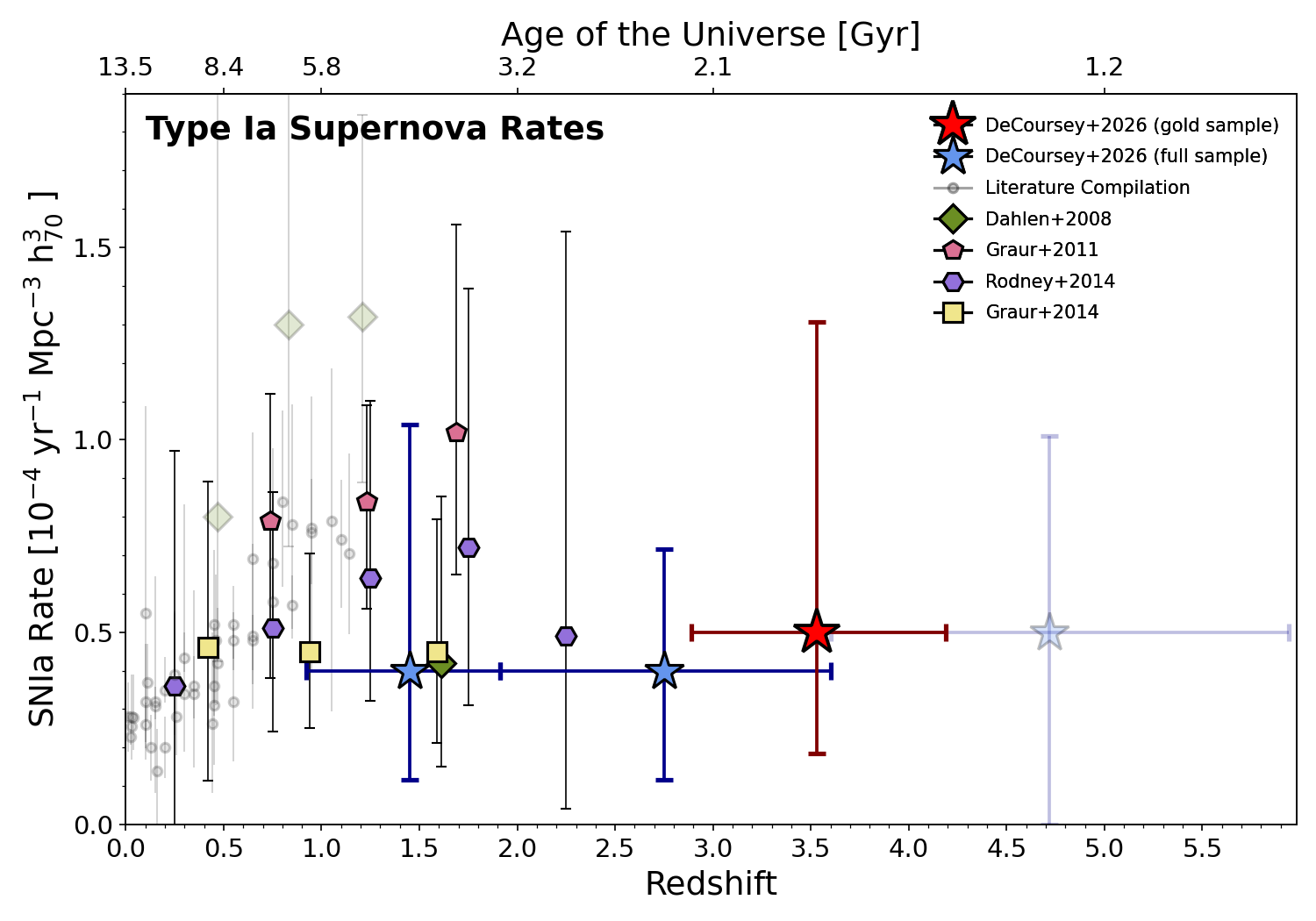}}
    \caption{The volumetric SN\,Ia rates derived from the JTS sample. The red stars and blue stars are the volumetric JTS SN\,Ia rates arising from the gold sample and full sample, respectively. The x-axis errorbars define the redshift bins, and the y-axis errorbars show the combined statistical and systematic SN\,Ia rate uncertainties. The highest-redshift blue star is translucent because we suspect this high-$z$ rate is contaminated with misclassified CC\,SNe, artificially raising the rate (Section \ref{subsec:ia_comparison}). As a comparison, we display rates from \citet{graur2014} as yellow squares, \citet{rodney2014} as purple hexagons, \citet{graur2011} as pink pentagons, and \citet{dahlen2008} as green diamonds. Lower redshift SN\,Ia rates are compiled from the literature and shown as small gray circles \citep{cappellaro1999, pain2002, madgwick2003, strolger2003, tonry2003, blanc2004, mannucci2005, neill2006, botticella2008, horesh2008, dilday2010, rodney2010, perrett2012, okumura2014, cappellaro2015, desai2024, desai2026}.}
    \label{fig:ia_rates}
\end{figure*}

We present the full sample and gold sample volumetric SN\,Ia rates in Figure \ref{fig:ia_rates} and Table \ref{tab:snia_rates}. The full-sample rates are broken up into three redshift bins: $z$\,$=$\,0.92--1.91, $z$\,$=$\,1.91--3.60, and $z$\,$=$\,3.60--5.94. The gold sample only has one redshift bin at 
$z$\,$=$\,2.89--4.19 due to the scarcity of high-$z$ SNe Ia with a multi-epoch light curve or spectrum. For each sample, we select the redshift bins to balance the total number of observed SNe\,Ia in each bin. The statistical uncertainties listed in Table \ref{tab:snia_rates} were computed in the same way as for the CC\,SN rates.
Systematic rate uncertainty estimates are also listed in Table \ref{tab:snia_rates} in parentheses, which account for the aforementioned misclassification biases (Appendix \ref{appendix:systematics}).

The gold and full samples together show flat SN Ia rates from $z$\,$\sim$\,1 to $\sim$\,6 although the error bars are large due to the limited sample size and misclassification systematics. We caution that the rate in the highest-redshift bin ($z$\,$=$\,3.60--5.94) may be artificially boosted due to CC\,SN contamination (see Section \ref{subsec:ia_comparison}).  Therefore, it is shown as a translucent symbol.

For comparison, we also plot in Figure~\ref{fig:ia_rates} previous SN\,Ia rates results from \citet{dahlen2008}, \citet{graur2011}, \citet{rodney2014}, and \citet{graur2014}.  Our rates in the two lower-redshift bins are broadly consistent with these previous measurements, but are at the lower end of the distribution.  At $z$\,$>$\,3, this work provides the first SN\,Ia rate measurements, so there is no existing result to compare directly.




\begin{deluxetable*}{cccccc}
\tablecaption{JTS Volumetric Type Ia Supernova Rates}
\tablehead{
\colhead{Redshift Range} & \colhead{$z^a_\mathrm{eff}$} & \colhead{Filter$^b$} & \colhead{N$^c_\mathrm{obs,tot}$} & \colhead{N$^d_\mathrm{obs,filt}$} & \colhead{R$^e_\mathrm{Ia}$} 
} 
\label{tab:snia_rates}
\startdata
\multicolumn{3}{l}{\textbf{Full Sample}} \\
0.92\,$<$\,$z$\,$\leq$\,1.91 & 1.45 & F115W & 2.2 & 2.2 & 0.4$^{+0.4(0.5)}_{-0.2(0.2)}$ \\ 
1.91\,$<$\,$z$\,$\leq$\,3.60 & 2.75 & F200W & 3.3 & 3.3 & 0.4$^{+0.3(0.1)}_{-0.2(0.2)}$ \\ 
3.60\,$<$\,$z$\,$\leq$\,5.94 & 4.72 & F277W & 3.0 & 3.0 & 0.5$^{+0.5(0.1)}_{-0.3(0.4)}$ \\ 
\hline
\multicolumn{3}{l}{\textbf{Gold Sample}} \\
2.89\,$<$\,$z$\,$\leq$\,4.19 & 3.53 & F277W & 1.6 & 1.6 & 0.5$^{+0.8(0.1)}_{-0.3(0.1)}$ \\ 
\enddata
\tablecomments{$^a$$z_\mathrm{eff}$ is the effective volume-weighted redshift of the redshift bin.}
\tablecomments{$^b$Observed NIRCam filter used for the visibility window calculation. See Appendix \ref{appendix:boomrate} for more details.}
\tablecomments{$^c$N$_\mathrm{obs,tot}$ is the total number of observed SNe\,Ia in that redshift bin, including sources not detected in the selected filter.}
\tablecomments{$^d$N$_\mathrm{obs,filt}$ is the number of observed SNe\,Ia in that redshift bin that were detected in the selected filter. This value is used in the rate calculation (see Equation \ref{eq:Ia_rate}).}
\tablecomments{$^e$In units of 10$^{-4}$ SNe\,Ia year$^{-1}$ comoving Mpc$^{-3}$ with statistical uncertainties listed first and systematic uncertainties listed in parentheses.}
\end{deluxetable*}

\section{Discussion} \label{sec:discussion}

\subsection{Comparing the JTS CC\,SN Rates to the Literature} \label{subsec:cc_comparison}


Both our 0.99\,$<$\,$z$\,$\leq$\,1.61 gold and 0.95\,$<$\,$z$\,$\leq$\,1.49 full sample CC\,SN rates show 1$\sigma$ agreement with the $z$\,$=$\,1.10 rate from \citet{dahlen2012}. Our rates also agree with both the $z$\,$=$\,1.06 and $z$\,$=$\,1.57 rate measurements from \citet{petrushevska2016}. However, the uncertainties associated with the rates from \citet{petrushevska2016} are incredibly large due to small sample size, so 1$\sigma$ agreement with these rates is not very constraining. Our lowest-$z$ rates also agree at the 1$\sigma$ level with the $z$\,$=$\,0.9--1.3 and $z$\,$=$\,1.3--1.7 CC\,SN rates reported by \citet{strolger2015}. Their $z$\,$=$\,1.7--2.1 rate agrees with our $z$\,$=$\,1.49--2.05 full sample and $z$\,$=$1.61--2.05 gold sample rate at the 1$\sigma$ level, and their $z$\,$=$\,2.1--2.5 rate agrees with our $z$\,$=$\,2.05--2.78 full sample and $z$\,$=$\,2.05--2.83 gold sample rates at the 1$\sigma$ level. However, both our uncertainties and those from \citet{strolger2015} are large due to small sample size, so the agreement is not very constraining.

\subsection{Comparing the JTS CC\,SN Rates to Expectations Based on Cosmic SFRD} \label{subsec:csfrd_comparison}

Building on the promises of \citet{dahlen1999}, \citet{dahlen2012}, \citet{strolger2015}, and \citet{cappellaro2015}, CC\,SN rates provide an alternative and largely independent mechanism for probing the volumetric cosmic SFRD, thus confirming the cosmic history of stellar births as we know it, albeit from a stellar deaths perspective. Because CC\,SNe have short lifetimes relative to galaxy-scale changes, CC\,SN rates and the instantaneous SFR can be related in the following simple way: 

\begin{equation} \label{eq:ccsn_sfr_relation}
    \mathrm{R}_\mathrm{CC}(z) = k_\mathrm{CC} \times \rho_\mathrm{SFR}(z)
\end{equation}
where $\rho_\mathrm{SFR}$ is the cosmic SFRD and $k_\mathrm{CC}$ represents the number of stars per unit mass of star formation that produce CC\,SNe (units of M$_\odot^{-1}$), or:

\begin{equation}
    k_\mathrm{CC} = \frac{\int^{m_u}_{m_l} \phi(m)dm}{\int^{m_\mathrm{max}}_{m_\mathrm{min}}m\phi(m)dm} 
\end{equation}
where $\phi(m)$ is the initial mass function (IMF), $m_l$ and $m_u$ denote the bounds of the CC\,SN progenitor mass range, and $m_\mathrm{min}$ and $m_\mathrm{max}$ represent the limits of the full mass range of the IMF. We use this relation to convert the cosmic SFRDs derived in \citet{madau2014} and \citet{harikane2022} to expected CC\,SN rates (hereafter M\&D14 and H22, respectively). In performing this conversion, we assume a Salpeter IMF \citep{salpeter1955}, a stellar mass range of 0.1--120M$_\odot$, and a CC\,SN progenitor mass range of 8--50M$_\odot$. These parameters yield a $k_\mathrm{CC}$ value of $\sim$0.007 M$_\odot^{-1}$. 


Several systematic effects, separate from the misclassification uncertainty (Appendix \ref{appendix:systematics}), may contribute to an underestimation of the JTS CC\,SN rates relative to SFRD-based expectations as seen in Figure~\ref{fig:cc_rates}. However, our errors are dominated by misclassification systematics and small number statistics. The effects of ``missing” SNe due to high extinction in dust obscured host galaxies \citep[e.g.,][]{mattila2012} are considered separately in the companion paper \citep{vassallo2026} and here we report the SN rates without including corrections for missing SNe.

It is possible that the JTS CC\,SN rates are somewhat underestimated due to an oversimplification of source detection efficiency in the visibility window calculation. As detailed in Appendix \ref{appendix:boomrate}, the detection efficiency of only one JTS NIRCam filter is considered in determining an SN's contribution to the visibility window given a specific set of conditions (i.e., subtype, host extinction, peak luminosity, etc.). However, the JTS sample was constructed with multi-band selection criteria (see Section 3.2 of \citealt{decoursey2025_jts}). Since the JTS sample required detections in multiple bands for candidate inclusion, the JTS SNe were subjected to stricter selection requirements than those considered in our single-filter visibility window calculations. As a result, the detectability of SNe may be overestimated in the visibility window calculation, causing a potential underestimation of rates. We plan to update the visibility window algorithm for future rate calculations so that it can exactly replicate the selection criteria used to generate the SN sample of interest.  

As mentioned in Section \ref{subsec:jades_sne}, 17 JTS sources lack classifications, and we omitted non-classified sources from the rate calculation. Of these sources, nine lie at $z$\,$<$\,0.7, so their omission had no effect on our derived rates. However, the omission of these other eight unclassified sources may systematically lower our derived rates, assuming they lie at $z$\,$>$\,0.7. \tr{102} lies at $z_\mathrm{phot}$\,$=$\,3.96 but lacked sufficient photometric data to be classified. If it was classified as a CC\,SN with P$_\mathrm{SD}$(CC)\,$=$\,1, the $z$\,$=$\,2.78--5.20 full sample rate would increase by $\sim$7\%. \tr{16} lies at $z_\mathrm{spec}$\,$=$\,1.771 but its photometry yielded an ambiguous classification (see Appendix \ref{appendix:tr16}). If it was classified as a CC\,SN with P$_\mathrm{SD}$(CC)\,$=$\,1, the $z$\,$=$\,1.49--2.05 full sample rate would increase by $\sim$9\%. The six other omitted sources could not be classified because their redshifts were either unknown or unconstrained. This makes it more difficult to precisely determine the extent to which each redshift bin's rate is underestimated. However, if we uniformly spread the contribution of these sources across the four full sample redshift bins and assume that they are all classified with P$_\mathrm{SD}$(CC)\,$=$\,1, then the full sample CC\,SN rates would increase by $\sim$12\% across each redshift bin. In the most extreme scenario where all six omitted sources fall into the same redshift bin and each is classified with P$_\mathrm{SD}$(CC)\,$=$\,1, then the largest change that could result in any of the full sample redshift bins would be a $\sim$45--55\% increase.

We also note that the HUDF (JADES Deep Field) is under-dense in galaxy number density and thus SFRD, suggested by the surveys of X-ray AGNs \citep[e.g.,][]{cowie2002, moretti03, bauer2004}, star-forming Lyman-break galaxies \citep[e.g.,][]{oesch2007} and dust-obscured star-forming galaxies with ALMA \citep[e.g.,][]{dunlop2017,fujimoto2024,hill2024,sunf2025}. These surveys collectively suggest that galaxy number counts in the HUDF are $\sim$30--50\% lower than the cosmic mean. Although it is difficult to quantify the effect of this under-density for our particular SN redshift bins, it could in principle result in a non-negligible deficit that may further contribute to the lower measured CC\,SN rates relative to the global SFRD expectations. The SFRD-based rate expectations themselves are also subject to uncertainties in the assumed CC\,SN progenitor mass range and should therefore not be taken as ground truth. 

Although our CC\,SN rate measurements do not show a statistically significant decline beyond cosmic noon, this trend may emerge more strongly as we increase the size of our high-$z$ CC\,SN sample with future JWST SN surveys. We suspect that many of the $z$\,$\gtrsim$\,3 single-SED sources that have been partially classified as SNe\,Ia are actually CC\,SNe (see Section \ref{subsec:ia_comparison}), but we tested that even if every $z$\,$\gtrsim$\,3 single-SED SN is assigned P$_\mathrm{SD}$(CC)\,$=$\,1, the $z$\,$=$\,2.78--5.20 full sample CC\,SN rate is still lower than the $z$\,$=$\,2.05--2.78 rate.

The relative decrease between the $z$\,$=$\,2.05--2.83 and $z$\,$=$\,2.83--4.45 gold sample rates may be systematically underestimated due to a bias introduced by the selective follow-up observations that define the gold sample. These observations preferentially targeted very high-$z$ SNe and thus artificially enhanced the contribution of $z$\,$\gtrsim$\,3 events relative to lower-$z$ events in the gold sample. Of the four JTS SNe at $z$\,$>$\,0.7 that were discovered in Epoch2 but received no follow-up (and hence were excluded from the gold sample), one fell in the 0.99\,$<$\,$z$\,$\leq$\,1.61 bin, two in the 1.61\,$<$\,$z$\,$\leq$\,2.05 bin, and one in the 2.05\,$<$\,$z$\,$\leq$\,2.83 bin. As a test, we recomputed the CC\,SN gold sample rates including these sources and assuming the full areal coverage of the JADES Deep Field. We found that the $z$\,$=$\,2.83--4.45 rate was $\sim$30\% lower than the $z$\,$=$\,2.05--2.83 rate (as opposed to $\sim$25\% lower when considering the true gold sample), indicating that the aforementioned high-$z$ selection preference indeed biases the shape of the gold sample rates.


Further analysis on the relation between our observed CC\,SN rates and cosmic SFRD is beyond the scope of this paper. However, \citet{vassallo2026} delves into significantly more detail regarding the comparison between the JTS CC\,SN rates and expected CC\,SN rates derived from cosmic SFRD measurements. They investigate the effects of the assumed IMF and CC\,SN progenitor mass range on the relationship between star formation and CC\,SN production efficiency, and they explore the fraction of SNe missed by JTS due to varying levels of dust obscuration as a function of redshift.


\subsection{Comparing the JTS SN\,Ia Rates to the Literature} \label{subsec:ia_comparison}


Our $z$\,$=$\,0.92--1.91 SN\,Ia rate shows 1$\sigma$ agreement with the $z$\,$=$\,1.61 measurement from \citet{dahlen2008}. It also agrees with the $z$\,$=$\,1.23 and $z$\,$=$\,1.69 rates from \citet{graur2011}, $z$\,$=$\,1.25 and $z$\,$=$\,1.75 rates from \citet{rodney2014}, and $z$\,$=$\,1.59 rate from \citet{graur2014} at the 1$\sigma$ level. The uncertainties on our $z$\,$=$\,0.92--1.91 SN\,Ia rate are quite large given the small number of SNe\,Ia in this bin and the misclassification systematics, allowing our rate to simultaneously agree with each of these literature measurements. The \citet{dahlen2008} rates show a sharp decrease in SN\,Ia rates between $z$\,$\sim$\,1 and $z$\,$\sim$\,1.6, whereas the \citet{graur2011}, 
\citet{rodney2014}, and \citet{graur2014} rates allow an increase or flattening of SN\,Ia rates in this redshift regime. Because our redshift bin encompasses this full range, we cannot distinguish between these two scenarios. However, over the course of its mission, Roman will discover tens of thousands of SNe\,Ia, many of which will be at $z$\,$>$\,1 \citep{rose2021, roman2025}. The Roman sample of $z$\,$>$\,1 SNe\,Ia will provide much stronger constraints on the behavior of SN\,Ia rates in the $z$\,$\sim$\,1--2 regime (see Section \ref{subsec:roman}).

Our $z$\,$=$\,1.91--3.60 SN\,Ia rate shows 1$\sigma$ agreement with the $z$\,$=$\,2.25 rate from \citet{rodney2014}, though the uncertainties of each measurement are large. Given the uncertainties, it is not possible to make any definitive statement on the behavior of SN\,Ia rates at $z$\,$>$\,2. However, there is a spectroscopically-classified SN\,Ia in the JTS sample at $z$\,$=$\,2.90 \citep{pierel2024} and there are multiple other 2\,$<$\,$z$\,$<$\,3 SNe\,Ia that have been spectroscopically-classified with NIRSpec (e.g., \citealt{pierel2025, siebert2025}), so the SN\,Ia rate must be nonzero in the 2\,$<$\,$z$\,$\lesssim$\,3 regime. This is reflected in our non-zero gold sample rate at $z$\,$=$\,2.89--4.19. However, this sample contains only 1.6 sources, so the statistical uncertainties on the rate measurement are too large to draw any meaningful conclusions.

There are several JTS SNe that are omitted from the rate calculations because they lack classifications (Section \ref{subsec:jades_sne}). While it is unlikely that all of these omitted SNe are SNe\,Ia, it is plausible a subset of them are. Increasing the number of SNe\,Ia in the full sample redshift bins would substantially increase the associated SN\,Ia rates, as there are very few SNe\,Ia in each bin. Adding just one SN\,Ia to each full sample redshift bin would cause a $\sim$25--50\% increase in each rate. Due to these omissions, the derived full sample SN\,Ia rates may be systematically underestimated. 

There is, however, a competing systematic effect in the highest-$z$ bin that may cause the $z$\,$=$\,3.60--5.94 SN\,Ia rate to be overestimated. \texttt{STARDUST2} assigned partial SN\,Ia classifications to the majority of the single-SED sources at $z$\,$>$\,3 (see Table \ref{tab:jades22}), suggesting that misclassification of CC\,SNe as SNe\,Ia may be driving the increase in the full sample SN\,Ia rate at $z$\,$=$\,3.60--5.94. If a significant fraction of these sources are truly CC\,SNe, the $z$\,$=$\,3.60--5.94 SN\,Ia rate would decrease substantially once the CC\,SN contamination is removed. Indeed, the combined statistical and systematic uncertainties allow the $z$\,$=$\,3.60--5.94 SN\,Ia rate to be zero at the 1$\sigma$ level. Accordingly, this rate measurement should be treated with caution.


Any analysis regarding the DTD that best fits our observed SN\,Ia rates is beyond the scope of this paper. Once a larger sample of more robustly-classified high-$z$ SNe\,Ia is constructed with future JWST and Roman observations, we can constrain the SN\,Ia progenitor scenario by fitting various DTDs/SFRD histories to the observed SN\,Ia rates \citep[e.g.,][]{palicio2024} and compare the observed SN\,Ia rates to those expected by various SN\,Ia models \citep[e.g.,][]{kobayashi2009, briel2022, yoon2026}.


\subsection{Ways to Improve Photometric Classification of High-$z$ Supernovae} \label{subsec:discussion_stardust_performance}

\texttt{STARDUST2} largely succeeded in correctly classifying true CC\,SNe from just one SED (Figures \ref{fig:confusion_matrices_zspec}--\ref{fig:classification_accuracy_vs_z}). However, multiple aspects of \texttt{STARDUST2}'s current design and template library limit its performance. Here, we break down the primary limitations and discuss potential improvements. 
Our classifier options for the JTS sample were constrained by NIRCam's wavelength regime, but we anticipate that data from Roman will enable new SN models and classifiers for high-$z$ SNe (see Section \ref{subsec:roman}).


\subsubsection{Extending \texttt{SALT3-NIR's} Rest-Frame Spectral and Phase Coverage} \label{subsubsec:salt3-nir}

\texttt{STARDUST2} struggled to correctly classify true SNe\,Ia with just one SED, especially at $z$\,$\lesssim$\,1.3 (Figure \ref{fig:classification_accuracy_vs_z}). As shown in Figure \ref{fig:salt3_wavelength_bounds}, the \texttt{SALT3-NIR} rest-frame spectral coverage is restricted to 2,000--20,000\AA, and \texttt{SALT3-NIR} loses rest-frame spectral coverage of F410M and F444W at $z$\,$\lesssim$\,1.3. This loss likely contributed to the significant low-$z$ decline in average P$_\mathrm{SD}$(Ia) output for input SNe\,Ia. 
To improve SN\,Ia classification accuracy, the \texttt{SALT3-NIR} spectral coverage must be extended beyond 2$\mu$m so that NIRCam LW observations can be better modeled at low-$z$. Similarly, expanding \texttt{SALT3-NIR's} spectral coverage into the rest-frame UV ($<$\,2000\AA) would allow NIRCam SW filters to be better modeled at high-$z$. 

Additionally, \texttt{SALT3-NIR}'s rest-frame phase coverage extends only to $+$50 rest-frame days beyond B-band peak, meaning that it cannot model SN\,Ia SEDs beyond this time. This is problematic because JWST can detect late-phase SNe\,Ia, and they would be systematically less likely to be correctly classified. 
We can improve SN\,Ia classification accuracy by extending the \texttt{SALT3-NIR} model beyond 50 rest-frame days postpeak. While \texttt{SALT3-NIR} has limitations, we selected it as the SN\,Ia model in our template library because \texttt{SALT3-NIR} can account for intrinsic color/shape differences, and \texttt{SALT3-NIR} is the standardization model with the widest wavelength and phase coverage at the moment. 



\subsubsection{Increasing the Diversity of the SN\,Ia Template Library} \label{subsubsec:snia_diversity}

The \texttt{SALT3-NIR} model was constructed using spectrophotometric information from typical low-$z$ SNe\,Ia \citep{pierel2022}. As a result, it struggled to replicate the light curves of abnormal high-$z$ SNe\,Ia, which made them more likely to be misclassified as CC\,SNe (e.g., see Section 
\ref{subsubsec:single_epoch_test} for a discussion of \tr{27}). 
It is plausible that there are abnormal high-$z$ SNe\,Ia in our sample that lack spectra and are thus being photometrically misclassified as CC\,SNe. To reduce SN\,Ia misclassification, it would be beneficial to expand and diversify the SN\,Ia template library with other SN\,Ia templates (e.g., \texttt{hsiao}, \texttt{BayeSN}; \citealt{hsiao2007, mandel2022, grayling2024}), including templates that characterize the various SN\,Ia subtypes and peculiar SNe\,Ia. Roman observations will significantly increase the amount of NIR spectrophotometric data available to train \texttt{SALT3-NIR} and other models, which will further improve their ability to fit SNe\,Ia.


\subsubsection{Enabling Classifications Other than ``CC\,SN" and ``SN\,Ia"} \label{subsubsec:no_other}
\texttt{STARDUST2} assigns classification probabilities in just three categories: SNe\,Ia, SNe\,II, or SNe\,Ib/c. There is no ``other" option. SNe\,Ib/c and SNe\,II exhibit diverse light curves following their wide variety of possible explosion energies, progenitor masses, envelope compositions, etc. SNe\,Ia, on the other hand, exhibit less diversity in their light curves. \texttt{STARDUST2} therefore has more flexibility in fitting the dozens of CC\,SN templates than the one SN\,Ia model in the library to exotic SNe and non-SN sources. We expect that, if there is an exotic SN or non-SN source contaminating the JTS sample, it would be photometrically classified as a CC\,SN regardless of how epochs populate its light curve. However, given the relatively small JTS survey area ($\sim$25 arcmin$^2$), it is unlikely that the sample contains any rare exotic SNe or non-SN sources. We are not concerned with active galactic nuclei (AGN) contamination because the sample has already been vetted for AGN \citep{decoursey2025_jts}. 

To further improve \texttt{STARDUST2's} ability to reliably identify true CC\,SNe and SNe\,Ia, it would be beneficial to add an ``other" category to its template library that contains templates for transient/variable sources that are neither CC\,SNe nor SNe\,Ia. For example, this category may include templates of AGN, tidal disruption events, and pair-instability SNe. With this ``other" category, the CC\,SN category will no longer be the ``default" for any input source that does not fit the SN\,Ia templates well, reducing potential contamination in CC\,SN samples.


\subsubsection{Moving Away from Classification ``Probabilities"} \label{subsubsec:probability_problem}

As shown in Table \ref{tab:jades22}, most of the single-SED JTS SNe were classified with P$_\mathrm{SD}$(CC)\,$=$\,1. This gives the false impression that these sources were classified as CC\,SNe with 100\% certainty based on just one SED. Rather, P$_\mathrm{SD}$(CC)\,$=$\,1 means that the CC\,SN templates provided significantly better fits than the SN\,Ia template but does not necessarily mean that they fit well to the data.

The ambiguity in the interpretation of the output “probabilities” motivates the need for clearer model comparison metrics. These updated metrics may include, for example, the mean, median, and standard deviation of the reduced $\chi^2$ distribution for the fits corresponding to each SN type. These metrics describe how well the templates for each SN type fit the data on their own terms, rather than in comparison to the other SN types. Detailed testing will be required to determine the best-suited metric for future SN classifiers.


\subsubsection{Enabling the Proper Incorporation of Upper Limits} \label{subsubsec:upper_limits}



As described in Section \ref{subsubsec:model_generation}, \texttt{STARDUST2} is not equipped to properly incorporate upper limits, necessitating an approximation that biases the fitting routine. 
Given the prevalence of non-detections in high-$z$ SN monitoring, it is crucial to develop a classifier that can meaningfully interpret upper limits and fully leverage all available observational constraints in high-$z$ SN classification.


\subsection{SN Rates with Roman} \label{subsec:roman}

Roman will fundamentally transform our ability to measure SN rates by assembling statistically powerful samples at high redshift, allowing substantially more robust constraints on SN demographics and their evolution across cosmic time.

As part of the HLTDS, Roman will enable the discovery of large samples of spectroscopically classified SNe\,Ia. Simulations show that spectroscopic redshifts can be obtained for $\sim$7$\times10^{3}$ SNe~Ia through a combination of direct SN spectroscopy and host-galaxy spectroscopy, with the redshift distribution extending to $z$\,$\sim$\,3 \citep{Chen25}. A spectroscopically classified sample of SN\,Ia sample of this size and redshift reach will dramatically improve the robustness of SN~\,Ia rate measurements at $z$\,$\gtrsim$\,1, where current samples are small and rely heavily on photometric classification.

Roman will also open a new window into the most luminous SLSNe-I, which can be detected out to higher redshifts than any other type of SN, making them uniquely valuable tracers of massive-star deaths at early times. SLSNe-I are a rare but extremely luminous subclass of CC\,SNe, meaning that their volumetric rates trace the same underlying population of massive stars. The Roman HLTDS is expected to discover up to $\sim$100 SLSNe over the mission lifetime, populating redshifts out to $z$\,$\sim$\,5 \citep{Gomez23}. This sample will allow the extension of CC\,SN rate measurements to $z$\,$\sim$\,6 with meaningful statistics, ushering in a new regime for high-$z$ CC\,SN rate studies. Additionally, the HLTDS is expected to discover tens of thousands of CCSNe out to $z$\,$\sim$\,3 \citep{Rose25}. Such a large volumetric sample will help anchor the SFRD measurements inferred from SLSNe rates down to more nearby redshifts.

Roman's HLTDS nominally includes plans for single-frame differencing. The survey sensitivity can be pushed 1--2 magnitudes deeper by stacking multiple single-frame images, which is now being planned by the Roman hIgh-redshift transient SciencE (RISE) Wide-Field Science team.\footnote{\href{https://science.nasa.gov/mission/roman-space-telescope/roman-for-scientists/}{Wide Field Science Teams Data Products User Documentation}\label{footnote1}} Such deep stacks, combined with the time-dilated and redshifted light curves at higher-$z$, will reveal a larger fraction of the 2\,$<$\,$z$\,$<$\,3 SN population, extend SN\,Ia and CCSN discoveries out to $z$\,$>$\,3, and, crucially, enable the detection of light curve evolution beyond peak brightness. Stacking also provides the hidden benefit of median-filtering out low-$z$~contaminants, which have shorter evolutionary timescales. This is especially powerful for selecting high-$z$~SNe whose light-curves have much longer time evolution because of time dilation. This capability will transform the high-redshift science of SN rates discussed above into a robust statistical discipline.


\section{Conclusions} \label{sec:conclusions}
Adopting the SN sample from the JTS \citep{decoursey2025_jts}, we compute volumetric CC\,SN and SN\,Ia rates out to $z$\,$\sim$\,5 for the first time. In order to disentangle the CC\,SNe and the SNe\,Ia in the sample, we ran a modified version of the \texttt{STARDUST2} classifier code to perform light curve fitting or SED fitting for each JTS SN. A subset of the JTS SNe were spectroscopically-classified with JWST program 6541. In these cases, the spectroscopic classification overrode the photometric classification. 

More than half of the JTS SNe were photometrically classified with just one SED. To address concerns regarding the reliability of these single-SED SN classifications, we simulated $\sim$23,000 mock SEDs of various SN subtypes (SNe\,Ia, SNe\,Ib/c, SNe\,IIP, SNe\,IIL, and SNe\,IIn) with a variety of redshift, phases, peak M$_\mathrm{B}$ values, and color excesses. We then ran each of these SN SEDs through the \texttt{STARDUST2} classifier to quantitatively characterize its ability to correctly identify CC\,SNe and SNe\,Ia based on one SED. We evaluated the CC\,SN vs SN\,Ia true positive classification rates as a function of redshift, yielding valuable insight into the regimes where \texttt{STARDUST2} performs well and where it breaks down. 

We computed CC\,SN and SN\,Ia rates using both the full JTS sample, which includes sources classified with single SEDs, and a ``gold" sample that was restricted to objects classified via spectroscopy or multi-epoch light curves. Classifications in the gold sample are more reliable, but the reduced sample size leads to larger rate uncertainties driven by small number statistics. The gold sample is also affected by possible systematic uncertainties regarding the selection of SNe that were followed-up.

Below, we list our conclusions from the CC\,SN rate analysis, and the SN\,Ia rate analysis, and the classification accuracy analysis:

\begin{itemize}
    


    \item We provide the first measurement of CC\,SN rates at $z$\,$\gtrsim$\,2.5. In units of 10$^{-4}$ CC\,SNe yr$^{-1}$ Mpc$^{-3}$, the full sample CC\,SN rates are 6.5$^{+2.3(0.5)}_{-1.7(0.3)}$ at 2.05\,$<$\,$z$\,$\leq$\,2.78 and 4.1$^{+1.4(1.0)}_{-1.1(0.3)}$ at 2.78\,$<$\,$z$\,$\leq$\,5.20. The CC\,SN rates arising from the gold sample are 6.8$^{+3.8(0.4)}_{-2.5(0.2)}$ at 2.05\,$<$\,$z$\,$\leq$\,2.83 and 5.2$^{+3.1(1.1)}_{-2.0(0.4)}$ at 2.83\,$<$\,$z$\,$\leq$\,4.45. The uncertainties listed outside of (within) the parentheses are statistical (systematic) in nature.

    \item Both sets of CC\,SN rates show a tentative decrease beyond cosmic noon, as predicted by galaxy luminosity-based measurements of cosmic SFRD. This is the first observational indication that CC\,SN rates, like the cosmic SFRD, decline beyond cosmic noon. A companion paper, \citet{vassallo2026}, compares the observed CC\,SN rates to the cosmic SFRD in more detail.

    \item In units of 10$^{-4}$ SNe\,Ia yr$^{-1}$ Mpc$^{-3}$, the full sample SN\,Ia rates are 0.4$^{+0.4(0.5)}_{-0.2(0.2)}$ at 0.92$<$\,$z$\,$\leq$\,1.91 and 0.4$^{+0.3(0.1)}_{-0.2(0.2)}$ at 1.91\,$<$\,$z$\,$\leq$\,3.60. The gold sample SN\,Ia rate is 0.5$^{+0.8(0.1)}_{-0.3(0.1)}$ at 2.89\,$<$\,$z$\,$\leq$\,4.19. The uncertainties listed outside of (within) the parentheses are statistical (systematic) in nature.

    \item Assuming \texttt{STARDUST2} is provided with (1) a spectroscopic or well-constrained photometric redshift in the 0.7\,$\leq$\,$z$\,$\leq$\,5 regime and (2) an SN SED with coverage in at least five JTS filters, it can effectively identify CC\,SNe with just one SED. Under the same conditions, \texttt{STARDUST2} is much less effective at identifying SNe\,Ia. \texttt{STARDUST2} cannot reliably disentangle CC\,SN types (SNe\,II vs SNe\,Ib/c) when provided with just one SED. 

\end{itemize}

To better constrain CC\,SN and SN\,Ia rates at high-$z$, we must continue to discover and systematically monitor distant SNe with multi-band and multi-epoch NIRCam observations. Additionally, we must continue to target sufficiently bright SNe with NIRSpec to produce more reliable SN type classifications. JWST program 8060 (Cycles 4--6; \citealt{egami2025}) is a multi-cycle high-$z$ transient program with 3 epochs of deep NIRCam observations and 2 epochs of NIRSpec follow-up per cycle in the JADES Deep Field. This program is specifically designed to increase both the quantity and quality of the high-$z$ SN sample. The Roman HLTDS will also provide a large and robust sample of high-$z$ SNe, enabling transformative advances in high-$z$ time-domain science.



\begin{acknowledgments}
We thank the anonymous referee whose comments and suggestions led to significant improvements in the paper. This work is based on observations made with the NASA/ESA/CSA James Webb Space Telescope. The data were obtained from the Mikulski Archive for Space Telescopes at the Space Telescope Science Institute, which is operated by the Association of Universities for Research in Astronomy, Inc., under NASA contract NAS 5-03127 for JWST. These observations are associated with program \#1180 and 6541. The specific JWST observations analyzed can be accessed via \dataset[DOI: 10.17909/c4qk-xv53]{https://doi.org/10.17909/c4qk-xv53}. Additionally, this work made use of the {\it lux} supercomputer at UC Santa Cruz which is funded by NSF MRI grant AST 1828315. The STScI TSST group acknowledges partial support from JWST-GO-06541, JWST-GO-06585, and JWST-GO-05324.

SM and CV acknowledge support from the Research Council of Finland project 350458. EE, ZJ, BDJ, BER, and CNAW acknowledge support from JWST/NIRCam contract to the University of Arizona, NAS 5-02105. AJB and AJC acknowledge funding from the ``FirstGalaxies" Advanced Grant from the European Research Council (ERC) under the European Union’s Horizon 2020 research and innovation programme (Grant agreement No. 789056). AJC gratefully acknowledges support from the Cosmic Dawn Center through the DAWN Fellowship. The Cosmic Dawn Center (DAWN) is funded by the Danish National Research Foundation under grant No. 140. DJE is supported as a Simons Investigator and by JWST/NIRCam contract to the University of Arizona, NAS 5-02105. DJE, BDJ, and BER acknowledge support from JWST Program \#3215. RH acknowledges funding provided by the Johns Hopkins University, Institute for Data Intensive Engineering and Science (IDIES). RM acknowledges support by the Science and Technology Facilities Council (STFC), by the ERC through Advanced Grant 695671 ``QUENCH”, and by the UKRI Frontier Research grant RISEandFALL. RM also acknowledges funding from a research professorship from the Royal Society. ST acknowledges support by the Royal Society Research Grant G125142. YZ acknowledges support from the MAOF grant 12641898 and visitor support from the Observatories of the Carnegie Institution for Science, Pasadena, CA, where part of this work was completed. The research of CCW is supported by NOIRLab, which is managed by the Association of Universities for Research in Astronomy (AURA) under a cooperative agreement with the National Science Foundation. The authors used an AI-based language tool to assist with grammar and stylistic improvements. The scientific content, analysis, and conclusions are entirely the authors’ own.
\end{acknowledgments}

\facilities{JWST(NIRCam and NIRSpec)}

\software{astropy \citep{astropy2013, astropy2018, astropy2022}}


\appendix

\onecolumngrid
\section{Updated JTS Classifications} \label{appendix:updated_classifications}

In Tables \ref{tab:jades23} and \ref{tab:jades22}, we present the updated JTS classifications resulting from the modified \texttt{STARDUST2} scheme that is described in Section \ref{subsec:jades_sne}. 
Table \ref{tab:jades23} also lists the spectroscopic classifications of the JTS SNe that were targeted with JWST program 6541 \citep{egami2023}. Notably, \tr{22} and \tr{24} were both spectroscopically-classified as SNe\,IIP and their light curve fitting results agree with P$_\mathrm{SD}$(CC)\,$=$\,1. However, their best-fit light curves have high reduced-$\chi^2$ values (34.6 and 268.4, respectively). These high values may have resulted from underestimated photometric uncertainties and a limited number of SN templates in the library, which are based on observed low-$z$ SNe that may plausibly differ from high-$z$ SNe. The \tr{22} and \tr{24} light curves were well reproduced by theoretical models of SNe\,II in \citet{moriya2025}. \tr{24} was estimated to have a large extinction in \citet{moriya2025}, which may have further contributed to its large reduced-$\chi^2$ in the \texttt{STARDUST2} fit.  \tr{26}, spectroscopically-classified as an SN\,Ic-BL \citep{siebert2024}, suffers from the same underestimated photometric uncertainties and limited \texttt{STARDUST2} template library, resulting in a high reduced-$\chi^2$ value for its best-fit light curve. The photometric uncertainty underestimation was more significant for brighter SNe that were more likely to be targeted with spectroscopy, resulting in systematically higher reduced-$\chi^2$ values for the spectroscopically-classified SNe. While light curve fitting alone favors an SN\,Ib/c classification for \tr{27}, it has been spectroscopically classified as an SN\,Ia \citep{pierel2024}, so we manually assign \tr{27} as an SN\,Ia with P(Ia)\,$=$\,1.


\begin{longrotatetable}
\begin{deluxetable}{ccccccCCccccc}
\tablecaption{JADES SN Classifications (2023 Sample)}
\tablehead{
\colhead{TNS ID} & \colhead{$z_\mathrm{host}$} & \colhead{$z_\mathrm{fit}$} & \colhead{P$_\mathrm{SD}$(Ia)$^a$} & \colhead{P$_\mathrm{SD}$(CC)$^a$} & \colhead{Best Model} & \colhead{Phase$^b$} & \colhead{Peak M$_\mathrm{B}$$^c$} & \colhead{E(B$-$V)$^d$} & \colhead{$\chi^2$/DOF$^e$} & \colhead{DOF} & \colhead{\# Epochs} & \colhead{Subtype$^f$}
} 
\label{tab:jades23}
\startdata
\tr{53}	    & 4.35 $\pm$ 0.04 &	4.36 	& 0	      &	1	    & snana-2004ib & 32	     & -18.1	& 0.1	  &	1.0	    & 13	  &	3 &	\nodata \\
\tr{50}	    & 4.117	          &	4.117	& 0	      &	1	    & snana-2006gq & 37	     & -17.5	& 0.1	  &	1.3	    & 23	  &	4 &	\nodata \\
\tr{88}	    & 3.74 $\pm$ 0.17 &	3.91 	& 0.34    &	0.66	& snana-2007pg & 8	     & -18.5	& 0.5	  &	0.7	    & 8	      &	2 &	\nodata \\
\tr{10}	    & 3.61	          &	3.61	& 0	      &	1	    & snana-2006ix & 7	     & -18.9	& 0.3	  &	7.4	    & 16	  &	3 &	II      \\
\tr{44}	    & 3.21 $\pm$ 0.58 &	2.94	& 0	      &	1	    & snana-2005gi & 23	     & -17.3	& 0.3	  &	1.7	    & 16	  &	4 &	\nodata \\
\tr{71}	    & 3.09	          &	3.09	& 0.24    &	0.76	& snana-2006kn & 32	     & -17.5	& 0.3	  &	1.6	    & 13	  &	3 &	\nodata \\
\tr{27}	    & 2.90	          &	2.90	& 0	      &	1	    & snana-2006ep & 34	     & -18.1	& 0.0	  &	2.3	    & 30	  &	4 &	Ia      \\
\tr{36}	    & 2.86 $\pm$ 0.10 &	2.69	& 0	      &	1	    & snana-2006kv & 11	     & -17.6	& 0.4	  &	1.3	    & 26	  &	4 &	\nodata \\
\tr{26}	    & 2.83	          &	2.83	& 0	      &	1	    & snana-2004gq & -12	 & -18.4	& 0.0	  &	41.8	& 30	  &	4 &	Ic-BL   \\
\tr{52}	    & 2.78 $\pm$ 0.12 &	2.64 	& 0	      &	1	    & snana-2006iw & 8	     & -16.3	& 0.1	  &	1.6	    & 15	  &	3 &	\nodata \\
\tr{15}$^g$	& 2.77 $\pm$ 0.86 &	\nodata	& \nodata &	\nodata	& \nodata	   & \nodata & \nodata	& \nodata &	\nodata	& \nodata &	3 &	\nodata \\
\tr{29}	    & 2.73	          &	2.73	& 0	      &	1	    & snana-2004gq & -1	     & -18.1	& 0.4	  &	16.4	& 21	  &	3 &	Ib/c    \\
\tr{6}	    & 2.623	          &	2.623	& 0	      &	1	    & snana-2004gq & -2	     & -17.2	& 0.0	  &	12.7	& 17	  &	3 &	\nodata \\
\tr{11}	    & 2.344	          &	2.344	& 0	      &	1	    & snana-2007iz & 16	     & -16.4	& 0.0	  &	2.6	    & 16	  &	3 &	\nodata \\
\tr{19}	    & 2.24 $\pm$ 0.13 &	1.92	& 0.03	  &	0.97	& snana-2007lj & 53	     & -16.9	& 0.1	  &	1.0	    & 5	      &	1 &	\nodata \\
\tr{7}	    & 2.06	          &	2.06	& 0	      &	1	    & snana-2007nv & 45	     & -16.9	& 0.0	  &	2.9	    & 17	  &	3 &	\nodata \\
\tr{28}	    & 1.94 $\pm$ 0.12 &	1.79	& 0	      &	1	    & snana-2006ep & 22	     & -17.3	& 0.0	  &	3.9	    & 29	  &	4 &	\nodata \\
\tr{87}	    & 1.932	          &	1.932	& 0	      &	1	    & snana-2006jo & 8	     & -16.7	& 0.8	  &	0.8	    & 11	  &	2 &	\nodata \\
\tr{60}	    & 1.912	          &	1.912	& 0	      &	1	    & snana-2007kw & 30	     & -16.9	& 0.1	  &	4.2	    & 6	      &	1 &	\nodata \\
\tr{5}	    & 1.86 $\pm$ 0.10 &	1.57	& 0	      &	1	    & snana-2004gq & 3	     & -16.8	& 0.2	  &	6.9	    & 11	  &	3 &	\nodata \\
\tr{9}	    & 1.854	          &	1.854	& 0	      &	1	    & snana-2004ib & 19	     & -17.2	& 0.0	  &	10.2	& 15	  &	4 &	\nodata \\
\tr{83}	    & 1.748	          &	1.748	& 0	      &	1	    & snana-2007kw & 43	     & -18.0	& 0.8	  &	1.5	    & 6	      &	1 &	\nodata \\
\tr{22}	    & 1.62	          &	1.62	& 0	      &	1	    & snana-2006iw & -1	     & -18.0	& 0.1	  &	34.6	& 30	  &	4 &	IIP     \\
\tr{35}	    & 1.50	          &	1.50	& 0	      &	1	    & snana-2004fe & 39	     & -17.2	& 0.9	  &	3.4	    & 10	  &	2 &	\nodata \\
\tr{30}	    & 1.19 $\pm$ 0.11 &	1.28	& 1	      &	0	    & salt3-nir	   & 18	     & -17.5	& \nodata &	20.8	& 1	      &	1 &	\nodata \\
\tr{81}	    & 1.171	          &	1.171	& 0	      &	1	    & snana-2006gq & 28	     & -16.1	& 0.5	  &	7.7	    & 8	      &	2 &	\nodata \\
\tr{48}	    & 1.16 $\pm$ 0.05 &	1.15	& 0	      &	1	    & snana-2006ep & 91	     & -16.4	& 0.9	  &	1.6	    & 8	      &	3 &	\nodata \\
\tr{45}	    & 1.139	          &	1.139	& 0	      &	1	    & snana-2004gv & 16	     & -16.8	& 0.4	  &	1.4	    & 2	      &	2 &	\nodata \\
\tr{24}	    & 1.01	          &	1.01	& 0	      &	1	    & snana-2005gi & 1	     & -17.8	& 0.4	  &	268.4	& 6	      &	2 &	IIP     \\
\tr{82}$^h$	& 0.665	          &	\nodata	& \nodata &	\nodata & \nodata	   & \nodata & \nodata	& \nodata &	\nodata	& \nodata &	1 &	\nodata \\
\tr{14}$^h$	& 0.657	          &	\nodata	& \nodata &	\nodata	& \nodata	   & \nodata & \nodata	& \nodata &	\nodata	& \nodata &	3 &	\nodata \\
\tr{90}$^h$	& 0.533	          &	\nodata	& \nodata &	\nodata	& \nodata	   & \nodata & \nodata	& \nodata &	\nodata	& \nodata &	1 &	\nodata \\
\tr{89}$^h$	& 0.21	          &	\nodata	& \nodata &	\nodata	& \nodata	   & \nodata & \nodata	& \nodata &	\nodata	& \nodata &	1 &	\nodata \\
\tr{25}$^g$	& \nodata	      &	\nodata	& \nodata &	\nodata	& \nodata	   & \nodata & \nodata	& \nodata &	\nodata	& \nodata &	1 &	\nodata \\
\sidehead{Marginal Detections}
\tr{84}	    & 1.86 $\pm$ 0.11 &	1.69	& 0	      &	1       & snana-2006jo & 11	     & -15.9	& 0.4	  &	1.6	    & 21	  &	4 &	\nodata \\
\tr{59}	    & 0.996	          &	0.996	& 0	      &	1       & snana-2007iz & 11	     & -16.7	& 0.8	  &	3.4	    & 5	      &	2 &	\nodata \\
\enddata
\tablecomments{$^a$P$_\mathrm{SD}$(Ia) and P$_\mathrm{SD}$(CC) are the \texttt{STARDUST2} outputs, where P$_\mathrm{SD}$(CC)\,$=$\,P$_\mathrm{SD}$(II)$+$P$_\mathrm{SD}$(Ib/c).}
\tablecomments{$^b$ ``Phase" is the rest-frame phase (days) of the first observation of the SN relative to maximum light, based on the best-fit light curve.}
\tablecomments{$^c$ ``Peak M$_\mathrm{B}$" is the peak B-band absolute magnitude associated with the best-fit light curve.}
\tablecomments{$^d$ ``E(B$-$V)" is the color excess associated with the best-fit light curve.}
\tablecomments{$^e$ ``DOF" stands for ``degrees of freedom." We do not account for model uncertainties in the $\chi^2/\mathrm{DOF}$ calculation, which may result in overfitting for sources with sparse light curve coverage, causing $\chi^2/\mathrm{DOF}$\,$<$\,1. }
\tablecomments{$^f$ The subtype column is based on spectroscopic classification from JWST program 6541 \citep{egami2023}.}
\tablecomments{$^g$We are unable to classify this source because its host photo-$z$ is either unknown or insufficiently constrained ($z$\,$-$\,3$\sigma_z$\,$<$\,0.7).}
\tablecomments{$^h$We are unable to classify these sources because they lie at $z$\,$<$\,0.7, where the \texttt{SALT3-NIR} model has insufficient rest-frame spectral overlap with the JTS NIRCam filters (see Section \ref{subsec:jades_sne}).}
\end{deluxetable}
\end{longrotatetable}
\begin{longrotatetable}
\begin{deluxetable*}{ccccccCCccc}
\tablecaption{JADES SN Classifications (2022 Sample)}
\tablehead{
\colhead{TNS ID} & \colhead{$z_\mathrm{host}$} & \colhead{$z_\mathrm{fit}$} & \colhead{P$_\mathrm{SD}$(Ia)$^a$} & \colhead{P$_\mathrm{SD}$(CC)$^a$} & \colhead{Best Model} & \colhead{Phase$^b$} & \colhead{Peak M$_\mathrm{B}$$^c$} & \colhead{E(B$-$V)$^d$} & \colhead{$\chi^2$/DOF$^e$} & \colhead{DOF}
} 
\label{tab:jades22}
\startdata
\tr{77}	     & 4.82 $\pm$ 0.05 & 4.81	 & 0.31	   & 0.69	 & snana-2007og	    & 27	  &	-17.7	& 0.1	  &	0.9	    & 3       \\
\tr{33}	     & 4.82 $\pm$ 0.49 & 4.20	 & 0.74	   & 0.26	 & salt3-nir	    & 35	  &	-20.5	& \nodata &	4.2	    & 2       \\
\tr{39}	     & 4.504	       & 4.504	 & 0.35	   & 0.65	 & snana-2007nv	    & 50	  &	-17.0   & 0.1	  &	2.0	    & 4       \\
\tr{93}	     & 4.471	       & 4.471	 & 0.38	   & 0.62	 & snana-2007og	    & 34	  &	-17.3	& 0.1	  &	1.4	    & 4       \\
\tr{107}	 & 4.24 $\pm$ 0.09 & 4.26	 & 0.19	   & 0.81	 & snana-2007og	    & 39	  &	-17.6	& 0.3	  &	0.2	    & 3       \\
\tr{102}$^i$ & 3.96 $\pm$ 0.14 & \nodata & \nodata & \nodata & \nodata	        & \nodata &	\nodata	& \nodata &	\nodata & \nodata \\
\tr{103}	 & 3.605	       & 3.605	 & 0.69	   & 0.31	 & salt3-nir	    & -18	  &	-20.0   & \nodata &	2.4	    & 5       \\
\tr{13}	     & 3.58 $\pm$ 0.14 & 3.30	 & 0	   & 1	     & snana-04d1la	    & -5	  &	-18.2	& 0.1	  &	3.1	    & 3       \\
\tr{38}	     & 3.166	       & 3.166	 & 0.85	   & 0.15	 & salt3-nir	    & 21	  &	-18.5	& \nodata &	1.9	    & 5       \\
\tr{55}	     & 2.79 $\pm$ 0.11 & 2.82	 & 0	   & 1	     & snana-2004ib	    & 15	  &	-17.5	& 0.3	  &	1.2	    & 4       \\
\tr{21}	     & 2.73 $\pm$ 0.39 & 1.96	 & 0	   & 1	     & snana-2007iz	    & 58	  &	-18.0   & 0.5	  &	3.8	    & 3       \\
\tr{100}$^g$ & 2.62 $\pm$ 1.54 & \nodata & \nodata & \nodata & \nodata	        & \nodata &	\nodata	& \nodata &	\nodata	& \nodata \\
\tr{79}	     & 2.617	       & 2.617	 & 0	   & 1	     & snana-2004fe	    & 8	      &	-17.6	& 0.8	  &	1.8	    & 6       \\
\tr{80}	     & 2.617	       & 2.617	 & 0	   & 1	     & snana-2004gq	    & 90	  &	-18.2	& 0.6	  &	0.4	    & 6       \\
\tr{95}	     & 2.56 $\pm$ 0.39 & 2.55	 & 0	   & 1	     & snana-2007ld	    & 37	  &	-17.7	& 0.5	  &	2.0	    & 3       \\
\tr{8}$^g$	 & 2.48 $\pm$ 0.68 & \nodata & \nodata & \nodata & \nodata	        & \nodata &	\nodata	& \nodata &	\nodata	& \nodata \\
\tr{34}	     & 2.323	       & 2.323	 & 0	   & 1	     & snana-2006gq	    & 45	  &	-17.7	& 0.1	  &	2.2	    & 6       \\
\tr{23}	     & 2.315	       & 2.315	 & 0	   & 1	     & snana-2004gq	    & 3	      &	-18.1	& 0.4	  &	9.5	    & 6       \\
\tr{66}	     & 2.29 $\pm$ 0.23 & 2.48	 & 0.01	   & 0.99	 & snana-2007ld	    & 14	  &	-17.1	& 0.3	  &	1.7	    & 4       \\
\tr{92}	     & 2.02 $\pm$ 0.29 & 1.98	 & 0	   & 1	     & snana-2006gq	    & 27	  &	-17.0	& 0.8	  &	2.0	    & 3       \\
\tr{37}	     & 2.01 $\pm$ 0.16 & 1.75	 & 0	   & 1	     & snana-sdss014475 & -10	  &	-18.1	& 0.5	  &	1.8	    & 4       \\
\tr{20}	     & 2.00 $\pm$ 0.37 & 2.08	 & 0	   & 1	     & snana-2006ep	    & -12	  &	-20.5	& 0.5	  &	66.8	& 15      \\
\tr{111}	 & 1.92	           & 1.92	 & 0.98	   & 0.02	 & salt3-nir	    & 40	  &	-18.1	& \nodata &	1.5	    & 5       \\
\tr{2}	     & 1.79 $\pm$ 0.26 & 1.67	 & 0	   & 1	     & snana-2007ms	    & -1	  &	-18.6	& 0.3	  &	1.2	    & 2       \\
\tr{16}$^j$	 & 1.771	       & \nodata & \nodata & \nodata & \nodata	        & \nodata &	\nodata	& \nodata &	\nodata	& \nodata \\
\tr{64}	     & 1.766	       & 1.766	 & 0.15	   & 0.85	 & snana-2007pg	    & 14	  &	-16.7	& 0.3	  &	1.9	    & 6       \\
\tr{1}	     & 1.688	       & 1.688	 & 1	   & 0	     & salt3-nir	    & -7	  &	-18.4	& \nodata &	22.6	& 6       \\
\tr{69}	     & 1.62 $\pm$ 0.07 & 1.58	 & 0	   & 1	     & snana-2007lj	    & 50	  &	-16.5	& 0.2	  &	0.4	    & 4       \\
\tr{12}	     & 1.567	       & 1.567	 & 0	   & 1	     & snana-2006iw	    & -1	  &	-17.6	& 0.3	  &	1.0	    & 5       \\
\tr{46}	     & 1.42 $\pm$ 0.11 & 1.37	 & 0	   & 1	     & snana-2007lj	    & 36	  &	-17.0	& 0.3	  &	1.0	    & 3       \\
\tr{65}	     & 1.415	       & 1.415	 & 0	   & 1	     & snana-2004hx	    & 59	  &	-16.5	& 0.5	  &	1.9	    & 5       \\
\tr{54}	     & 1.36 $\pm$ 0.21 & 1.48	 & 0	   & 1	     & snana-2007lx	    & -15	  &	-18.2	& 0.2	  &	2.5	    & 1       \\
\tr{67}	     & 1.294	       & 1.294	 & 0	   & 1	     & snana-2004hx	    & 53	  &	-16.5	& 0.4	  &	3.4	    & 5       \\
\tr{56}	     & 1.244	       & 1.244	 & 0	   & 1	     & snana-2007lx	    & 28	  &	-16.0	& 0.1	  &	8.3	    & 5       \\
\tr{68}	     & 1.114	       & 1.114	 & 0	   & 1	     & snana-2007kw	    & -7	  &	-16.4	& 0.0	  &	6.2	    & 4       \\
\tr{18}	     & 1.094	       & 1.094	 & 0	   & 1	     & snana-2006kn	    & 39	  &	-17.8	& 0.5	  &	1.6	    & 3       \\
\tr{17}	     & 0.996	       & 0.996	 & 0	   & 1	     & snana-2004gv	    & 16	  &	-17.2	& 0.0	  &	23.5	& 3       \\
\tr{31}	     & 0.953	       & 0.953	 & 0	   & 1	     & snana-2006lc	    & 38	  &	-17.2	& 1.0	  &	22.9	& 3       \\
\tr{61}$^h$	 & 0.669	       & \nodata & \nodata & \nodata & \nodata	        & \nodata &	\nodata	& \nodata &	\nodata	& \nodata \\
\tr{109}$^h$ & 0.669	       & \nodata & \nodata & \nodata & \nodata	        & \nodata &	\nodata	& \nodata &	\nodata	& \nodata \\
\tr{3}$^h$	 & 0.665	       & \nodata & \nodata & \nodata & \nodata	        & \nodata &	\nodata	& \nodata &	\nodata	& \nodata \\
\tr{4}$^h$	 & 0.665	       & \nodata & \nodata & \nodata & \nodata	        & \nodata &	\nodata	& \nodata &	\nodata	& \nodata \\
\tr{110}$^h$ & 0.54	           & \nodata & \nodata & \nodata & \nodata	        & \nodata &	\nodata	& \nodata &	\nodata	& \nodata \\
\tr{101}$^g$ & \nodata	       & \nodata & \nodata & \nodata & \nodata	        & \nodata &	\nodata	& \nodata &	\nodata	& \nodata \\
\tr{32}$^g$	 & \nodata	       & \nodata & \nodata & \nodata & \nodata	        & \nodata &	\nodata	& \nodata &	\nodata	& \nodata \\
\sidehead{Marginal Detections}
\tr{96}	     & 3.913	       & 3.913	 & 0.10	   & 0.90	 & snana-2007iz	    & 12	  &	-17.0	& 0.2	  &	2.5	    & 5       \\
\tr{94}	     & 2.67 $\pm$ 0.20 & 2.26	 & 0.10	   & 0.90	 & snana-2007iz	    & 21	  &	-16.7	& 0.7	  &	2.3	    & 4       \\
\enddata
\tablecomments{$^{a-h}$ Same as Table \ref{tab:jades23}}
\tablecomments{$^i$ \tr{102} is unclassified because it lacks F200W, F335M, and F356W NIRCam coverage, and F090W is out of \texttt{SALT3-NIR}'s rest-frame spectral coverage at $z$\,$=$\,3.96\,$\pm$\,0.14. }
\tablecomments{$^j$\tr{16} is unclassified because it fits poorly to every template in the library. See Appendix \ref{appendix:tr16} for details.}
\end{deluxetable*}
\end{longrotatetable}


\onecolumngrid
\section{Excluding \tr{16} from the Rates and \texttt{snana-2006ez} from the Template Library} \label{appendix:tr16}

\begin{figure*}
    \centering
 {\includegraphics[width=0.5\linewidth]{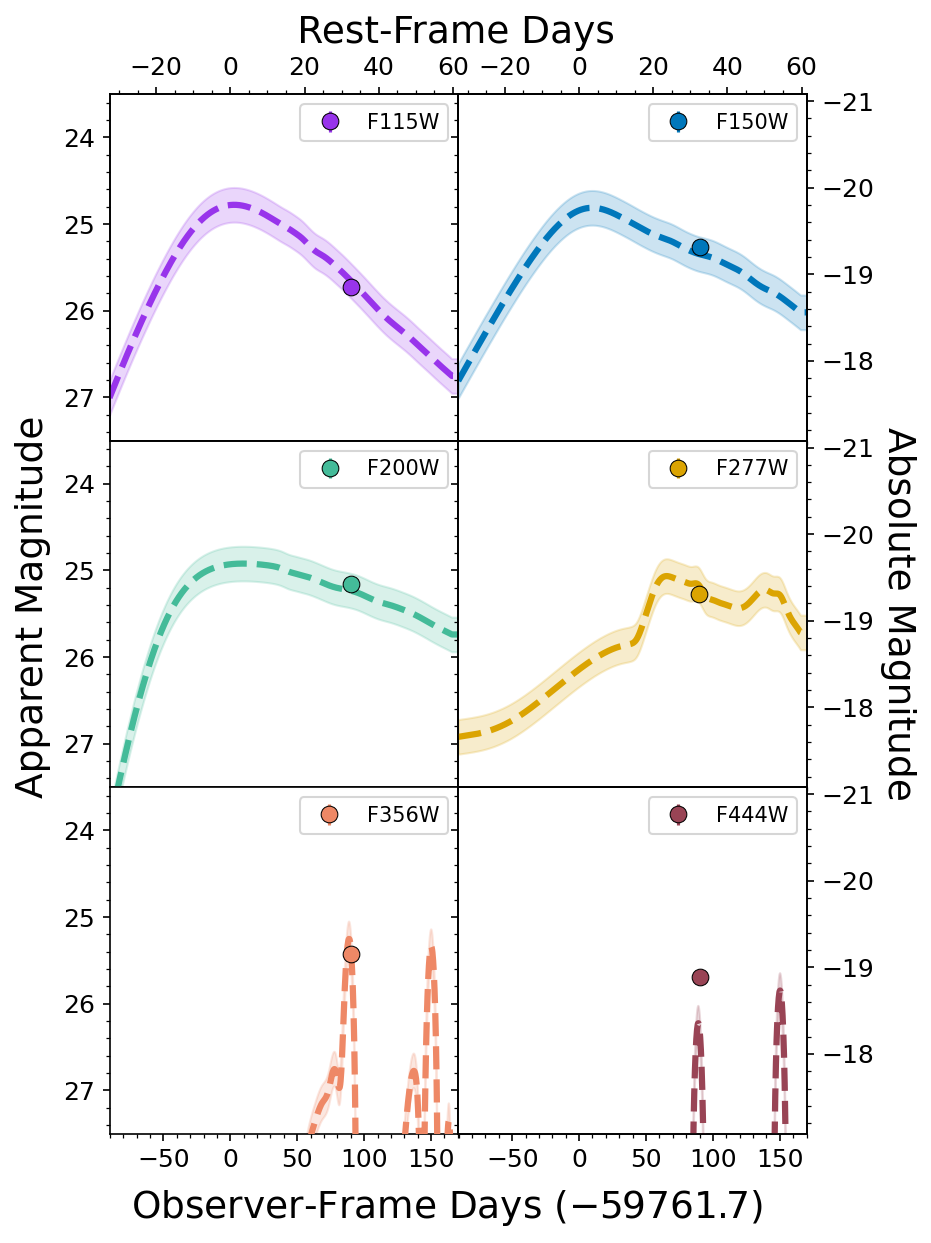}}
    \caption{The best-fit F115W, F150W, F200W, F277W, F356W, and F444W light curves to \tr{16}'s photometry. The \texttt{snana-2006ez} template (Type IIn) provided the best fit, but its F356W and F444W light curves are unphysical. We removed the \texttt{snana-2006ez} template from the template library due to its unphysical LW light curves, and we removed \tr{16} from the rate calculation because we could not determine if it was a CC\,SN, SN\,Ia, or something else.}
    \label{fig:tr16}
\end{figure*}

\tr{16} is a JTS source whose host resides at $z_\mathrm{spec}$\,$=$\,1.771. As detailed in Table \ref{tab:jades22}, it only has one observed SED and its classification could not be reliably determined. When it was passed through the \texttt{STARDUST2} classifier as described in Section \ref{subsec:jades_sne}, it was assigned P$_\mathrm{SD}$(Ia)\,$=$\,1. However, its reduced $\chi^2$ was extremely high ($\chi^2$\,$>$\,2000), indicating an exceedingly poor fit to the \texttt{SALT3-NIR} model. Prior to the removal of the Type IIn \texttt{snana-2006ez} template from the template library, it was classified with P$_\mathrm{SD}$(CC)\,$=$\,1, meaning the \texttt{snana-2006ez} fit to \tr{16} was better than the \texttt{SALT3-NIR} fit. We show the \texttt{snana-2006ez} fit to \tr{16}'s observed photometry in Figure \ref{fig:tr16}. As shown in the bottom two panels, the \texttt{snana-2006ez} F356W and F444W light curves are unphysical, prompting the removal of this template from the template library. With no other CC\,SN template producing a comparatively better fit to \tr{16}'s photometry than the Type Ia \texttt{SALT3-NIR} model, \texttt{STARDUST2} assigned \tr{16} with P$_\mathrm{SD}$(Ia)\,$=$\,1. Because removing one unphysical model caused \tr{16}'s classification to change so drastically and no other template produced a reasonable fit to \tr{16}'s photometry, we removed \tr{16} from the rate calculation. We could not reliably classify \tr{16} as a CC\,SN, SN\,Ia, or other type of source.

\onecolumngrid
\section{Exploring where \texttt{STARDUST2} Succeeds and Fails} \label{appendix:stardust_success_failure}

We have already investigated the ability of \texttt{STARDUST2} to accurately classify single-SED SNe as a function of redshift in Section \ref{subsec:results_individual_subtype}. However, we must still explore how other SN parameters correlate with classification accuracy for each SN subtype. These parameters include rest-frame phase, color excess, and peak M$_\mathrm{B}$. In Section \ref{subappendix:accuracy_vs_parameter}, we inspect how these parameters affect CC\,SN and SN\,Ia classification, and in Section \ref{subappendix:deriving_parameters}, we explore common misclassification scenarios for the various SN subtypes.

\subsection{Single-SED Classification Accuracy vs Phase, E(B$-$V), and Peak M$_B$} \label{subappendix:accuracy_vs_parameter}

\begin{figure*}
    \centering
 {\includegraphics[width=0.49\linewidth]{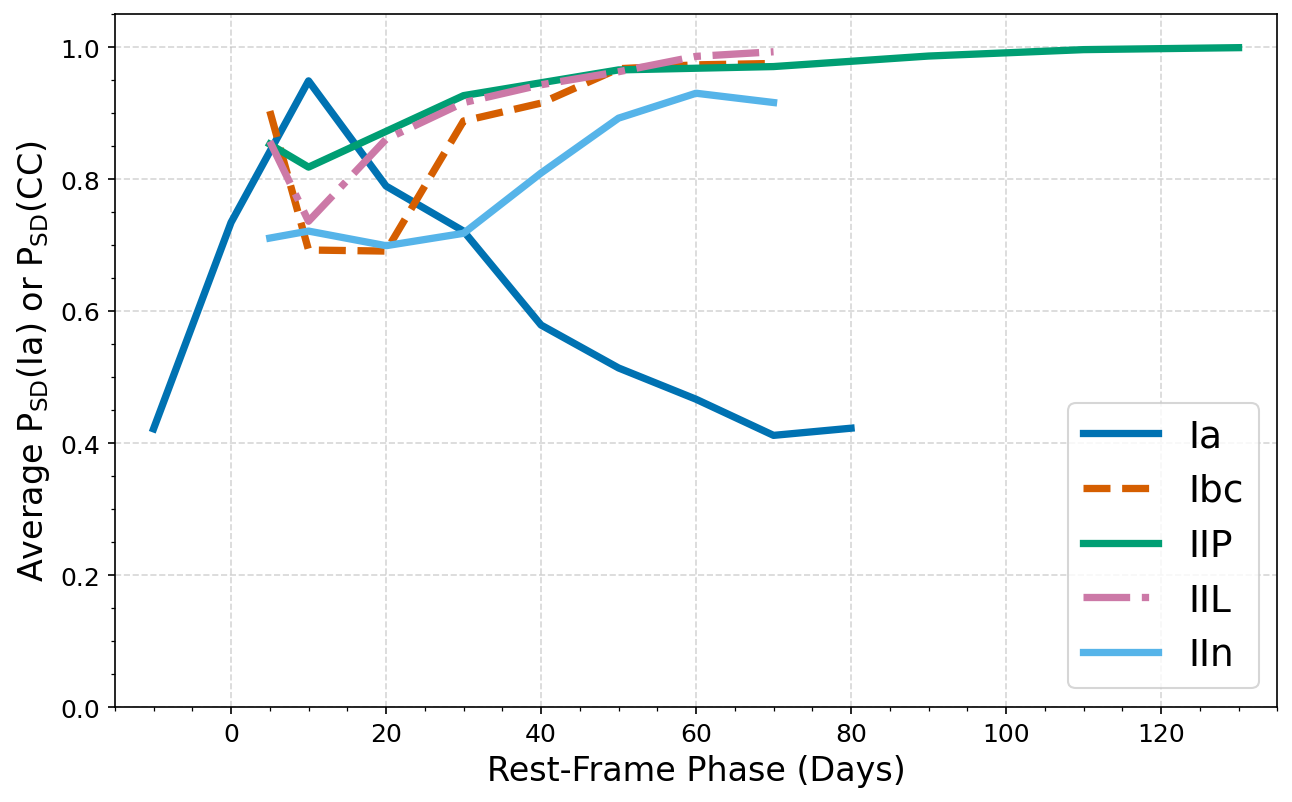}}
{\includegraphics[width=0.49\linewidth]{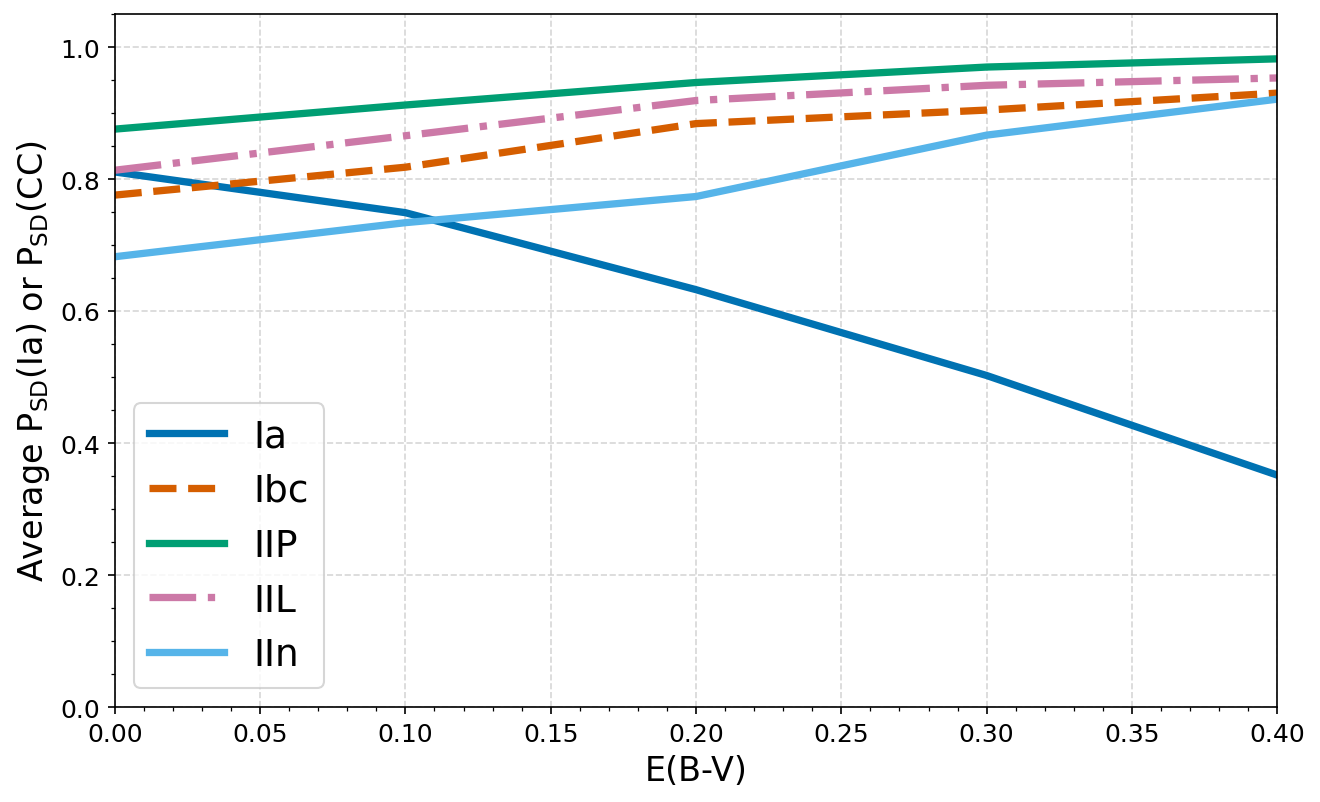} }
{\includegraphics[width=0.49\linewidth]{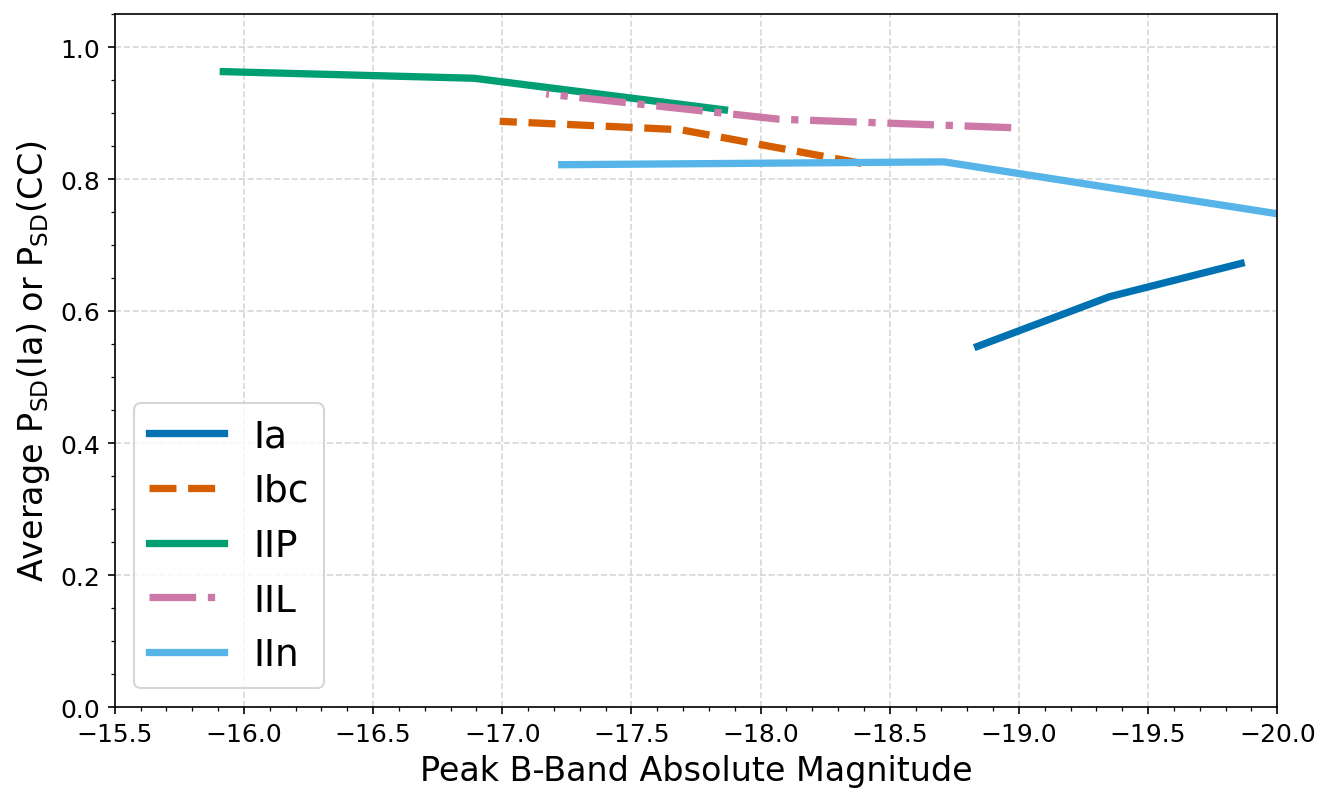} }
    \caption{Average output P$_\mathrm{SD}$(Ia) or P$_\mathrm{SD}$(CC) for input SNe\,Ia or CC\,SNe, respectively, as a function of input rest-frame phase (top left), E(B$-$V) (top right), and peak M$_\mathrm{B}$ (bottom). Phase is relative to explosion for CC subtypes and relative to peak for SNe\,Ia. Here, we show results from the mock SN sample with spectroscopic redshifts. The mock SN sample with photometric redshifts exhibits the same basic trends. 
    }
    \label{fig:classification_accuracy_vs_phase_ebv_peak}
\end{figure*}

Figure \ref{fig:classification_accuracy_vs_phase_ebv_peak} shows the average P$_\mathrm{SD}$(Ia) outputs for input SNe\,Ia and the average P$_\mathrm{SD}$(CC) outputs for each input CC\,SN subtype as a function of rest-frame phase (top left panel), E(B$-$V) (top right panel), and peak M$_\mathrm{B}$ (bottom panel). As seen in the top left panel, phase has a major effect on the average P$_\mathrm{SD}$(Ia) outputs for input SNe\,Ia. Average P$_\mathrm{SD}$(Ia) rises from P$_\mathrm{SD}$(Ia)\,$\sim$\,0.40 at $\sim$10 rest-frame days prepeak to P(Ia)$_\mathrm{SD}$\,$\sim$\,0.95 at $\sim$10 rest-frame days postpeak (phase\,$=$\,0 corresponds to peak for SNe\,Ia in Figure \ref{fig:classification_accuracy_vs_phase_ebv_peak}). Average P$_\mathrm{SD}$(Ia) declines beyond $\sim$10 days postpeak, dropping below P$_\mathrm{SD}$(Ia)\,$\sim$0.60 around $\sim$40 rest-frame days postpeak. These results are not surprising, as SNe\,Ia generally achieve a higher brightness at and around their peak than most CC\,SNe reach at any phase of their evolution (except for SNe\,IIn), making SNe\,Ia more uniquely identifiable at and around their peak. Average P$_\mathrm{SD}$(Ia) drops at phases when the SNe\,Ia are fainter and within the typical brightness ranges for normal CC\,SNe. Another reason why average P$_\mathrm{SD}$(Ia) drops at later phases is that the \texttt{SALT3-NIR} model (the only SN\,Ia model in our template library) extends only to 50 rest-frame days postpeak. This means that an input SN\,Ia that is beyond 50 rest-frame days postpeak could not be fit as an SN\,Ia with the correct phase, making it systematically less likely to be correctly classified as an SN\,Ia (Section \ref{subsubsec:salt3-nir}).

There are also some notable trends regarding the average P$_\mathrm{SD}$(CC) outputs for input CC\,SNe. SNe\,Ib/c are most commonly misclassified as SNe\,Ia (i.e., their average P$_\mathrm{SD}$(CC) output is lowest) at $\sim$10--20 days post-explosion, which roughly corresponds to peak (phase\,$=$\,0 corresponds to explosion time for CC\,SNe in Figure \ref{fig:classification_accuracy_vs_phase_ebv_peak}). SNe\,II are also most often misclassified as SNe\,Ia when they are around their peak brightness, and their average P$_\mathrm{SD}$(CC) output rises at later phases. 

As seen in the top right panel of Figure \ref{fig:classification_accuracy_vs_phase_ebv_peak}, average P$_\mathrm{SD}$(Ia) output for input SNe\,Ia decreases strongly and approximately linearly with increasing E(B$-$V). Input SNe\,Ia that are not reddened are classified as SNe\,Ia with an average output of P$_\mathrm{SD}$(Ia)\,$\sim$\,0.80, whereas SNe\,Ia experiencing moderate reddening (E(B$-$V)\,$=$\,0.4) are classified as SNe\,Ia with an average output of only P$_\mathrm{SD}$(Ia)\,$\sim$\,0.35. CC\,SNe show the opposite trend. For each CC\,SN subtype, average P$_\mathrm{SD}$(CC) output increases roughly linearly with increasing E(B$-$V). Both CC\,SNe and SNe\,Ia are increasingly likely to be classified (or misclassified) as CC\,SNe as their color excess increases.

The bottom panel of Figure \ref{fig:classification_accuracy_vs_phase_ebv_peak} shows that average P$_\mathrm{SD}$(CC) output for input CC\,SNe is only slightly affected by the input CC\,SN's peak M$_\mathrm{B}$, with average output P$_\mathrm{SD}$(CC) decreasing for brighter peaks. The average P$_\mathrm{SD}$(Ia) output for input SNe\,Ia, on the other hand, increases for brighter peaks. This can be related back to the earlier discussion regarding average P$_\mathrm{SD}$(Ia) output vs phase: when SNe\,Ia are at their brightness phases, they are classified most accurately because SNe\,Ia are generally brighter than their CC\,SN counterparts. Similarly, SNe\,Ia that achieve higher peak brightnesses are more distinct from their fainter CC\,SN counterparts, making them more likely to be classified as SNe\,Ia. 

Figure \ref{fig:classification_accuracy_vs_phase_ebv_peak} clearly demonstrates how SN characteristics like rest-frame phase, E(B$-$V), and peak M$_\mathrm{B}$ affect \texttt{STARDUST2's} ability to accurately classify SNe\,Ia and CC\,SNe. In the next section, we explore common misclassification scenarios for single-SED SNe in more depth.


\subsection{Common Single-SED Misclassification Scenarios} \label{subappendix:deriving_parameters}

For transient searches like JTS where many SNe are observed just once, we can obtain robust constraints on SN redshift from their assigned hosts. However, it is very difficult to obtain robust constraints on SN phase because determining an SN's phase generally requires a well-sampled light curve. Here, we explore how accurately \texttt{STARDUST2} derived phase for our mock single-SED SNe, and we examine common phases for which CC\,SN and SN\,Ia SEDs are misclassified, with examples shown.

\begin{figure*}
    \centering
{\includegraphics[width=1\linewidth]{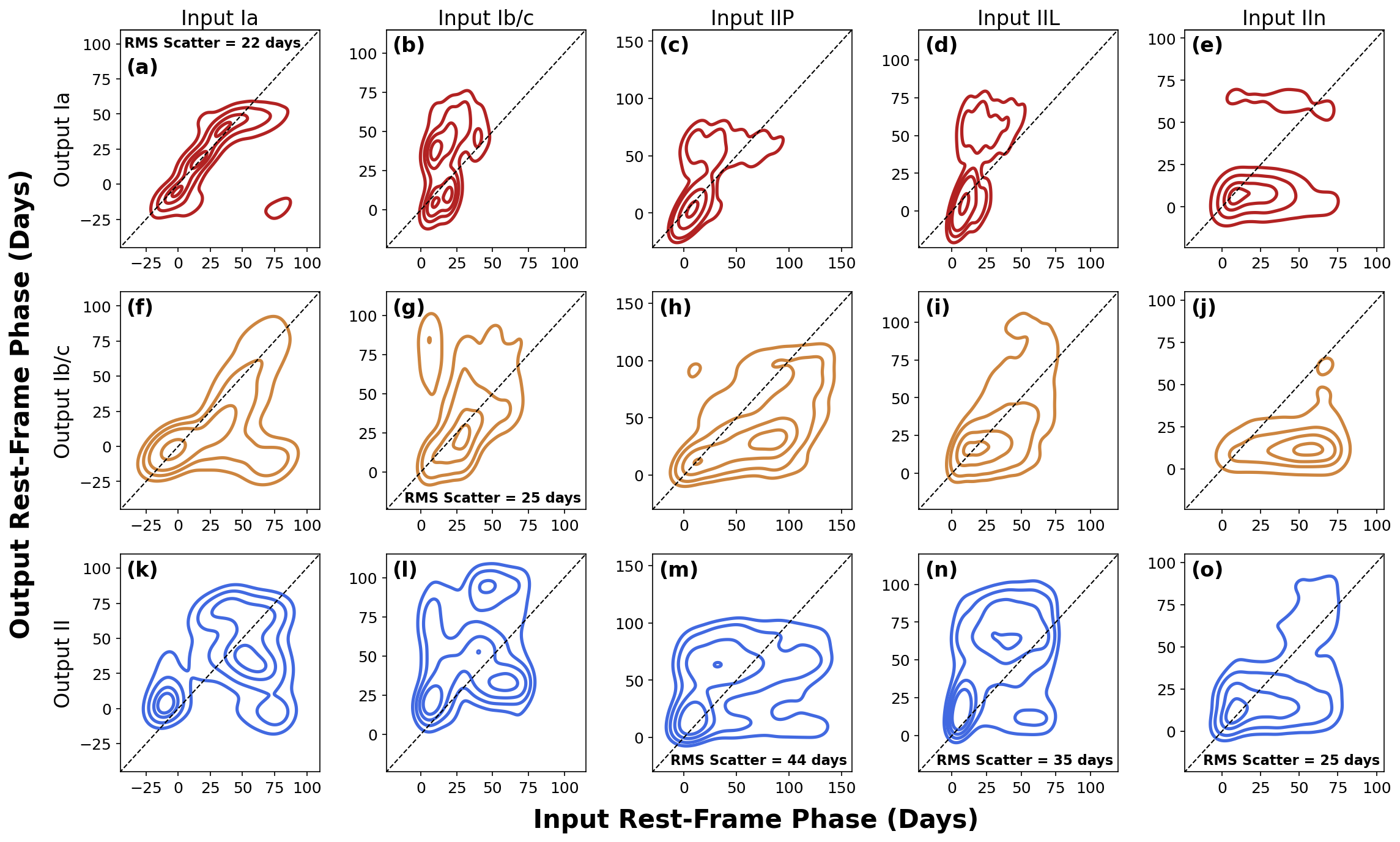} }
    \caption{Contours of the output vs input rest-frame phase (days) for the mock SNe. Each column shows the results for a different input SN subtype (from left to right: input SNe\,Ia, SNe\,Ib/c, SNe\,IIP, SNe\,IIL, SNe\,IIn), and each row shows the results for a different output SN type (from top to bottom: SNe\,Ia, SNe\,Ib/c, SNe\,II). Here, we define the ``output" as the SN type for which \texttt{STARDUST2} returned the highest probability, and we use the phase associated with the best-fit model of the output SN type. 
    Output SN\,Ia, SN\,Ib/c, and SN\,II contours are shown in red, gold, and blue, respectively. The panels that show the same input vs output SN type display the root-mean-square (RMS) scatter around the 1:1 input vs output phase line as a proxy for how well \texttt{STARDUST2} constrains SN phase. The contours levels are 10\%, 30\%, 60\%, and 90\%. For the ``input Ia" panels (leftmost column), the input and output phases are relative to peak. For the ``input CC" panels (rightmost four columns), the input and output phases are relative to explosion time. In some cases, the CC phases go below 0 (i.e., before explosion time) due to contour smoothing; it is not possible for the CC fits to actually have negative phases relative to explosion time.
    }
    \label{fig:input_vs_output_phase}
\end{figure*}

\begin{figure*}
    \centering
{\includegraphics[width=1\linewidth]{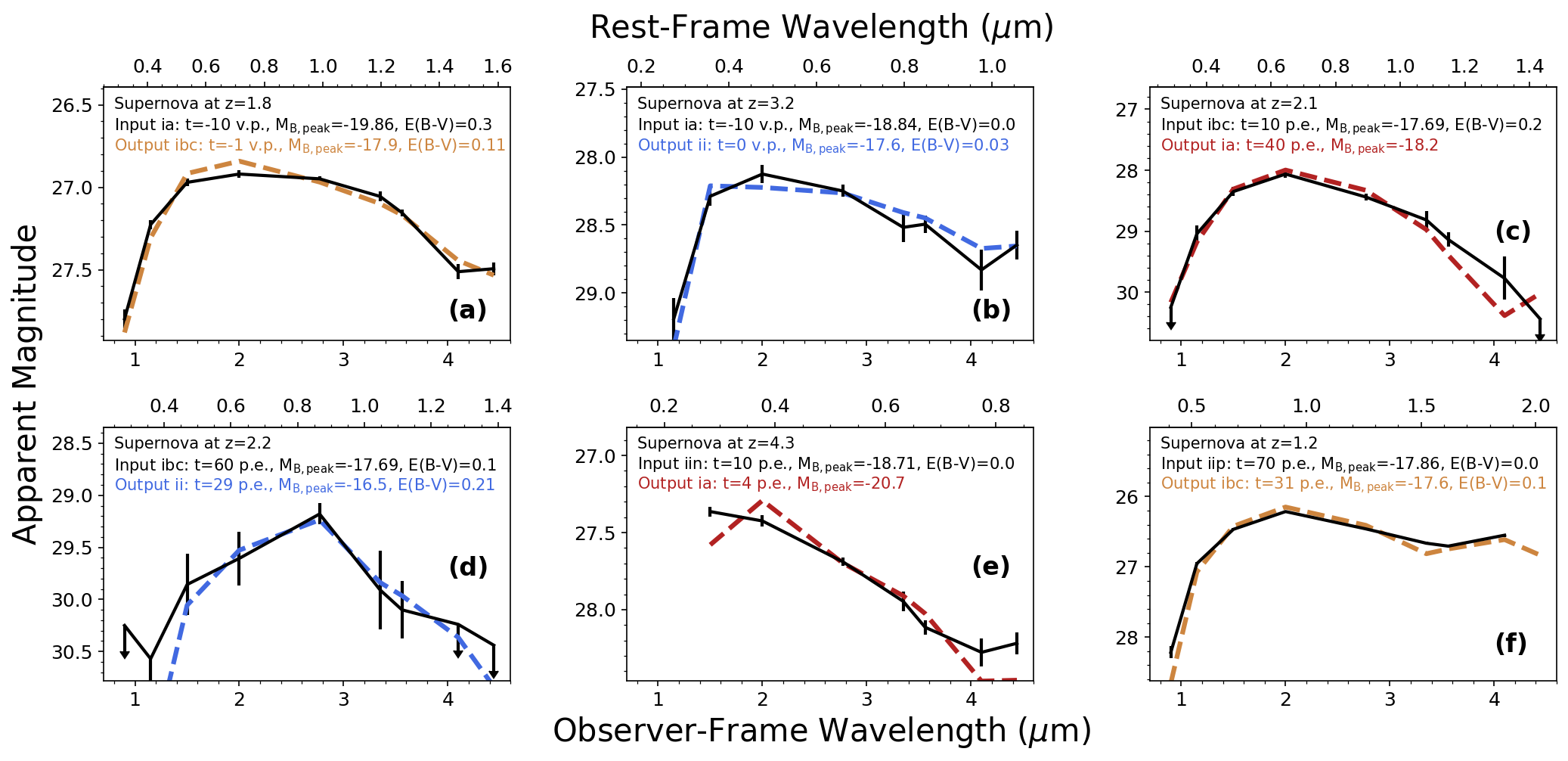} }
    \caption{Examples of the best-fit output SED compared to the input SED for common SN misclassification scenarios in our \texttt{STARDUST2} analysis. The solid black lines show the input SED (where downward arrows indicate upper limits and vertical lines indicate photometric uncertainty), and the colored dashed lines show the best-fit output SED. Output SN\,Ia, SN\,Ib/c, and SN\,II SEDs are shown in red, gold, and blue, respectively. In each panel, we list the input SN type, phase, peak M$_\mathrm{B}$, and E(B$-$V) in black, and we list the output SN parameters in the color corresponding to the output type. With regards to phase, ``v.p." stands for ``versus peak" and ``p.e." stands for ``postexplosion." The following common misclassification scenarios are shown: (a) prepeak SN\,Ia misclassified as a near-peak SN\,Ib/c, (b) prepeak SN\,Ia misclassified as an SN\,II at peak, (c) near-peak SN\,Ib/c misclassified as a postpeak SN\,Ia, (d) late-phase SN\,Ib/c misclassified as a postpeak SN\,II, (e) prepeak SN\,IIn misclassified as a bright prepeak SN\,Ia, (f) late-phase SN\,IIP misclassified as a postpeak SN\,Ib/c.
    }
    \label{fig:input_vs_output_seds}
\end{figure*}

Figure \ref{fig:input_vs_output_phase} displays contours of the input vs output rest-frame phase for every combination of input subtype (SN\,Ia, SN\,Ib/c, SN\,IIP, SN\,IIL, SN\,IIn) and output type (SN\,Ia, SN\,Ib/c, SN\,II). For example, panel f shows the phases of input mock SNe\,Ia on the x-axis and corresponding output SN\,Ib/c phases on the y-axis. The only SNe contributing to these contours are input SNe\,Ia that were classified with the highest probability as SNe\,Ib/c, and the corresponding SN\,Ib/c phases come from the best SN\,Ib/c fits. Below, we briefly discuss some of the input vs output scenarios displayed in Figure \ref{fig:input_vs_output_phase}. In some cases, we refer to Figure \ref{fig:input_vs_output_seds}, which shows examples of input SED versus best-fit output SED for common misclassification scenarios.

\begin{itemize}

\item \textit{Input SN\,Ia, output SN\,Ia (Figure \ref{fig:input_vs_output_phase}a)}: When \texttt{STARDUST2} correctly classifies input SNe\,Ia, the input and best-fit output phases are linearly correlated with a root-mean-square (RMS) scatter of $\sim$22 rest-frame days. 

\item \textit{Input SN\,Ia, output SN\,Ib/c (Figure \ref{fig:input_vs_output_phase}f)}: Prepeak/near-peak SNe\,Ia are commonly misclassified as near-peak SNe\,Ib/c. We show an example of an input reddened prepeak SN\,Ia SED that \texttt{STARDUST2} misclassified as a near-peak SN\,Ib/c in Figure \ref{fig:input_vs_output_seds}a.

\item \textit{Input SN\,Ia, output SN\,II (Figure \ref{fig:input_vs_output_phase}k)}: 
SNe\,Ia are most often misclassified as SNe\,II when they are prepeak. Figure \ref{fig:input_vs_output_seds}b shows an example of a prepeak SN\,Ia SED being misclassified as an SN\,II at peak.

\item \textit{Input SN\,Ib/c, output SN\,Ib/c (Figure \ref{fig:input_vs_output_phase}g)}: When \texttt{STARDUST2} correctly classifies SNe\,Ib/c, the RMS scatter of the input vs best-fit output phase is $\sim$25 rest-frame days. 

\item \textit{Input SN\,Ib/c, output SN\,Ia (Figure \ref{fig:input_vs_output_phase}b)}: Near-peak SNe\,Ib/c are commonly misclassified as near-peak or postpeak SNe\,Ia. Figure \ref{fig:input_vs_output_seds}c shows a near-peak SN\,Ib/c SED that was misclassified as a postpeak SN\,Ia. 

\item \textit{Input SN\,Ib/c, output SN\,II (Figure \ref{fig:input_vs_output_phase}l)}: The top right panels of Figures \ref{fig:confusion_matrices_zspec} and \ref{fig:confusion_matrices_zphot} show that \texttt{STARDUST2} misclassified roughly half of the input mock SNe\,Ib/c as SNe\,II. As seen in Figure \ref{fig:input_vs_output_phase}l, near-peak/postpeak SNe\,Ib/c can be confused with near-peak/postpeak SNe\,II, with an example shown in Figure \ref{fig:input_vs_output_seds}d. In this example, \texttt{STARDUST2} misclassifies a slightly reddened SN\,Ib/c that is 60 rest-frame days postexplosion as a moderately reddened SN\,II that is 29 rest-frame days postexplosion. 

\item \textit{Input SN\,II, output SN\,II (Figure \ref{fig:input_vs_output_phase}m, n, o)}: When \texttt{STARDUST2} correctly classifies SNe\,IIP, SNe\,IIL, and SNe\,IIn, the RMS scatter on the input vs best-fit output phases are 44, 35, and 25 days, respectively. In all three cases, \texttt{STARDUST2} derives phase most accurately when the input SN\,II is near or at peak, although there are cases where near-peak SNe\,IIP and SNe\,IIL are mistaken as postpeak SNe\,II. For all three subtypes, the best-fit output phase is largely unconstrained when the input SN\,II is late-phase.

\item \textit{Input SN\,II, output SN\,Ia (Figure \ref{fig:input_vs_output_phase}c, d, e)}: It is generally uncommon for input SNe\,IIP and SNe\,IIL to be misclassified as SNe\,Ia, but SNe\,IIn are more often misclassified as SNe\,Ia (see the bottom panel of Figures \ref{fig:confusion_matrices_zspec} and \ref{fig:confusion_matrices_zphot}). Figure \ref{fig:input_vs_output_phase}e shows contours for the input SN\,IIn phase vs best-fit output SN\,Ia phase (relative to explosion time). The most common misclassification scenario involves prepeak/near-peak SNe\,IIn being misclassified as prepeak/near-peak SNe\,Ia. We show an example of this in Figure \ref{fig:input_vs_output_seds}e. Here, the input SED is an SN\,IIn that is 10 days postexplosion, and the best-fit output SED is a bright SN\,Ia that is 4 days postexplosion.

\item \textit{Input SN\,II, output SN\,Ib/c (Figure \ref{fig:input_vs_output_phase}h, i, j)}: It is common for input SNe\,II to be misclassified as SNe\,Ib/c (see top right panel of Figures \ref{fig:confusion_matrices_zspec} and \ref{fig:confusion_matrices_zphot}). Figure \ref{fig:input_vs_output_phase}h compares input SN\,IIP phases to the corresponding best-fit output SN\,Ib/c phases. Late-phase SNe\,IIP are commonly misclassified as postpeak SNe\,Ib/c, and an example of this is shown in Figure \ref{fig:input_vs_output_seds}f. In this example, the SED of an SN\,IIP that is 70 days postexplosion is misclassified as a slightly reddened SN\,Ib/c that is 31 days postexplosion. 

\end{itemize}

\onecolumngrid
\section{Correcting for Single-SED Misclassification: Bayesian Method} \label{appendix:bayesian_method}

Here, we explain the first method we used to derive and apply misclassification correction factors to the classification probabilities of the single-SED JTS sources. As explained in Section \ref{subsec:results_cc_vs_ia}, we constructed four-component confusion matrices detailing CC\,SN vs SN\,Ia classification accuracy. In order to characterize how CC\,SN vs SN\,Ia classification accuracy changes with redshift, we generated redshift-binned confusion matrices in the 0.7\,$\leq$\,$z$\,$\leq$\,5 range with steps of $\delta_z$\,$=$\,0.1. 
To account for potential misclassification of single-SED JTS sources, we use Bayes's Theorem to calculate updated probabilities for these sources that are informed by our redshift-binned confusion matrices:



\begin{equation} \label{eq:ia}
    \mathrm{P(}\mathrm{Ia}\vert \mathrm{Ia'}\mathrm{)} = \frac{\mathrm{TPR}_\mathrm{Ia}\, \mathrm{P}_\mathrm{SD} \mathrm{(Ia})}{\mathrm{P}_\mathrm{SD}\mathrm{(Ia)} \, \mathrm{TPR}_\mathrm{Ia} \, + \, \mathrm{[}1 - \mathrm{P}_\mathrm{SD} \mathrm{(Ia)]} \, \mathrm{FPR}_\mathrm{Ia}}
\end{equation}

\begin{equation} \label{eq:cc}
    \mathrm{P(}\mathrm{CC}\vert \mathrm{CC'}\mathrm{)} = \frac{\mathrm{TPR}_\mathrm{CC}\, \mathrm{P}_\mathrm{SD} \mathrm{(CC})}{\mathrm{P}_\mathrm{SD}\mathrm{(CC)} \, \mathrm{TPR}_\mathrm{CC} \, + \, \mathrm{[}1 - \mathrm{P}_\mathrm{SD} \mathrm{(CC)]} \, \mathrm{FPR}_\mathrm{CC}}
\end{equation}
where
\begin{itemize}
    \item P(Ia$\vert$Ia$^\prime$) is the probability that a source classified as an SN\,Ia by \texttt{STARDUST2} is a true SN\,Ia.
    \item P$_\mathrm{SD}$(Ia) is the output SN\,Ia probability from \texttt{STARDUST2}.
    \item TPR$_\mathrm{Ia}$ is the ``true positive rate" of SN\,Ia classification by \texttt{STARDUST2}. This is the ``assigned Ia given Ia" component of the CC\,SN vs SN\,Ia confusion matrix.
    \item FPR$_\mathrm{Ia}$ is the ``false positive rate" of SN\,Ia classification by \texttt{STARDUST2}. This is the ``assigned Ia given CC" component of the CC\,SN vs SN\,Ia confusion matrix. It is equivalent to the ``false negative rate" of input CC\,SNe (1$-$TPR$_\mathrm{CC}$).
    \item P(CC$\vert$CC$^\prime$) is the probability that a source classified as a CC\,SN by \texttt{STARDUST2} is a true CC\,SN.
    \item P$_\mathrm{SD}$(CC) is the output CC\,SN probability from \texttt{STARDUST2}.
    \item TPR$_\mathrm{CC}$ is the ``true positive rate" of CC\,SN classification by \texttt{STARDUST2}. This is the ``assigned CC given CC" component of the CC\,SN vs SN\,Ia confusion matrix.
    \item FPR$_\mathrm{CC}$ is the ``false positive rate" of CC\,SN classification by \texttt{STARDUST2}. This is the ``assigned CC given Ia" component of the CC\,SN vs SN\,Ia confusion matrix. It is equivalent to the ``false negative rate" of input SNe\,Ia (1$-$TPR$_\mathrm{Ia}$).
\end{itemize}

The four components of the redshift-binned confusion matrices are TPR$_\mathrm{Ia}$, FPR$_\mathrm{Ia}$, TPR$_\mathrm{CC}$, and FPR$_\mathrm{CC}$. For each single-SED JTS source, we 
identify the two confusion matrices associated with the redshifts nearest to the source redshift. We then linearly interpolate the confusion matrix components to the exact source redshift. 
These four components are then input into Equations \ref{eq:ia} and \ref{eq:cc} to compute misclassification-corrected P(Ia$\vert$Ia$^\prime$) and P(CC$\vert$CC$^\prime$) values. These updated probabilities should be interpreted as the probability that the source is truly an SN\,Ia given that \texttt{STARDUST2} classifies it as an SN\,Ia and the probability that the source is truly a CC\,SN given that \texttt{STARDUST2} classifies it as a CC\,SN, respectively. We then renormalize P(Ia$\vert$Ia$^\prime$) and P(CC$\vert$CC$^\prime$) to sum to 1 to preserve the total number of observed SNe included in the combined SN\,Ia and CC\,SN rates. 


\onecolumngrid
This Bayesian method of misclassification correction, however, was not very effective because most of the corrected CC\,SN and SN\,Ia probabilities ended up being the same as their corresponding initial probabilities. This was because most of the prior probabilities (i.e., output probabilities from \texttt{STARDUST2}) were P$_\mathrm{SD}$(Ia)\,$=$\,0 and P$_\mathrm{SD}$(CC)\,$=$\,1 (or vice-versa). Bayes's Theorem cannot mathematically change 0\% and 100\% results, so it was ineffective in providing ``corrected" probabilities. This is not a shortcoming of Bayes's Theorem, but rather it points to an issue with our prior probabilities. It is concerning that most of the single-SED sources were classified with P$_\mathrm{SD}$(CC)\,$=$\,1. We discuss this in detail in Section \ref{subsubsec:probability_problem}. Because Bayes's Theorem could not provide updated probabilities for most of the single-SED JTS sources, we do not present the ``misclassification-corrected" rates that arise from this correction method.


\onecolumngrid
\section{Correcting for Single-SED Misclassification: Frequentist Method} \label{appendix:frequentist_method}

Upon discovering the issues with the Bayesian single-SED classification correction method, we explored a frequentist method of misclassification correction. This method used the average TPR of CC\,SN and SN\,Ia classification (TPR$_\mathrm{CC,avg}$ and TPR$_\mathrm{Ia,avg}$, respectively) associated with the redshift bins used for the rate calculations, listed in Tables \ref{tab:ccsn_rates} and \ref{tab:snia_rates}.
We applied the following system of linear equations to each redshift bin to convert the observed number of single-SED CC\,SNe and SNe\,Ia in that bin to the true number of single-SED CC\,SNe and SNe\,Ia that would have existed in that bin in the absence of misclassification:

\begin{equation}\label{eq:freq_cc}
\mathrm{O}_{\mathrm{CC}} = \mathrm{TPR}_\mathrm{CC,avg} \times \rm{T}_{\rm{CC}} + (1-\mathrm{TPR}_\mathrm{Ia,avg}) \times \mathrm{T}_{\mathrm{Ia}}
\end{equation}
\begin{equation} \label{eq:freq_ia}
\mathrm{O}_{\mathrm{Ia}} = (1-\mathrm{TPR}_\mathrm{CC,avg}) \times \mathrm{T}_{\mathrm{CC}} + \mathrm{TPR}_\mathrm{Ia,avg} \times \mathrm{T}_{\mathrm{Ia}}
\end{equation}
Here, O$_{\mathrm{CC}}$ and O$_{\mathrm{Ia}}$ denote the observed numbers of single-SED CC\,SNe and SNe\,Ia in a given redshift bin, respectively. The quantities T$_{\mathrm{CC}}$ and T$_{\mathrm{Ia}}$ represent the underlying true numbers of single-SED CC\,SNe and SNe\,Ia that would exist in the redshift bin in the absence of misclassification.
Equation \ref{eq:freq_cc} therefore states that the observed single-SED CC\,SN count is composed of correctly classified single-SED CC\,SNe (TPR$_\mathrm{CC,avg} \times \mathrm{T}_{\mathrm{CC}}$) combined with contributions from misclassified single-SED SNe\,Ia ($(1-\mathrm{TPR}_{\mathrm{Ia,avg}}) \times \mathrm{T}_{\mathrm{Ia}}$). Equation \ref{eq:freq_ia} similarly expresses the observed single-SED SN\,Ia count as the sum of misclassified single-SED CCSNe ($(1-\mathrm{TPR}_{\mathrm{CC,avg}}) \times \mathrm{T}_{\mathrm{CC}}$) and correctly classified single-SED SNe\,Ia (TPR$_{\mathrm{Ia,avg}} \times \mathrm{T}_{\mathrm{Ia}}$). Solving this system of equations allows the true underlying numbers of single-SED CC\,SNe and SNe\,Ia (T$_\mathrm{CC}$ and T$_\mathrm{Ia}$) to be recovered from the observed counts and the average classification TPRs in each redshift bin.


In some of the CC\,SN redshift bins, the solution for T$_{\mathrm{Ia}}$ becomes negative, which is not physically meaningful. This occurs when the observed CC\,SN count and the derived TPR$_\mathrm{CC,avg}$ imply that more CCSNe have been misclassified as SNe\,Ia than there are observed SNe\,Ia in that redshift bin. In such cases the equations formally require a negative number of true SNe\,Ia in order to satisfy the system.

To estimate statistical uncertainties, we assigned Poisson uncertainties to O$_{\mathrm{Ia}}$ and O$_{\mathrm{CC}}$ following \citet{gehrels1986}. These propagated into very large statistical uncertainties in the inferred true counts.
Additional systematic uncertainty arose from the variations in the TPR$_\mathrm{CC}$ and TPR$_\mathrm{Ia}$ values across each redshift bin. In our initial calculations, we assumed the average TPR$_\mathrm{CC}$ and TPR$_\mathrm{Ia}$ values for each redshift bin, but some redshift bins exhibit significant TPR variations (see Figure \ref{fig:classification_accuracy_vs_z}). To estimate the impact of these variations, we computed T$_{\mathrm{Ia}}$ and T$_{\mathrm{CC}}$ using the minimum and maximum TPR$_{\mathrm{Ia}}$ and TPR$_{\mathrm{CC}}$ values in each redshift bin, with the resulting range providing an estimate of the systematic uncertainty. In all cases, both the statistical and systematic uncertainties were comparable to or larger than the derived true counts themselves.
Therefore, the frequentist misclassification correction did not provide useful constraints on the CC\,SN and SN\,Ia classifications. For this reason, we do not report the misclassification-corrected rates derived using this method.


\onecolumngrid
\section{Details of the Visibility Window Calculation} \label{appendix:boomrate}

\begin{figure*}
    \centering
{\includegraphics[width=16cm]{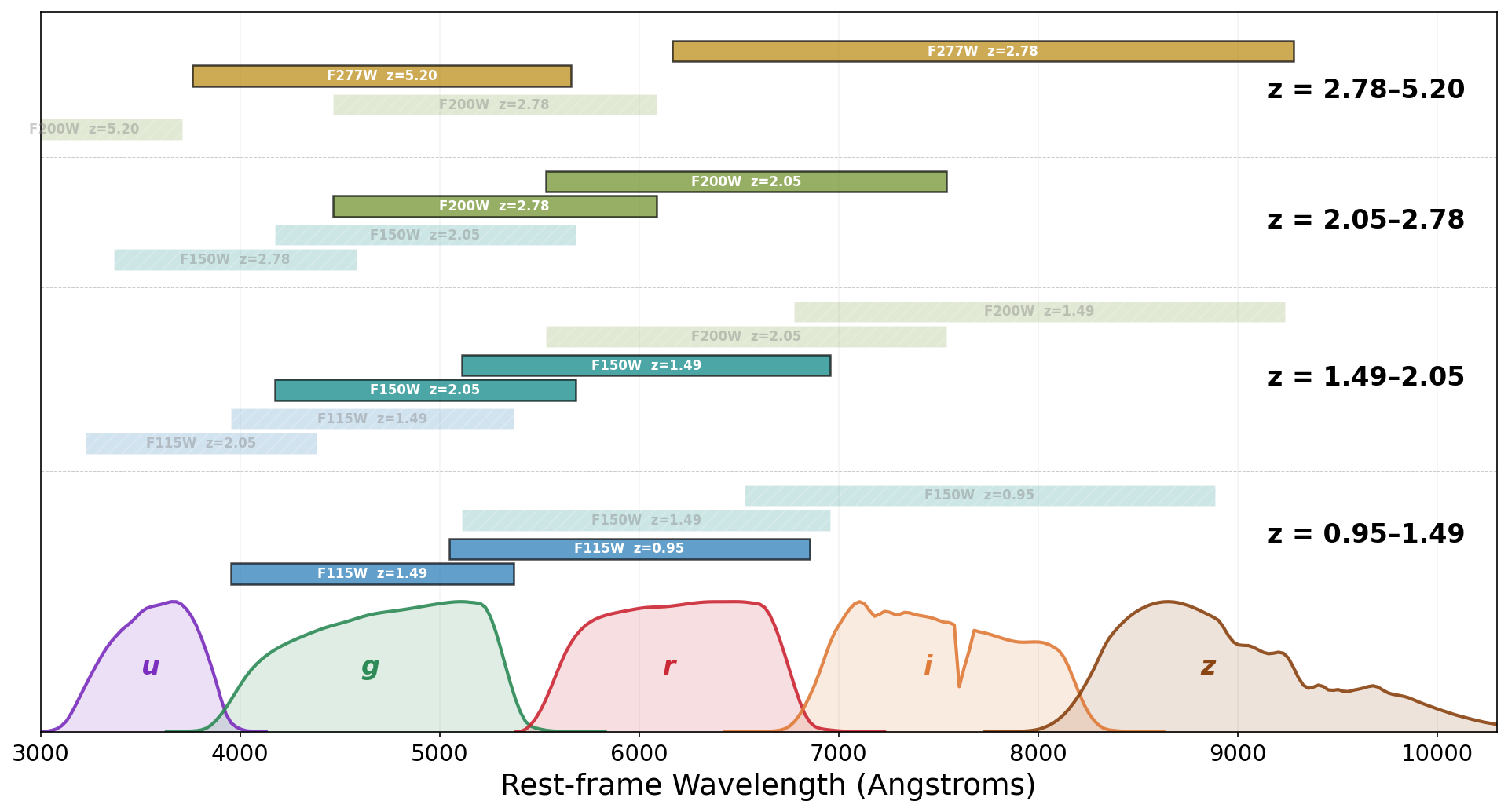}}
\caption{The rest-frame coverage of the observed NIRCam filters at the redshift bin edges for which they were evaluated in the CC\,SN full sample visibility window calculation. 
The NIRCam filters used for the calculation are shown as dark colored bars, whereas NIRCam filters that were considered for but not used in the calculation are shown as faint bars. The redshift bins corresponding to the colored and faint bars are shown to the right of the bars, as well as within the bars. The rest-frame SDSS filter coverages are shown as colored curves along the bottom of the plot, with the rest-frame wavelengths labeled on the x-axis.}
\label{fig:filter_coverage}
\end{figure*}

In the visibility window calculation, a synthetic light curve is generated that represents an SN of a given subtype at redshift $z$. The light curve is only created in one filter, so the observer-frame filter for which the light curve is generated is selected such that its rest-frame coverage closely corresponds a rest-frame Sloan Digital Sky Survey (SDSS) filter. For example, F115W closely corresponds to rest-frame $r$-band at $z$\,$=$\,0.95 and rest-frame $g$-band at $z$\,$=$\,1.49, so we generate F115W light curves for the $z$\,$=$\,0.95--1.49 bin. Figure \ref{fig:filter_coverage} shows the observer-frame NIRCam filters that have been selected for each redshift bin for the CC\,SN full sample. It compares the rest-frame coverage of the NIRCam filters at the redshift bin edges to the rest-frame coverage of the SDSS filters, justifying the NIRCam filter selection for each bin. We selected NIRCam filters that most closely match rest-frame $g$- or $r$-band at the redshift bin edges. This was possible for every redshift bin except for the $z$\,$=$\,2.78--5.20 bin, where F277W most closely matches $i$-band in the rest frame at $z$\,$=$\,2.78. Similar choices were made for the CC\,SN gold sample and SN\,Ia full and gold samples. The observer-frame filters selected for every redshift bin for the CC\,SN (SN\,Ia) full and gold samples are listed in Table \ref{tab:ccsn_rates} (Table \ref{tab:snia_rates}).

The CC\,SN and SN\,Ia light curves are initially constructed in the rest-frame using the \texttt{nugent} SN models \citep{gilliland1999, nugent2002, levan2005}, in the SDSS filter that most closely corresponds to the observed NIRCam filter at redshift $z$. Each light curve is then anchored to the peak absolute magnitude for its subtype \citep{richardson2014}. Because these peak absolute magnitudes are reported in the B-band rather than the SDSS filters we use, a color correction is required. We measure the peak color difference between the relevant SDSS filter and the B-band from the \texttt{nugent} model SEDs at peak, apply this offset to the \citet{richardson2014} B-band value, and anchor the SDSS light curve to the resulting color-corrected peak.

A k-correction is then applied to convert the rest-frame SDSS light curve to the observed JWST filter at redshift $z$, after host galaxy extinction is applied to the rest-frame light curve. We step through this observed light curve from explosion to 400 rest-frame days postexplosion to determine the SN brightness at each phase. We compare this SN flux to an empirically-derived detection efficiency curve for the given NIRCam filter to determine the likelihood of detecting the SN under those conditions. This likelihood, along with the likelihood of the sampled host galaxy extinction and the sampled peak magnitude, is used to iteratively determine the visibility window for the SN subtype at the given redshift (see Equation \ref{eq:visibility_window_dm_da}).

\begin{figure*}
    \centering
{\includegraphics[width=12cm]{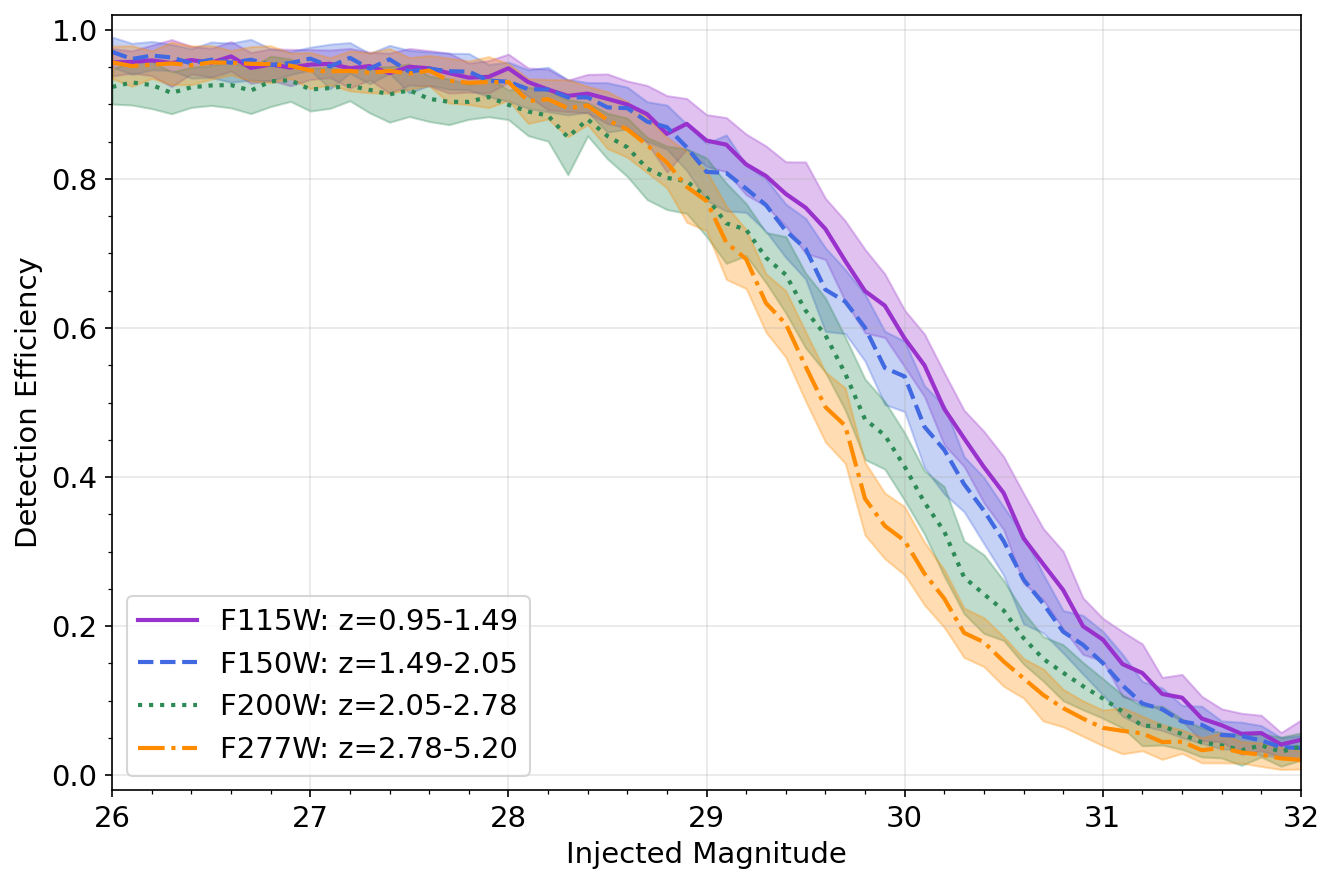}}
\caption{Average detection efficiency as a function of injected point source magnitude for the specific observed NIRCam filter/redshift bin combinations corresponding to the CC\,SN full sample. Similar detection efficiency curves exist for the filter/redshift bin combinations associated with the CC\,SN gold sample, SN\,Ia full sample, and SN\,Ia gold sample. The shaded regions indicate the 1$\sigma$ uncertainties on the average detection efficiencies.}
\label{fig:detection_efficiency}
\end{figure*}

The detection efficiency curves were generated from point source injection/recovery simulations in the JTS difference images. We built detection efficiency curves for every NIRCam filter and redshift bin combination that existed among the CC\,SN full, CC\,SN gold, SN\,Ia full, and SN\,Ia gold samples. For example, for the $z$\,$=$\,0.95--1.49 CC\,SN full sample redshift bin, we generated an F115W detection efficiency curve by injecting and recovering point sources around randomly-selected $z$\,$=$\,0.95--1.49 galaxies in the JTS F115W difference image. The detection efficiency curves corresponding to the four CC\,SN full sample redshift bins are shown in Figure \ref{fig:detection_efficiency}.

We randomly assign CC\,SN point source injection positions relative to the host galaxy center using an exponential decay probability distribution with $\lambda$\,$=$\,1.50 half-light radii, truncated at 6 half-light radii. This places most injected sources near bright host centers, where we expect high levels of star formation. We also expect poorer image subtractions near host centers, causing an artificial but non-negligible decrease in detection efficiency. These poor host subtractions are more prevalent for bright and complex galaxies at low-$z$, underscoring the importance for independently evaluating detection efficiency for each redshift bin considered in the rate calculation. For the SN\,Ia point source injections, we randomly assign injection positions relative to host center with an n\,$=$\,2 S\'ersic profile, truncated at 6 half-light radii. We recover the point sources with \texttt{DAOStarFinder} with the parameters described in Section 3 of \citet{decoursey2025_jts}. This injection/recovery scheme is conducted for $m$\,$=$\,26--32, with $\Delta m$\,$=$\,0.1. For each magnitude, we perform 20 iterations of 100 point sources injections/recoveries and adopt the average recovery fraction as the detection efficiency. For the visibility window calculation, we parameterize the detection efficiency curves as sigmoid functions:

\begin{equation} \label{eq:detection_efficiency}
\epsilon = \frac{T}{1+e^\frac{m - m_{50}}{S}},
\end{equation}
where $T$ is the maximum detection efficiency achieved at the bright end of the magnitude distribution, $m$ is the source magnitude, $m_{50}$ is the magnitude for which 50\% of injected point sources are recovered, and $S$ is the scale parameter that controls the steepness of the sigmoid transition. We adopt the best-fit $T$, $m_{50}$, and $S$ parameters from the empirically-measured detection efficiency curves for each filter/redshift combination in the visibility window calculation.

\onecolumngrid
\section{Estimating Rate Systematic Uncertainties} \label{appendix:systematics}

Misclassification between CC\,SNe and SNe\,Ia is a major source of systematic uncertainty in our rate measurements. Because the rates in Tables \ref{tab:ccsn_rates}--\ref{tab:snia_rates} are not corrected for misclassification, they scale with the observed numbers of classified SNe rather than the true numbers. We estimate the resulting systematic uncertainty by forward-modeling the observed counts with Equations \ref{eq:freq_cc} and \ref{eq:freq_ia} over plausible ranges of classification accuracy and intrinsic SN\,Ia and CC\,SN fractions.

We derive the classification accuracy from the CC\,SN vs SN\,Ia confusion matrices of the mock SN sample (top left panels of Figures \ref{fig:confusion_matrices_zspec}--\ref{fig:confusion_matrices_zphot}), binned in redshift over 0.7\,$\leq$\,$z$\,$\leq$\,5.0 with $\delta_z$\,$=$\,0.1. These binned TPRs are analogous to the curves in Figure \ref{fig:classification_accuracy_vs_z}, except that all CC\,SN subtypes are combined into a single class, weighted by their volume-limited fractions (Section \ref{subsec:results_classification}). For each redshift bin used in the rate calculations, we adopt the minimum and maximum TPR$_\mathrm{CC}$ and TPR$_\mathrm{Ia}$ within the bin, including values linearly interpolated to the bin edges. For bins extending beyond $z$\,$=$\,5.0, we adopt the $z$\,$=$\,5.0 values. The spec-$z$ and photo-$z$ confusion matrices are combined, weighted by the fractions of the photometrically-classified SNe in the bin with spectroscopic and photometric host redshifts, respectively.

The intrinsic SN\,Ia fraction, $f_\mathrm{Ia} \equiv \mathrm{T}_\mathrm{Ia}/(\mathrm{T}_\mathrm{CC} + \mathrm{T}_\mathrm{Ia})$, is unconstrained at these redshifts. We therefore consider $0.05 \leq f_\mathrm{Ia} \leq 0.25$. The upper bound approximates the local volume-limited fraction \citep{li2011_fractions}, and the lower bound approximates the ratio of our SN\,Ia and CC\,SN rates at $z$\,$>$\,2. We expect $f_\mathrm{Ia}$ to decline toward high redshift, as the age of the Universe limits the delay times available to SN\,Ia progenitors.

Normalizing $\mathrm{T}_\mathrm{CC} + \mathrm{T}_\mathrm{Ia} = 1$, we evaluate O$_\mathrm{CC}$ and O$_\mathrm{Ia}$ at all eight combinations of the minimum and maximum TPR$_\mathrm{CC}$, TPR$_\mathrm{Ia}$, and $f_\mathrm{Ia}$. For each combination, we compute the fractional bias $b = \mathrm{T}_\mathrm{CC}/\mathrm{O}_\mathrm{CC} - 1$
($\mathrm{T}_\mathrm{Ia}/\mathrm{O}_\mathrm{Ia} - 1$) for the CC\,SN (SN\,Ia) rates, which is the fractional amount by which the true rate exceeds ($b$\,$>$\,0) or falls below ($b$\,$<$\,0) the reported rate. Because $b$ varies monotonically with each parameter, the minimum and maximum values over these combinations, $b_\mathrm{min}$ and $b_\mathrm{max}$, bound the full parameter space. Spectroscopically-classified SNe are not subject to this uncertainty, so we scale $b$ by the photometrically-classified fraction of the bin. In every redshift bin, $b_\mathrm{min}$\,$<$\,0\,$<$\,$b_\mathrm{max}$, so the upper and lower systematic uncertainties are the rate multiplied by the scaled $b_\mathrm{max}$ and $|b_\mathrm{min}|$, respectively. These values are listed in parentheses in Tables \ref{tab:ccsn_rates}--\ref{tab:snia_rates}.

These estimates are subject to three caveats. First, the confusion matrices are derived from single-SED classifications, whereas we apply them to all photometrically-classified SNe, including those with multi-epoch light curves. Multi-epoch classifications draw on more information and therefore should be more accurate, so the resulting uncertainties are conservative for these sources; deriving dedicated multi-epoch confusion matrices is beyond the scope of this work. Second, this method considers only the classified SNe in the rate sample. The effect of omitting unclassified sources, which would increase the rates, is discussed in Sections \ref{subsec:csfrd_comparison}--\ref{subsec:ia_comparison}. Third, these uncertainties account only for misclassification. They do not include other systematic effects, such as the single-filter approximation in the visibility window calculation or the under-density of the JADES Deep Field (Section \ref{subsec:csfrd_comparison}), which cannot be reliably quantified for our redshift bins. The effect of dust-obscured SNe missed by optical/NIR surveys is explored in \citet{vassallo2026}.

\bibliography{references}{}
\bibliographystyle{aasjournalv7}

\end{document}